\documentclass[twocolumn,twocolappendix]{aastex63}
\usepackage[caption=false]{subfig}

\usepackage{savesym}
\savesymbol{tablenum}
\usepackage{siunitx}
\usepackage{comment}
\defcitealias{Trueba2020}{T20}	 	
\restoresymbol{SIX}{tablenum}
\def\kms{$\text{km}$ $\text{s}^{-1}$ }
\def\gmcc{$GM/{c}^{2}$ }
\def\rce{${R}_{CE}$ }
\def\dvce{${\delta V}_{CE}$ }
\submitjournal{ApJ}
\accepted{November 6th, 2021}
\shortauthors{N. Trueba et al.}

\shorttitle{Central Engine Size Scales in Neutron Stars}

\begin{document}

\title{A Spectroscopic Angle on Central Engine Size Scales in Accreting Neutron Stars}

\correspondingauthor{Nicolas Trueba}
\email{ntrueba@umich.edu}

\author{Nicolas Trueba}
\affiliation{Department of Astronomy, University of Michigan, Ann Arbor, MI}
\author{J.~M.~Miller}
\affiliation{Department of Astronomy, University of Michigan, Ann Arbor, MI}
\author{A.~C.~Fabian}
\affiliation{Institute of Astronomy, University of Cambridge, Madingley Road, Cambridge CB3 OHA, UK}
\author{J.~Kaastra}
\affiliation{SRON, Netherlands Institute for Space Research, Sorbonnelaan 2, 3584 CA Utrecht, The Netherlands}
\author{T.~Kallman}
\affiliation{NASA Goddard Space Flight Center, Code 662, Greedbelt, MD 20771, USA}
\author{A.~Lohfink}
\affiliation{Department of Physics, Montana State University, Bozeman, MT 59717-3840, USA}
\author[0000-0002-8961-939X]{R.~M.~Ludlam}\thanks{NASA Einstein Fellow}
\affiliation{Cahill Center for Astronomy and Astrophysics, California Institute of Technology, Pasadena, CA 91125, USA}
\author{D.~Proga}
\affiliation{Department of Physics, University of Nevada, Las Vegas, Las Vegas, NV 89154, USA}
\author{J.~Raymond}
\affiliation{Harvard-Smithsonian Center for Astrophysics, 60 Garden Street, Cambridge, MA 02138, USA}
\author{C.~Reynolds}
\affiliation{Institute of Astronomy, University of Cambridge, Madingley Road, Cambridge CB3 OHA, UK}
\author{M.~Reynolds}
\affiliation{Department of Astronomy, University of Michigan, Ann Arbor, MI}
\author{A.~Zoghbi}
\affiliation{Department of Astronomy, University of Michigan, Ann Arbor, MI}
\affiliation{Department of Astronomy, University of Maryland, College Park, MD 20742, USA}
\affiliation{CRESST II, NASA Goddard Space Flight Center, Greenbelt, MD 20771, USA}

\begin{abstract}
Analyses of absorption from disk winds and atmospheres in accreting compact objects typically treat the central emitting regions in these systems as point sources relative to the absorber. This assumption breaks down if the absorbing gas is located within $few \times 1000\cdot GM/{c}^{2}$, in which case a small component of the absorber's Keplerian motion contributes to the velocity-width of absorption lines. Here, we demonstrate how this velocity-broadening effect can be used to constrain the sizes of central engines in accreting compact objects via a simple geometric relationship, and develop a method for modeling this effect. We apply this method on the Chandra/HETG spectra of three ultra-compact and short period neutron star X-ray binaries in which evidence of gravitationally redshifted absorption, owing to an inner-disk atmosphere, has recently been reported. The significance of the redshift is above $5\sigma$ for XTE J1710$-$281 (this work) and 4U 1916$-$053, and is inconsistent with various estimates of the relative radial velocity of each binary. For our most sensitive spectrum (XTE J1710$-$281), we obtain a 1$\sigma$ upper bound of 310 $\text{km}$ $\text{s}^{-1}$ on the magnitude of this geometric effect and a central engine of size ${R}_{CE} < 60 ~ GM/{c}^{2}$ (or, $< 90 ~ GM/{c}^{2}$ at the $3\sigma$ level). These initial constraints compare favorably to those obtained via microlensing in quasars and approach the sensitivity of constraints via relativistic reflection in neutron stars. This sensitivity will increase with further exposures, as well as the launch of future microcalorimeter and grating missions.
\end{abstract}

\section{Introduction}\label{sec:intro}

Robust observational constraints on the physical size of the central emitting regions of accreting compact objects remain key to much of our understanding of both the nature of the compact object itself (be it a black hole or neutron star) and the accretion process in these systems. Prior to the first detection of gravitational wave emission from a black hole merger \citep{Abbott2016}, observational studies of black holes relied primarily on electromagnetic emission from gas near the event horizon and therefore studies were (and predominantly still are) limited to those actively accreting. Indeed, the compactness of the emitting regions in active galactic nuclei (or, AGN) was key in establishing that AGN emission is powered by accreting supermassive black holes \citep{Lynden1969,Rees1984}. Compactness also played a role in identifying compact object accretion as the mechanisms responsible for the emission in multiple galactic X-ray sources in early X-ray observations \citep{Shklovsky1967}.

The increased sensitivity of X-ray observatories (spectroscopy and timing alike) has led to the development of sophisticated analytical tools capable of constraining the size of these central emitting regions, hereafter referred as central engines, to scales of a few gravitational radii ($GM/{c}^{2}$). Fluorescent X-ray emission lines originating from surface gas in the inner radii of the disk, for instance, are subject to Doppler shifts and boosting due to the Keplerian motion of the disk, as well as general relativistic effects that can be used to map these regions \citep{Fabian1989,George1991}. This technique has been successful in measuring the spins of several stellar mass black holes \citep[e.g.][]{Brenneman2006,MM2015,Draghis2020} by constraining the size of the innermost stable circular orbit (or, ISCO) via spectral fitting of the resulting relativistic profile of Fe fluorescent lines. These are complemented by independent constraints based on fitting the continuum with relativistic disk emission models and sophisticated radiative transfer calculations through the surface atmosphere of the disk \citep{McClintock2014,Miller2009,Reynolds2020}.

Despite the success of these techniques, important questions remain unanswered at the limit of our current sensitivity. Our understanding of the nature of the X-ray corona in both X-ray binary systems and AGN remains, in many respects, limited. Constraining the geometry of the corona will likely be key in determining much about its physical origin \citep{Merloni2001}; although studies suggest the corona is compact \citep{Fabian2015,Fabian2017}, a robust scientific case still requires further constraints from a diverse sample of source types, luminosity/accretion states, and independent observational techniques.    

In this work, we present a novel approach to measuring the size of central engines in accreting compact objects via absorption line widths. Especially in low mass X-ray binaries (or, LMXBs), disk winds and disk atmospheres have been observed in multiple high-inclination black hole (or, BH) and neutron star (or, NS) systems. Based on their ionization and densities, these are known to emerge from the surface of the disk and may originate as close as $few\times1000~GM/{c}^{2}$ from the compact object (see \citealt{Miller2008, Neilsen2012, Miller2015, Trueba2019} for disk winds in BH LMXBs; \citealt{Trueba2020} and \citealt{Ponti2018a} for disk atmospheres in NS ultra-compact X-ray binaries; see \citealt{Roz2011} for numerical models) and, by necessity, retain much of their Keplerian motion as they intercept the observer's line-of-sight. As the separation between the absorber and central engine decreases, the orbital motion of the absorber becomes non-negligible: parts of the emitting area are absorbed by gas in which the Keplerian motion is not entirely orthogonal to the line-of-sight. As a result, parts of the emitting area are absorbed by gas that is slightly \emph{redshifted}, while others are absorbed by slightly \emph{blueshifted} gas.

This differential absorption effect results in a specific form of line broadening, the degree of which depends solely\footnote{There is also a dependence on inclination; however, this effect is minimal in high-inclination sources such as the dipping and eclipsing systems analyzed in this work.} on the orbital radius of the absorber, the local Keplerian velocity, and the size of the central engine. If constraints can be placed on the orbital quantities of the absorber (e.g. via the photoionization parameter, P-Cygni profiles, or gravitational redshift), the size of the central engine can be constrained provided the spectrum is sensitive enough to measure the degree of velocity broadening. 

This technique was first mentioned in \cite{Trueba2020} on the NS ultra-compact X-ray binary (or, UCXB) 4U 1916$-$053, and was facilitated by the independent constraint on the absorbing radius via the gravitational redshift in the disk atmosphere. We note that including the effects of a finite rather than point-like source has been common in other facets of astronomy, such as studies of young stars (e.g. \citealt{Calvet1993}). Similar geometric effects on the continuum in NS binaries have also been subject to study \citep[e.g][]{Roz2018}. 

The best Chandra/HETG spectra of absorption phenomena in LMXB systems, both in terms of signal-to-noise and prominence of absorption lines, are those of disk winds in black hole LMXBs; however, these tend to display complex absorption from multiple distinct absorption zones \citep[see][]{Miller2015} that complicate this type of analysis. Although these sources may ultimately prove to be excellent laboratories for applying this technique in the future, the aforementioned complexities coupled with the intricacies of relying on the photoionization parameter to derive the geometry of the system mean that these are not ideal candidates for developing and initial testing of this new approach. 

This work was motivated by the discovery of redshifted absorption lines in the Chandra/HETG spectrum of the neutron star UCXB 4U 1916$-$053 in \citet[hereafter \citetalias{Trueba2020}]{Trueba2020}, where the bulk of the absorption was found to be redshifted. The analysis suggests that the absorption is consistent with a nominally static inner disk atmosphere and the observed redshift is argued to be gravitational in origin. Although absorption from static disk atmospheres are commonly detected in these ultra-compact and short-period sources \citep{Ponti2014, Ponti2018a,Ponti2018b, Younes2009, Gavriil2012}, these are poorly understood and subject to limited observation, especially with the Chandra/HETG. \citetalias{Trueba2020} also identifies archival Chandra/HETG spectra for two other short-period NS sources with prominent absorption lines redshifted by a comparable magnitude, XTE J1710$-$281 and AX J1745.6$-$2901, suggesting a physical connection.

We present results on the Chandra/HETG spectra of the three ultra-compact and short period neutron star LMXBs identified in \citetalias{Trueba2020} as sources harboring redshifted disk atmospheres: 4U 1916$-$053, XTE J1710$-$281, and AX J1745.6$-$2901. In Section \ref{sec:ind} we perform spectral fits to photoionized disk atmosphere absorption in these sources and discuss the possible physical origin of the observed redshift in each. In Section \ref{sec:RCE} we explain in detail the mechanics of the central engine velocity-broadening effect on absorption lines and demonstrate why the specific analysis techniques we developed are required to properly model this effect. We conclude by discussing the constraints we obtain from these sources and compare the performance of this method in Section \ref{sec:discussions}.
 
This new method for constraining the size scales of central engines in accreting compact objects is not limited to this specific class of objects and can be applied to a much more diverse sample (e.g. black hole LMXBs with absorbing disk winds); however, as will be discussed in Section \ref{sec:RCE}, gravitationally redshifted absorption is particularly sensitive to these techniques due to (a) its proximity to the central engine, and (b), the significantly reduced uncertainty involved in measuring the orbital radius of the absorber compared with commonly used techniques. 

\section{Sources, Observations, and Data Reduction}\label{sec:data}

\begin{figure}
\centering
\subfloat{\includegraphics[width=0.47\textwidth,angle=0]{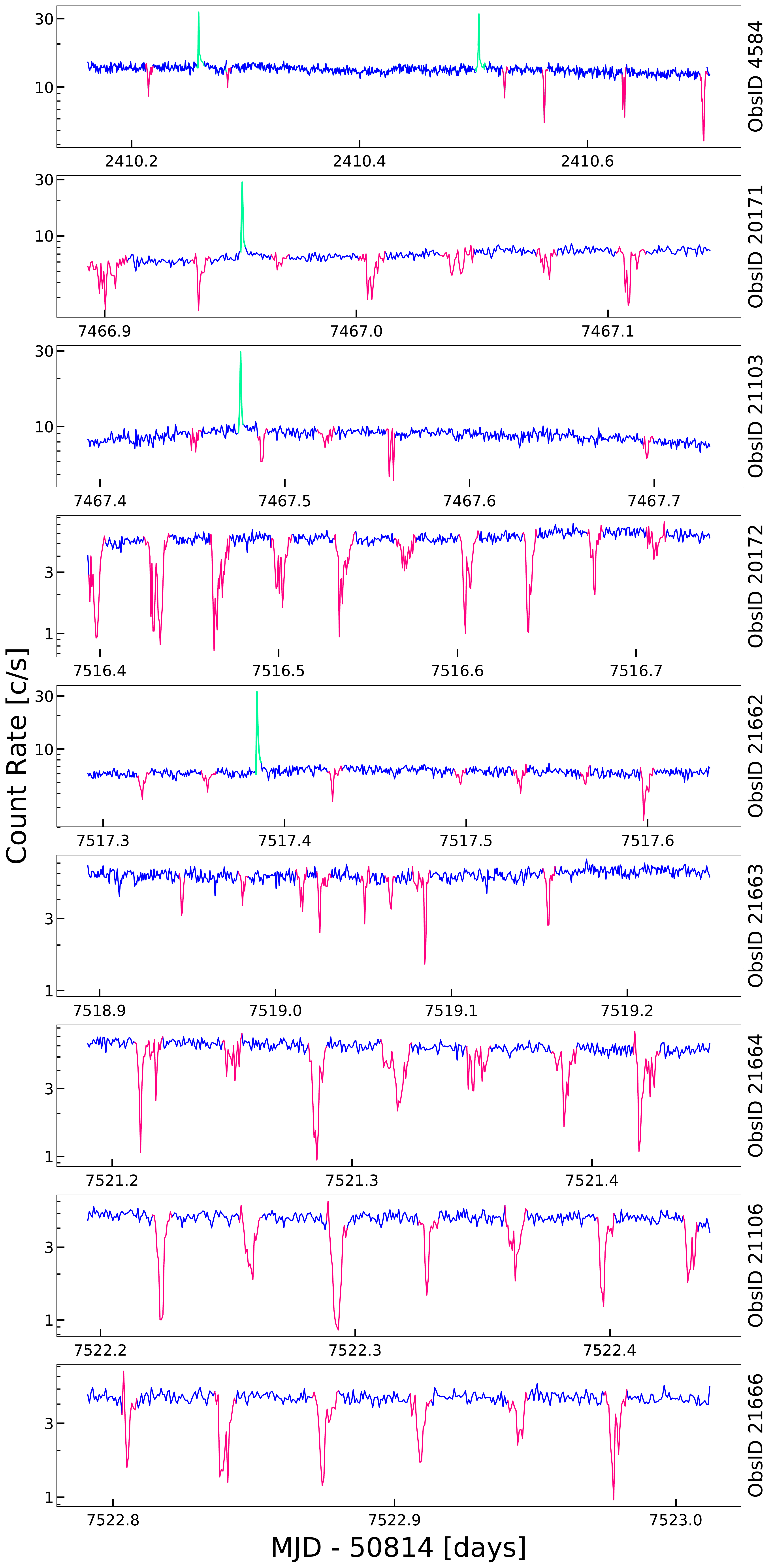}}
\caption{\footnotesize Chandra/HETG lightcurves of 4U 1916$-$053, observations with visible dips (ObsID 21104 and 21105 do not display noticeable dips). Time intervals corresponding to dipping periods (plotted in magenta) and X-ray bursts (green) were removed before spectral extraction. The absorption found during persistent periods is nearly constant throughout all observations (see \citetalias{Trueba2020}). Note that the scale of the x-axes is not consistent between panels.} \label{fig:lc_1916}
  
\end{figure}

\begin{table}[t]
\renewcommand{\arraystretch}{1.1}
\caption{Sources and Observations}
\vspace{-1.0\baselineskip}
\begin{footnotesize}

\begin{center}
\begin{tabular*}{0.47\textwidth}{ l @{\extracolsep{\fill}}   c cc   }
\tableline
\tableline

Obs.ID & Exp. time & Count Rate &  Date\\
 & (ks) & (counts~${\text s}^{-1}$) & (YYYY/MM/DD\\

\tableline 
\multicolumn{4}{c}{XTE J1710$-$281}\\
\tableline

${12468}^{X}$ & 73.3 & 7.02 & 2011/07/23\\
12469 & 73.3 & 7.86 & 2011/08/07\\

\tableline 
\multicolumn{4}{c}{4U 1916$-$053}\\ 
\tableline 
4584 & 46.06 & 19.79 & 2004/08/07 \\
20171 & 21.57 & 13.09 & 2018/06/11 \\
21103 & 28.41 & 15.40 & 2018/06/12\\
${21104}^{Y}$ & 22.54 & 16.15 & 2018/06/13\\
${21105}^{Y}$ & 21.17 & 16.81 & 2018/06/15 \\
20172 & 29.87 & 11.19 & 2018/07/31\\
21662 & 28.89 & 12.50 & 2018/08/01\\
21663 & 29.83 & 11.93 & 2018/08/02\\
21664 & 21.90 & 12.50 & 2018/08/05\\
21106 & 20.59 & 10.47 & 2018/08/06\\
21666 & 18.64 & 11.93 & 2018/08/06\\
\\ [-3.0ex]
\tableline 
\multicolumn{4}{c}{AX J1745.6$-$2901}\\ 
\tableline 
17857 & 117.2 & 8.28 & 2015/08/11\\

\\ [-3.0ex]
\tableline 
\tableline

\end{tabular*}
\vspace*{-1.0\baselineskip}~\\ \end{center} 
\tablecomments{Sources and Chandra/HETG observations used in this work. All observations were made using the ACIS-S timed exposure (or, TE) mode and FAINT data mode.${}^{X}$Observation 2 of XTE J1710$-$281, not used in main analysis.${}^{Y}$The two spectra of 4U 1916$-$053 which do not display noticeable dips in their lightcurves.} 
\end{footnotesize}
\label{tab:obs}
\end{table}

All three sources analyzed in this work are ultra-compact or short-period neutron star low-mass X-ray binaries. These are observed at high-inclinations, where both dips and (except for 4U 1916$-$052) eclipses from the companion star can be found in their X-ray lightcurves. In addition, the presence of type-I X-ray bursts in all three sources typically implies a low magnetic field strength for the central neutron star \citep{Galloway2017}. 

Strictly speaking, 4U 1916$-$053 is the only ``true'' ultra-compact X-ray binary in our sample (${P}_{orbital} < 1$ hour) with a $\sim50$-minute orbital period \citep{Walter1982}. It accretes persistently from a helium-rich donor: the partially degenerate core of a low-mass star, much like a white dwarf \citep{Heinke2013, Nelemans2006}. At the time of writing, there are 11 Chandra/HETG observations of the system totalling $\sim 300$~ks of combined exposure. As described in \citetalias{Trueba2020}, the character of the absorption found in these observations during persistent periods (i.e. removing dipping and bursting events) remains nearly constant in terms of absorbing column, ionization, and velocity shift. All 11 HETG observations of this source were utilized in this work and are listed in Table \ref{tab:obs}.

\begin{figure}
\centering
\subfloat{\includegraphics[width=0.47\textwidth,angle=0]{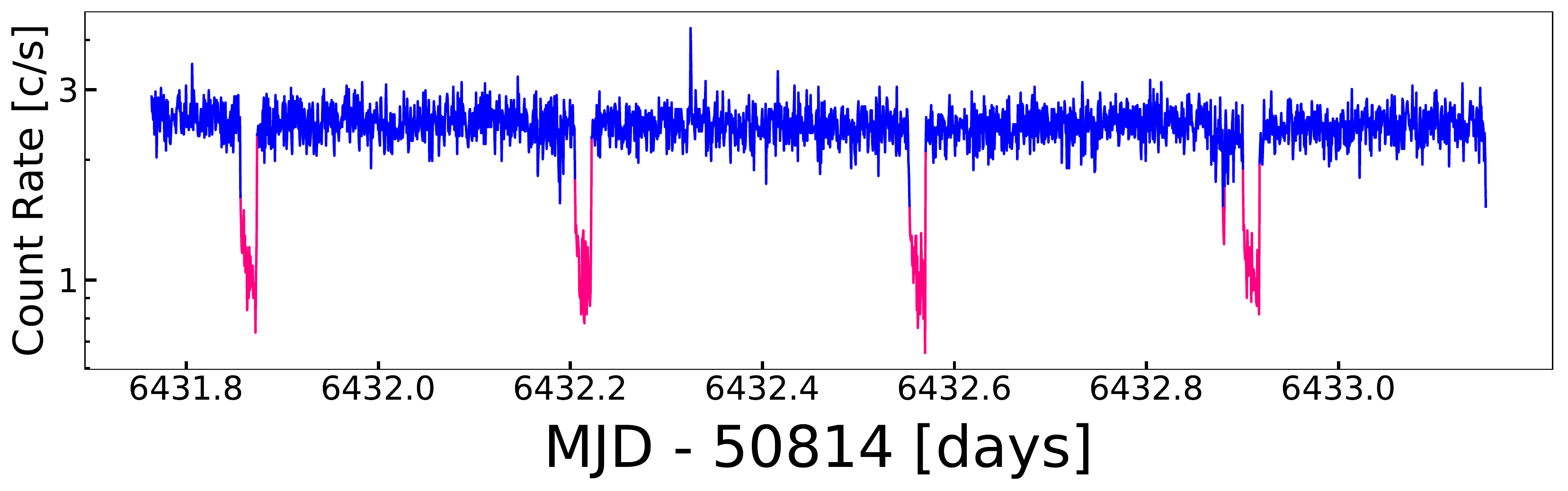}}
\caption{\footnotesize Chandra/HETG lightcurves of AX J1745.6$-$2901. Time intervals corresponding to dipping periods (plotted in magenta) were removed before spectral extraction.}\label{fig:lc_AX}
\end{figure}

\begin{figure}
\subfloat{\includegraphics[width=0.47\textwidth,angle=0]{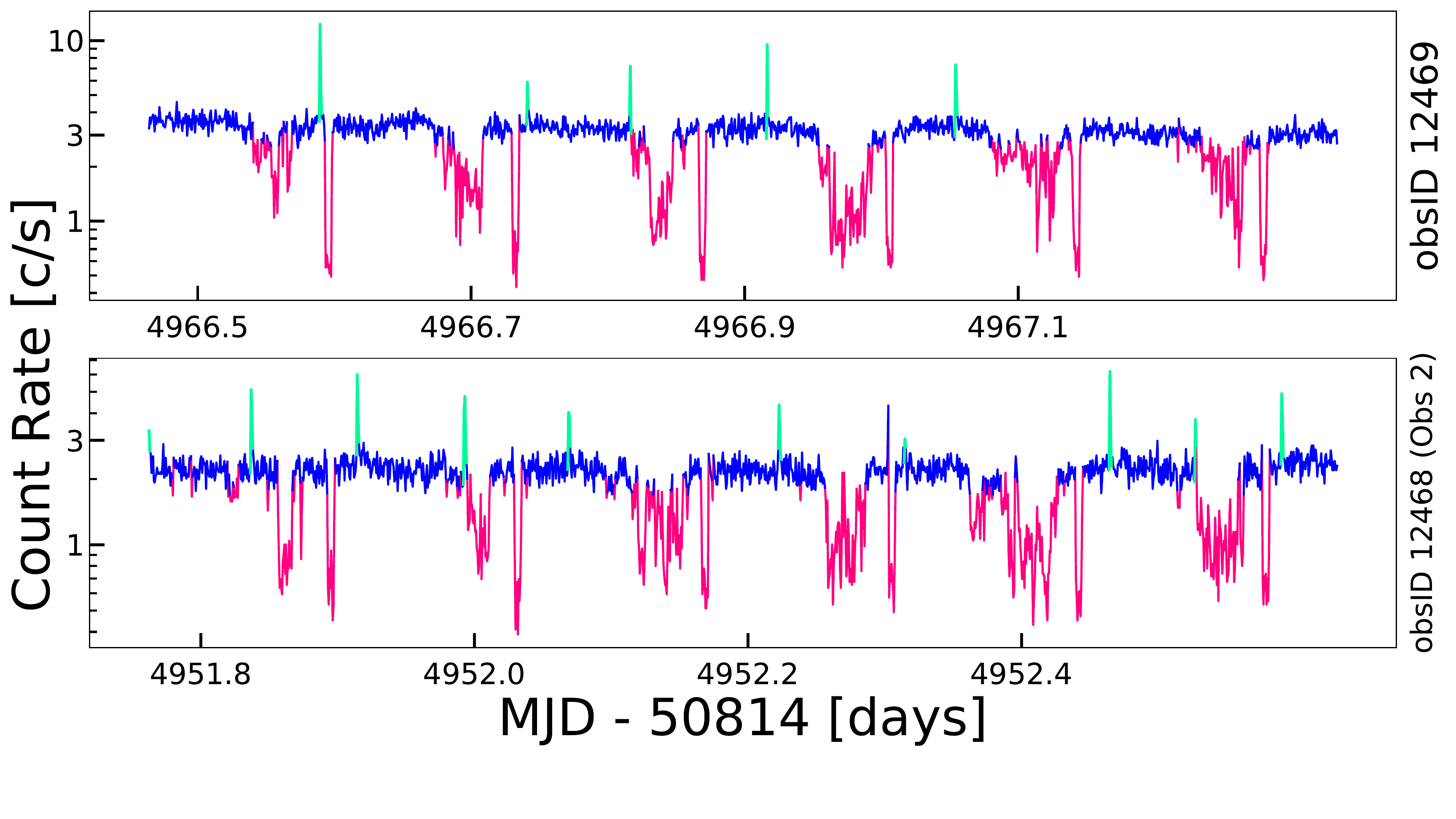}}
\vspace{-0.1in}
\caption{\footnotesize Chandra/HETG lightcurves of XTE J1710$-$281, observations with visible dips. Time intervals corresponding to dipping periods (plotted in magenta) and X-ray bursts (green) were removed before spectral extraction.}\label{fig:lc_XTE}
\end{figure}

With an orbital period of 3.28 hours \citep{Younes2009}, the neutron star X-ray binary XTE J1710$-$053 is more accurately described as a short-period LMXB, a class that shares many characteristics with ultra-compact systems. In addition to X-ray bursts, the system displays X-ray dips as well as eclipses from its stellar companion. Though the spectral type of the donor remains unknown \citep{Jain2011}, there is no evidence to suggest it is helium-rich. Estimates based on type-I X-ray bursts by \cite{Jonker2004} place the system at a distance of $17.3\pm2.5$ kpc, consistent with both previous and subsequent analyses \citep{Markwardt2001, Galloway2008}. At this distance, its galactic coordinates (356.3571 +06.9220) suggest the system is, to first order, located about 8-9 kpc opposite the galactic center and is very likely part of the thick-disk stellar population. As a result, the \emph{mean} expected radial velocity of the system relative to the local standard of rest is negligible and, instead, the local thick-disk velocity dispersion is a more appropriate point of comparison. 

There are two $\sim75$~ks archival Chandra/HETG observations of XTE J1710$-$281, separated by $\sim2$ weeks (see Table \ref{tab:obs}). This interval roughly corresponds to half the length of the $\sim30$ ~ day duty-cycle in which the source flux is observed to vary between 2 to 10 mCrab \citep{Markwardt2001, Galloway2008,Younes2009}. Highly significant redshifted absorption lines were identified in \citetalias{Trueba2020} for one observation (ObsID. 12469) while in the other (ObsID. 12468) these features are significantly less prominent, seeming to correspond to different phases throughout the putative duty-cycle. Both observations are included in this work. 

Finally, AX J1745.6$-$2901 is a neutron star LMXB with an orbital period of $\sim8.35$ hours ($30063.74 \pm 0.14$ s; \citealt{Ponti2017}), accreting from a class III giant stellar companion \citep{Ponti2018a}. It is located in the galactic center region, separated from Sgr A* by 1.45 arcmin at an estimated distance of at least 8 kpc \citep{Ponti2017,Jin2017}, and is absorbed by a large neutral ISM column of ${N}_{H,ISM}\sim30\times{10}^{22}$ ${\text{cm}}^{-2}$ \citep{Ponti2018a}. There is a single Chandra/HETG observation of this source (ObsID. 17857), which was triggered as part of a larger joint NuSTAR+XMM campaign \citep{Ponti2018a}. The source was observed with NuSTAR concurrently to the Chandra/HETG observation; however, this simultaneous observation is not used in this work. The presence of redshifted absorption lines in this observation was noted in \citetalias{Trueba2020}.

\subsection{Data Reduction}\label{sec:data_red}

The data for all Chandra/HETG observations considered in this work (11 for 4U 1916$-$053, 2 for XTE J1710$-$281, and 1 for AX J1745.6$-$2901) were reduced using CIAO version 4.9 and CALDB version 4.7.6. All observations were performed in the ACIS-S timed exposure (TE) mode and FAINT data mode. First-order HEG and MEG spectra for each observation, as well as their corresponding redistribution matrix and ancillary response files, were extracted using the CIAO routines {\it tg\_findzo}, {\it tg\_create\_mask}\footnote{The width\_factor\_hetg parameter in this routine was set to 10 in order to reduce contamination from MEG photons onto the HEG, thereby greatly increasing the sensitivity of the HEG in the Fe~K band and above.}, {\it tg\_resolve\_events}, {\it tgextract}, {\it mkgrmf}, and {\it fullgarf}. 

Separate plus and minus orders were combined using {\it combine\_grating\_spectra}. The same routine was used to combine the 11 separate 4U 1916$-$053 observations into a single high signal-to-noise spectrum. Time filtering of spectra (including the removal of dips, eclipses, and X-ray bursts) was done using the {\it dmcopy} routine. 
For 4U 1916$-$053, bursts and dips were filtered using the same technique as in \citetalias{Trueba2020}: Bursts were readily identified as large positive flux spikes with fast rises and exponential decays,  dips were identified as sharp negative flux variations at intervals of 3000 seconds (the orbital period) and were removed if they departed from a running average flux by more than 20 percent. This was possible due to the well-defined nature of dipping events in the lightcurves for this source (see Figure \ref{fig:lc_1916}), as was the case for the dips observed for AX J1745.6$-$2901 (Figure \ref{fig:lc_AX}). Per Figure \ref{fig:lc_XTE}, a large fraction of the dips observed in the lightcurves for the pair of XTE J1710$-$281 observations were not as clearly defined; therefore, we generated GTIs by filtering time bins in the X-ray lightcurves for ObsID 12469 in which the count rate fell below 2.5 counts per second, where the average persistent rate was $\sim 4$ counts per second. For ObsID 12468, these rates are 1.7 and 2.2 counts per second, respectively. 

\section{Baseline fits}\label{sec:ind}

Before implementing our central engine model (Section 
\ref{sec:RCE}), it was necessary to first perform spectral fits to the photoionized absorption in our spectra \emph{without} taking this specific line-broadening effect into account. As will be discussed in Section \ref{sec:rce_model}, standard analysis tools are ill-equipped to account for this effect; the model we developed required the use of multiple photoionized absorbers, and the fits obtained in this section serve as the baseline for this more complex model. In addition, the Chandra/HETG spectra of XTE J1710$-$281 and AX J1745.6$-$2901 have not been subject to detailed photoionization modeling in the literature and, with the exception of \citetalias{Trueba2020}, there are no published constraints of the redshift in their disk atmospheres.

\begin{table*}[t]
\renewcommand{\arraystretch}{1.1}
\caption{Parameters for Best-Fit Absorption Models}
\vspace{-1.0\baselineskip}
\begin{footnotesize}
\begin{center}
\begin{tabular*}{\textwidth}{l c c l @{\extracolsep{\fill}}  c  c   c }
\tableline
\tableline

Source &Spectrum & Model & Parameter & Zone 1 & Zone 2  & $\chi^{2}/\nu$\\

\\ [-3.0ex]
\tableline 
\tableline 
\\ [-3.0ex]

4U 1916$-$053 & Combined & 1 Zone & ${N}_{He}$ (${10}^{22} {\text{cm}}^{-2}$)  & ${50}_{-37}^{\ddagger}$ & $-$  & 2162.7/2090 = 1.03\\
&& & log ${\xi}$ & ${4.5}_{-0.5}^{+0.1}$ & $-$ & $-$ \\
&& & ${v}_{z}$ ($\text{km}$ $\text{s}^{-1}$) &${200} \pm 50 $ & $-$ & $-$  \\
&& & ${v}_{turb}$ ($\text{km}$ $\text{s}^{-1}$) &${160}_{-30}^{+40}$ & $-$ & $-$  \\

&&&${L}_{phot}$ (${10}^{36}$ erg/s) &  9.0 $\pm$ 2.2 & $-$ & $-$ \\

\\ [-3.0ex]
\tableline 
\tableline 
\\ [-3.0ex]

&& 2 Zone & ${N}_{He}$ (${10}^{22}$ ${\text{cm}}^{-2}$) & ${50}_{-16}^{\ddagger}$ & ${5}^{+35}_{-3}$  & 2127.2/2087 = 1.02\\
&&& log $\xi$  & ${4.8}_{-0.7}^{\ddagger}$ & ${3.8}^{+0.7}_{-0.3}$  & $-$ \\
&&& ${v}_{z}$ ($\text{km}$ $\text{s}^{-1}$) & ${490}_{-150}^{+160}$ & ${0}^{\dagger}$  &$-$\\
&&& ${v}_{turb}$ ($\text{km}$ $\text{s}^{-1}$) & ${100}_{-50\ddagger}^{+130}$ & ${70}^{+70}_{-20\ddagger}$  & $-$ \\
&&&${L}_{phot}$ (${10}^{36}$ erg/s) & 9.3 $\pm$ 2.3 & $-$  & $-$ \\

\\ [-3.0ex]
\tableline 
\\ [-3.0ex]
\tableline 
\\ [-3.0ex]

XTE J1710$-$281 & Obs 1 &1 Zone &${N}_{H}$ (${10}^{22} {\text{cm}}^{-2}$)  &${7}_{-3}^{+8}$ & $-$  & 2226/1975 = 1.13\\
&& & log ${\xi}$ & ${3.0} \pm 0.1$ & $-$ & $-$ \\
&&& ${v}_{z}$ ($\text{km}$ $\text{s}^{-1}$) &${310} \pm 50 $ & $-$ & $-$  \\
&&& ${v}_{turb}$ ($\text{km}$ $\text{s}^{-1}$) &${90}_{-20}^{+30}$ & $-$ & $-$  \\
&&&${L}_{phot}$ (${10}^{36}$ erg/s) &  5.6 $\pm$ 1.4 & $-$ & $-$ \\

\\ [-3.0ex]
\tableline 
\tableline 
\\ [-3.0ex]

&&2 Zone& ${N}_{H}$ (${10}^{22} {\text{cm}}^{-2}$)  & ${50}_{-40}^{+50\ddagger}$ & ${0.6}^{+0.4}_{-0.3}$  & 2158/1971 = 1.09\\
&&& log $\xi$  & ${3.5} \pm 0.3$ & ${2.6}_{-0.2}^{+0.1}$ & $-$ \\
&&& ${v}_{z}$ ($\text{km}$ $\text{s}^{-1}$) & ${300} \pm 100$  & ${250} \pm 280$  & $-$\\
&&& ${v}_{turb}$ ($\text{km}$ $\text{s}^{-1}$)  & ${50}_{\ddagger}^{+40}$ & ${500}_{-210}^{\ddagger}$  &$-$ \\
&&&${L}_{phot}$ (${10}^{36}$ erg/s) & 8.0 $\pm$ 2.0 & $-$  & $-$ \\

\\ [-3.0ex]
\tableline 
\tableline 
\\ [-3.0ex]

&Obs 2 & 1 Zone & ${N}_{H}$ (${10}^{22} {\text{cm}}^{-2}$)  &${100}_{-87}^{\ddagger}$& $-$   & 2195.6/1946 = 1.13\\
&&& log $\xi$&${4.2}_{-0.2}^{+0.3}$ &  $-$  & $-$  \\
&&& ${v}_{z}$ ($\text{km}$ $\text{s}^{-1}$) &${-20}^{+120}_{-180}$ & $-$  & $-$  \\
&&& ${v}_{turb}$ ($\text{km}$ $\text{s}^{-1}$) & ${50}_{\ddagger}^{+170}$ & $-$   &$-$ \\
&&&${L}_{phot}$ (${10}^{36}$ erg/s) & 5.3 $\pm$ 1.3& $-$ &  $-$  \\

\\ [-3.0ex]
\tableline 
\\ [-3.0ex]
\tableline 
\\ [-3.0ex]

AX J1745.6$-$2901 & &1 Zone &${N}_{H}$ (${10}^{22} {\text{cm}}^{-2}$)  & ${96}_{-56}^{+4\ddagger}$ & $-$  & 497/432 = 1.15\\
&& & log ${\xi}$ & ${4.0} \pm 0.2$ & $-$ & $-$\\
&& & ${v}_{z}$ ($\text{km}$ $\text{s}^{-1}$) &${270}^{+240}_{-230} $ & $-$ & $-$  \\
&& & ${v}_{turb}$ ($\text{km}$ $\text{s}^{-1}$) &${100}_{-30}^{+40}$ & $-$ & $-$  \\

&&&${L}_{phot}$ (${10}^{36}$ erg/s) &  8.9 $\pm$ 2.2 & $-$ & $-$ \\

\\ [-3.0ex]
\tableline 
\tableline

\end{tabular*}
\vspace*{-1.0\baselineskip}~\\ \end{center} 
\tablecomments{Best-fit parameter values for the absorption model that best describes each spectrum. All errors are at the 1$\sigma$ level. Errors truncated by the parameter fitting range are marked with a $\ddagger$ symbol, frozen parameters are marked with a $\dagger$ symbol (see text for more detail). }
\end{footnotesize}
\label{tab:all_fits}
\end{table*}

Spectral modeling was performed using SPEX version 3.05.00 \citep{Kaastra1996,Kaastra2018} and SPEXACT version 3.05.00 atomic database and associated routines. All data were binned using the ``obin'' command, which re-bins a spectrum to the optimal bin size given the particular statistics of a given source (for more detail, see \citealt{Kaastra2016}). Although other binning schemes were tested in order to boost the signal-to-noise (or, S/N) ratio of some of our spectra, we ultimately decided against it as the aims of this work require the highest resolution possible. All fits and errors were obtained using ${\chi}^{2}$ statistics.

The spectra of both XTE J1710$-$281 and 4U 1916$-$053 contain multiple absorption lines from H and He-like ions in the 1 to 2 keV energy band, as well as H and He-like Fe absorption in the Fe K band. Our initial scheme was based on utilizing the higher effective area of the MEG at lower energies (despite its lower resolution compared to the HEG) to fit 1 to 4.5 keV portion of the spectrum, while using the HEG from 4.5 to 8.5 keV where its effective area is higher. We modified this scheme by removing the 2.1 to 3 keV region of the MEG, a band that contains some lines but is significantly affected by instrumental features. AX J1745.6$-$2901 is much more heavily absorbed by neutral gas, meaning much of the flux in the lower energy band is lost, and there is little evidence of absorption outside of the Fe~K band. We therefore fit this spectrum in the 5 to 9 keV range using only the HEG.

All spectra were fit with a simple continuum model consisting of both blackbody and disk blackbody additive components (BB and DBB in SPEX), modified by neutral ISM absorption (ABSM in SPEX). In the case of AX J1745.6$-$2901, the disk blackbody component was not needed. The photoionized absorption was modelled using PION, a self-consistent X-ray photoionized absorption model in SPEX \citep{Mehdipour2016,Kaastra2018}. PION calculates a new ionization balance with each iteration of a fit based on the changing continuum, therefore the model was constructed so that each PION component uses the naked continuum model to calculate the ionization balance \emph{before} being absorbed by the ISM. For each PION component, we fit the equivalent hydrogen column (${N}_{H}$), the photoionization parameter (log $\xi$), the average systematic velocity of the absorber (${v}_{z}$), and its turbulent velocity ($v$ in SPEX; hereafter ${v}_{turb}$).

The best-fit models to photoionized absorption in our spectra are found in Table \ref{tab:all_fits}. Quoted errors are at the 1$\sigma$ level. Although PION does produce some continuum absorption, our continuum parameters were set as free parameters in order to ensure a good continuum fit regardless of how the parameter space of the photoionized absorber is being sampled. As is often the case for sources with simple continua in the narrow Chandra band, these changes in the underlying continua have little to no effect on the quality of the fit. We therefore treated these as nuisance parameters and based our confidence regions on the number of free parameters in our absorption components. Please see Appendix \ref{sec:cont_par} for more details.

\subsection{4U 1916-053}\label{sec:f1916}

\begin{figure}
\centering
\subfloat{\includegraphics[width=0.46\textwidth,angle=0]{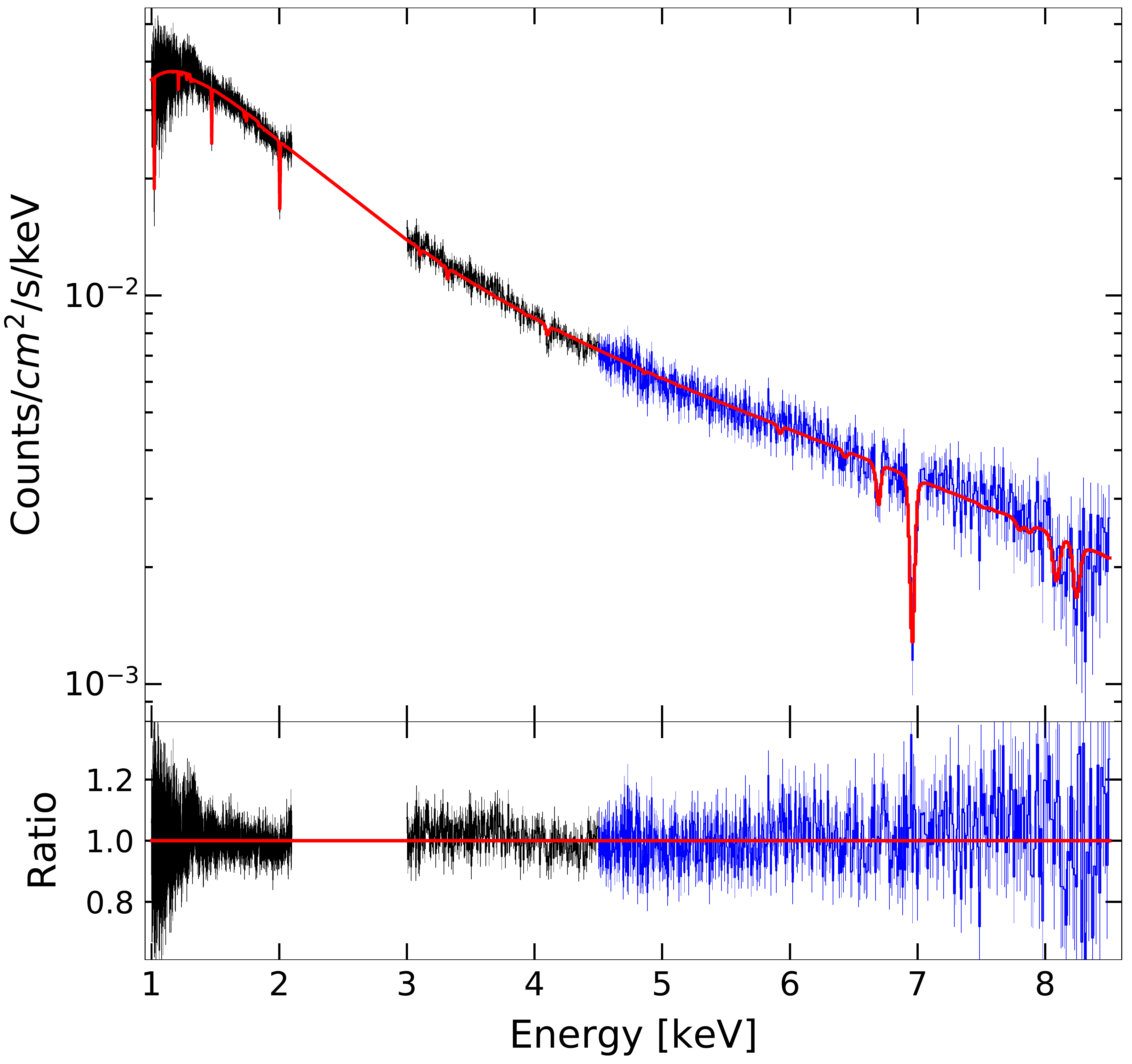}}
  \caption{\footnotesize Chandra/HETG spectrum of 4U 1916$-$053, single-zone model. The MEG portion of the spectrum (1-4.5 keV) is plotted in black, while the HEG portion (4.5-8.5 keV) is plotted in blue here to highlight where each arm of the HETG was used to fit each band (2.1-3 keV was omitted due to instrumental features, see text).} \label{fig:1916_1zone}
\end{figure}

\begin{figure*}
\subfloat{\includegraphics[width=0.99\textwidth,angle=0]{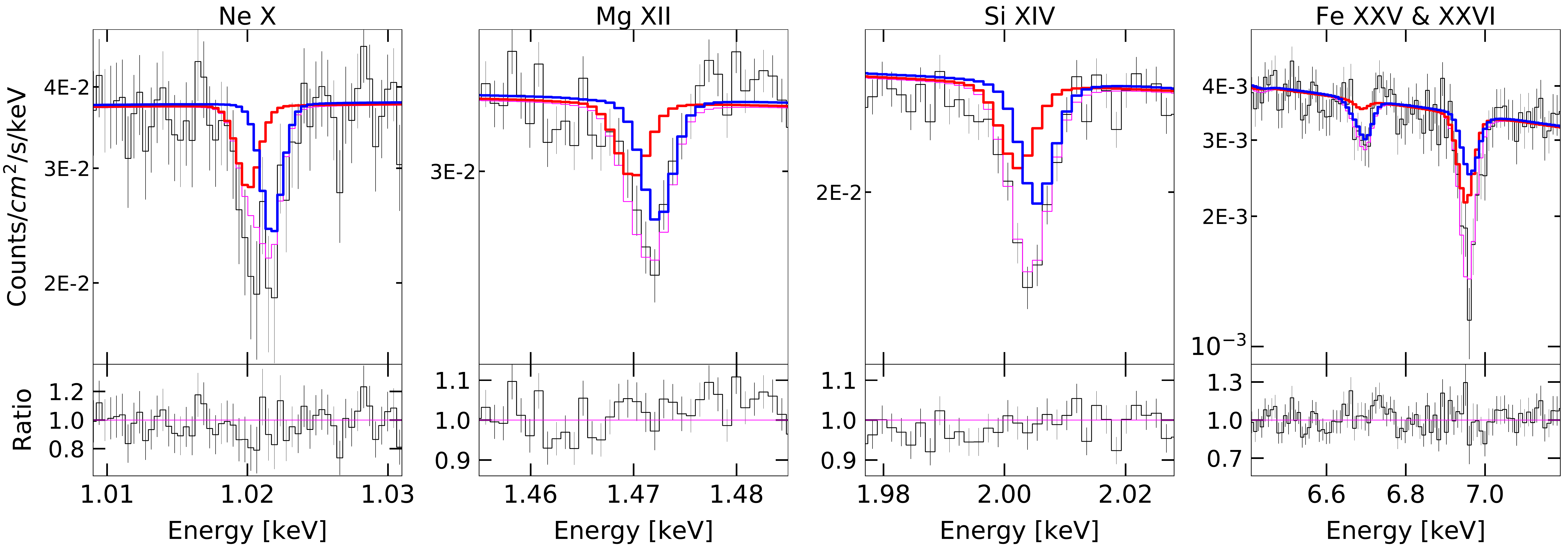}}

\centering
  \caption{\footnotesize The 2-zone model for the Chandra/HETG spectrum of 4U 1916$-$053, focusing on four of the most prominent absorption features. The best-fit model is plotted in magenta, while the contribution from the individual absorption zones are plotted in red and blue for Zone 1 and Zone 2, respectively. Starting from the left, the first three panels plot the MEG spectrum, showing comparable contributions from Zones 1 and 2 in Ne X (1.02 keV), Mg XII (1.47 keV), and Si XIV (2.01 keV), though the lower-ionization component (Zone 2) is more prominent. The rightmost panel shows the HEG spectrum of the Fe~K band. The redshifted Zone 1 makes up the majority of the Fe XXVI absorption (6.97 keV), while Fe~XXV is captured almost entirely by Zone 2.}\label{fig:1916_2zone}
\end{figure*}

A similar analysis of the photoionized absorption in the neutron star UCXB 4U 1916$-$053 was performed in \citetalias{Trueba2020}, where the nearly 300~ks worth of Chandra/HETG data of the source was grouped into three separate spectra corresponding to the epoch of observation and broadly similar continuum. Fits to these spectra revealed redshifted absorption from a disk atmosphere where the absorbing column, ionization, and velocity shift were nearly constant and within their 1$\sigma$ confidence regions. The lower energy portions of the MEG in these three spectra, however, had modest sensitivity (which worsened with decreasing photon energy) and therefore the prior analysis did not include many bins below $\sim 2$ keV (the rest energy of Si XIV). In order to utilize the higher spectral resolution portions of our spectra, eventually maximizing the sensitivity of our central engine model, we decided to co-add all observations in an effort to improve the S/N of the combined spectrum.

As was the case with the three spectra in \citetalias{Trueba2020}, the continuum of the combined spectrum can be described using both a blackbody and disk blackbody additive components. Although the best-fit continuum parameters are within a physically acceptable range (${T}_{bb} = {1.8} \pm 0.1$ keV, ${T}_{dbb} = {0.80}\pm 0.05$ keV, and normalizations that suggest radii in the 5-10 km range), we do not claim these to be an accurate physical description of the underlying continuum. A physical treatment would explicitly include Comptonization and other processes. Rather, this continuum is reasonably simple, and flexible enough that unmodeled Comptonization or a non-thermal power-law component will simply be accounted for by, e.g., a slightly higher blackbody temperature and flux.  This continuum is therefore suited to the purpose of establishing an ionization balance.

The best-fit, single absorption zone model (1-zone model) for 4U 1916$-$053 is shown in Figure \ref{fig:1916_1zone}, with best-fit parameters listed in Table \ref{tab:all_fits}. Note that the column parameter is listed as ${N}_{He}$ for this source, as it is accreting from a helium-rich donor and therefore we report an equivalent helium column. As a consequence, we set the upper bound of ${N}_{He}$ when fitting at the helium Compton-thick limit of $5\times {10}^{23}$ ${\text{cm}}^{-2}$, which is half that of the Hydrogen limit because a Helium gas contains twice as many electrons \emph{per ion} compared to a Hydrogen gas. See \citetalias{Trueba2020} for details about how abundances were modified to properly model this helium-rich absorption. We briefly note the Compton-thick limit for a hydrogen gas is quoted in the literature as either ${10}^{24}$ ${\text{cm}}^{-2}$ or, more strictly, closer to $1.5\cdot{10}^{24}$ ${\text{cm}}^{-2}$ (${\tau}_{e} \sim 1$); the more conservative limit we adopted should be considered a soft-limit. Indeed, the best-fit model prefers a column of ${5.0}^{\ddagger}_{-3.7}\times {10}^{23}$ ${\text{cm}}^{-2}$, though the large minus error-bar suggests the spectrum may still be consistent with absorption below the Compton-thick regime. The best-fit values for ${N}_{He}$, log $\xi$, and ${v}_{turb}$ are consistent to those obtained by fitting each spectrum separately in \citetalias{Trueba2020}. 

Despite its limited statistical significance, we briefly highlight a shift in ${v}_{z}$ from ${260}\pm 80$ $\text{km}$ $\text{s}^{-1}$ \citepalias{Trueba2020} to ${200} \pm 50$ $\text{km}$ $\text{s}^{-1}$ (this work) as it coincides specifically with the inclusion of MEG bins which contain absorption lines that typically correspond to ions with a lower degree of ionization. Per the definition of the ionization parameter, $\xi = \frac{L}{n\cdot{r}^{2}}$, and the approximation for gravitational redshift at large radii, ${z}_{grav} = {(r/\frac{GM}{{c}^{2}})}^{-1}$ (or, $c \cdot {(r/{r}_{g})}^{-1}$ in velocity space), this shift may simply represent an additional absorber located at larger radii, where both the ionization and redshift are lower. For example, this absorber could represent the outermost portion of the redshifted inner disk atmosphere, where we could reasonably expect changes in $\xi$ to be mostly sensitive to the distance $r$ from the photoionizing source, barring a dramatically steep density gradient. Alternatively, absorption arising in the outer disk could also produce low-ionization absorption with no noticeable velocity shift even with a significantly lower density.

In order to test these alternatives, we performed a two absorption zone (2-zone) fit where two PION components were used to model the inner and outer absorbers. The best-fit 2-zone model can be found in Figure \ref{fig:1916_2zone}, where the contribution of each absorption zone is plotted in red and blue, respectively. The fits and errors were obtained using the same methodology as the 1-zone fits, with the exception of how ${v}_{z}$ was treated for the outer zone. Although our fits were performed treating ${N}_{He}$, log $\xi$, ${v}_{z}$, and ${v}_{turb}$ for both zones as free parameters, performing error searches required ${v}_{z}$ to be frozen at its best-fit value of zero for the outer zone. This merely serves to prevent Zone 1 and Zone 2 from switching their proximity to the central engine within the error scans, which is extremely inefficient. The zones represent two distinct regions of parameter space and, as shown in \citetalias{Trueba2020}, absorption in the outer disk is consistent with zero shift. 

The best-fit 2-zone model yields $\chi^{2}/\nu = 2127.2/2087 = 1.02$; via an F-test, this represents a $5\sigma$ improvement (although the models are not nested). Zone 1 (the redshifted atmosphere) prefers the maximum value of ${N}_{He}$ and log $\xi$, the former corresponding to the Compton-thick soft limit, while the latter is typically the limit at which log $\xi$ becomes degenerate with ${N}_{He}$. In addition, the best-fit value of ${v}_{z} = {490}_{-150}^{+160}$ $\text{km}$ $\text{s}^{-1}$ is now significantly larger and statistically distinct at the 1$\sigma$ level from the results of the single-zone fit. Originally, we hypothesized a scenario in which the dominant absorption originated in the redshifted atmosphere and where the influence of an additional static absorber was limited. The 2-zone fit would suggest a significant contribution from this outer component in most lines (see Figure \ref{fig:1916_2zone}). Again, the 1-zone fit still prefers a redshifted absorber at the 5$\sigma$ level.

Using the best-fit parameters from Table \ref{tab:all_fits}, we can derive some basic physical properties of the disk atmosphere: If the density of the absorbing gas can be measured independently, the orbital radius of the disk atmosphere can be obtained using the definition of the photoionization parameter, $r = \sqrt{L/n\xi}$. As is often the case, we do not have independent constraints on the gas density of the disk atmosphere. However, we can relate the density ($n$) to the observed column density (${N}_{H}$) via ${N}_{H} = \Delta r\cdot n$, where $\Delta r$ is the thickness of the absorber along the line-of-sight. Combing this with the ionization parameter, we obtain the expression
\begin{equation}
\xi = \frac{L}{{N}_{H}{r}^{2}\frac{1}{\Delta r}},
\end{equation}
which can be can be re-arranged into
\begin{equation}
r = \frac{L}{{N}_{H}\xi}\cdot\frac{\Delta r}{r},
\end{equation}
and finally
\begin{equation}
r = \frac{L}{{N}_{H}\xi}\cdot f. \label{eq:rup}
\end{equation}

Here, we obtain an expression for the distance between the absorber and photoionizing source scaled by the filling factor ($f$), a scaling parameter which corresponds to the degree of clumping in a gas and where $0 < f \leq 1.0$. The quantity $r = \frac{L}{{N}_{H}\xi}$ uses the maximal filling factor value of 1 and is, therefore, an upper-limit on the orbital radius of an absorber. In this short discussion, we assume a neutron star mass of $1.4{M}_{\odot}$ when converting distances to units of gravitational radii ($GM/{c}^{2}$), as well as assume an arbitrary error on the photoionizing luminosity of $25\%$ (comparable to the error in various distance estimates) of the best-fit value (as reported in Table \ref{tab:all_fits}) and therefore we advise caution when interpreting these estimates.

We also note that there is evidence of some correlation between ${N}_{H}$ and $\xi$ in some of our fits (see Appendix \ref{sec:cont_par}), associated with large $\xi$ values and poorly constrained ${N}_{H}$ errors. In these specific cases, we explicitly report errors ${R}_{atm}$ assuming \emph{perfect} correlation between ${N}_{H}$ and $\xi$, along with their uncorrelated errors. Note that the difference in errors does not exceed 40\%.

For the 1-zone fit, we get a disk atmosphere orbital radius of ${R}_{atm} = f\cdot{2800}^{+5300}_{-1000\ddagger}$ (${}^{+4000}_{-1000\ddagger}$ if uncorrelated) $GM/{c}^{2}$, where the $\ddagger$ sign indicates that an error was calculated by propagating a parameter error which was truncated by its allowed fitting range. In this case the upper error on ${N}_{He}$ was truncated by the Compton-thick soft limit of $5\times {10}^{23}$ ${\text{cm}}^{-2}$, which corresponds to the lower error when used to calculate the radius. Using the maximum filling factor value of unity ($f = 1$), which assumes an absorber that is both perfectly homogeneous (i.e. no clumps) and where its width ($\Delta R$) is comparable to its distance to the photoionizing source, we obtain a minimum gravitational redshift of ${z}_{grav} \cdot c > {110}_{-70}^{+50\ddagger}$ $\text{km}$ $\text{s}^{-1}$. At this limit, the 1$\sigma$ errors overlap
with those obtained from the measured redshift in the atmosphere of ${v}_{z} = {200} \pm 50 $ $\text{km}$ $\text{s}^{-1}$.  

We also note that, in addition to the gravitational redshift, ${z}_{grav} = {(r/\frac{GM}{{c}^{2}})}^{-1}$, there is an additional term corresponding to the transverse Doppler effect that arises due to the orbital motion of the gas. This effect produces an additional redshift of magnitude ${z}_{TDE} = 0.5\cdot {v}_{\perp}^{2}/{c}^{2}=0.5\cdot{z}_{grav}$, and therefore the total redshift at radius r becomes ${z}_{total} = 1.5 \cdot {(r/\frac{GM}{{c}^{2}})}^{-1}$, where ${v}_{\perp}^{2}$ is the Keplerian velocity of the absorbing gas, orthogonal to the line-of-sight. The aforementioned minimum redshift, therefore, becomes ${z}_{total} \cdot c > {170}_{-110}^{+70\ddagger}$ $\text{km}$ $\text{s}^{-1}$, placing the measured and radius-derived redshifts in better agreement.

In the case of our 2-zone fit, we obtain a value of ${R}_{atm} = f\cdot{1400}^{+2700}_{-350\ddagger}$ (${}^{+2400}_{-350\ddagger}$ if uncorrelated) $GM/{c}^{2}$, and corresponding ${z}_{grav}\cdot c > {210}_{-140}^{+90\ddagger}$ $\text{km}$ $\text{s}^{-1}$ and ${z}_{total} \cdot c > {320}_{-210}^{+130\ddagger}$ $\text{km}$ $\text{s}^{-1}$. Though the upper error for ${z}_{grav}$ is truncated by the fitting range for $\xi$ and ${N}_{He}$, the $1\sigma$ confidence regions of redshift implied by the photoionization radius suggests a minimum redshift on the order of what was found in \citetalias{Trueba2020}, while ${z}_{total}$ is in 1$\sigma$ agreement with the best-fit value of ${v}_{z} = {490}^{+160}_{-150}$ $\text{km}$ $\text{s}^{-1}$ even when assuming a filling factor of unity. In the case of the outer absorption zone (Zone 2), we obtain a radius of ${R}_{outer} \approx f\cdot ({1.4}^{+1.4}_{-1}) \times {10}^{5}$ $GM/{c}^{2}$. Adopting a maximal filling factor of unity, these values seem more consistent with absorption from the outer disk; however, we cannot rule out smaller filling factors and therefore absorption in the inner few $\times1000$ $GM/{c}^{2}$. Using the orbital period of dips in the outer disk and (again) assuming a neutron star mass of $1.4{M}_{\odot}$, we obtain an outermost disk radius of $1.7 \times {10}^{5}$  $GM/{c}^{2}$. This value is within the poorly constrained $1\sigma$ errors of ${R}_{outer}$.

\subsection{XTE J1710-281}\label{sec:fXTE}

\begin{figure}
\centering
\subfloat{\includegraphics[width=0.46\textwidth,angle=0]{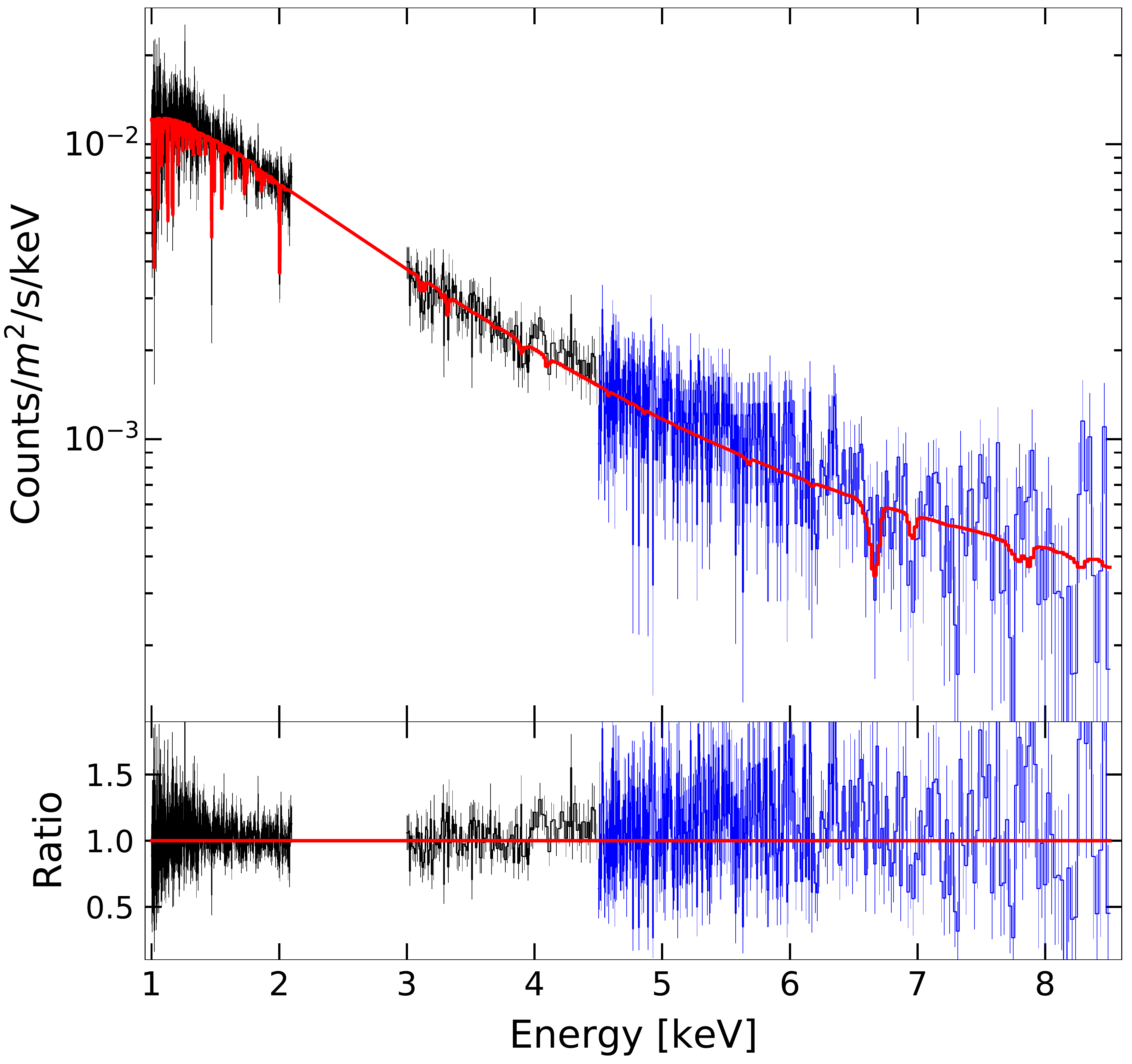}}
  \caption{\footnotesize Chandra/HETG spectrum of XTE J1710$-$281, single-zone model. The MEG portion of the spectrum (1-4.5 keV) is plotted in black, while the HEG portion (4.5-8.5 keV) is plotted in blue here to highlight where each arm of the HETG was used to fit each band (2.1-3 keV was omitted due to instrumental features, see text).}\label{fig:XTE_1zone}
\end{figure}

The two Chandra/HETG spectra of XTE J1710$-$281 have been the subject of spectral analysis only once in \cite{Raman2018}, an analysis that focuses mostly on the nature of the absorption during dipping periods. Although a photoionization analysis was performed using the XSPEC model {zxipcf}\footnote{This model used the XSTAR grid model used in \cite{Reeves2008}, assuming a simple powerlaw continuum with $\Gamma = 2$. A major drawback is that this grid utilizes only 12 points to sample the ionization parameter over 9 orders of magnitude, and therefore it likely does not capture the full information in a rich, high resolution absorption spectrum \citep{Reynolds2012}.} (an XSTAR-based partial covering pre-calculated absorption model), the work makes no mention of the highly significant redshift in prominent lines below 3 keV that we found in both our own reduction of the data (as will be discussed in this section) and via TGCat, and instead focused on the discovery of ionized Fe absorption. In this work we focus primarily on the observation with most prominent absorption, which we refer to as Obs 1 (ObsID 12469). We also present our analysis of the more tenuous absorption in Obs 2 (ObsID 12468) in order track changes in the absorbing disk atmosphere.

\begin{figure*}
\centering
\subfloat{\includegraphics[width=0.97\textwidth,angle=0]{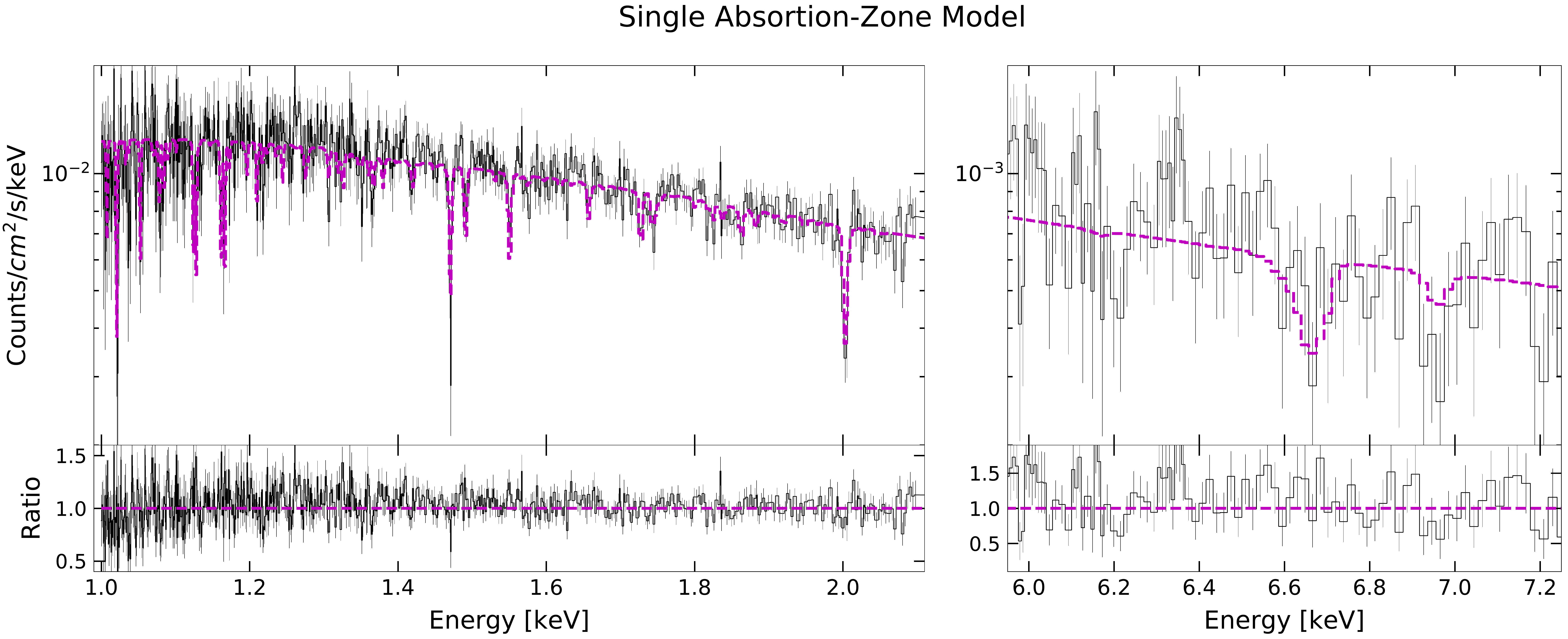}}\\
\vspace{-0.15in}
\subfloat{\includegraphics[width=0.97\textwidth,angle=0]{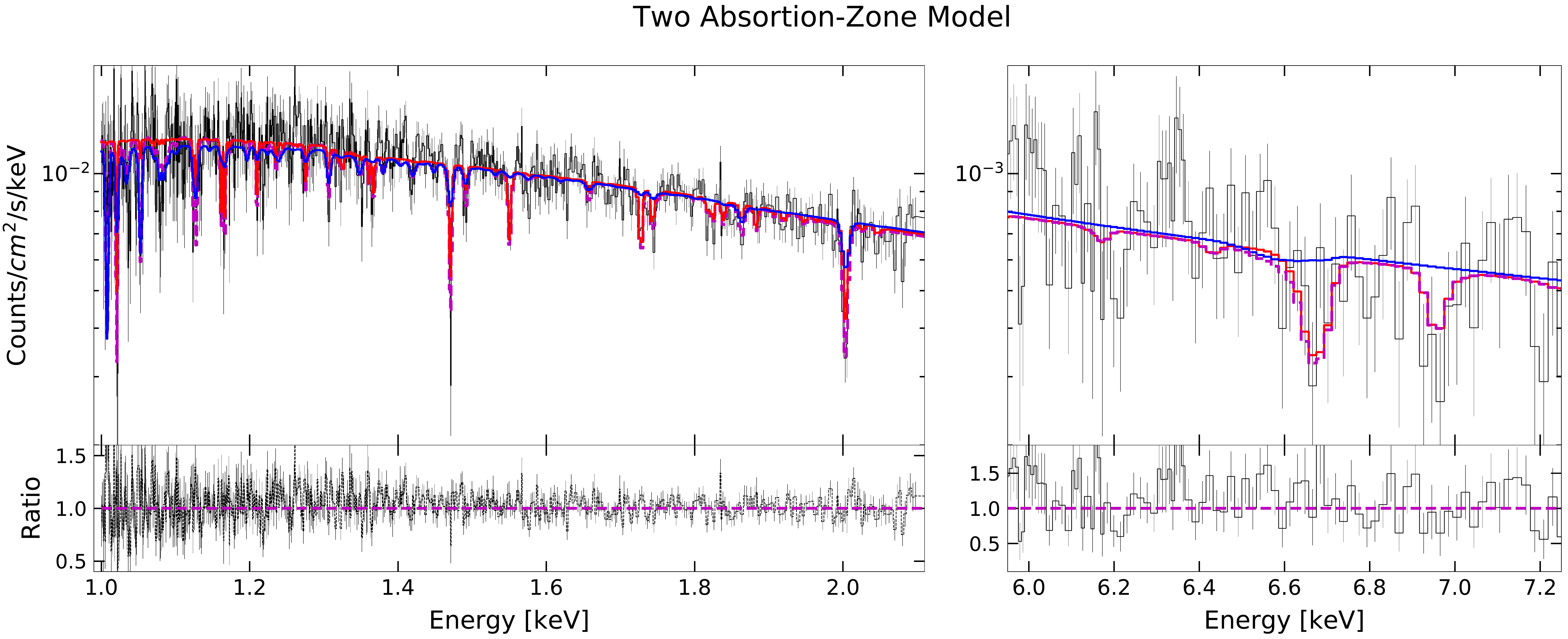}}
  \caption{\footnotesize The 1-zone (top) and 2-zone (bottom) models for the Chandra/HETG spectrum of XTE J1710$-$281 (observation 1). The MEG portion is plotted from 1 to 2.1 keV in the left panels, while the Fe~K region of the HEG spectrum is plotted in the right panels. The best-fit model is plotted with a dashed magenta line, while the separate Zone 1 and 2 components (of the 2-zone model) are plotted in red and blue, respectively. Here, the data is binned to a minimum S/N of 1 (using the ``vbin'' command in SPEX) for plotting purposes only - fitting was done at a lower level of binning (via the ``obin'' command, see text).}\label{fig:XTE_1_and_2}
\end{figure*}

As with our fits to 4U 1916$-$053, the continuum for both Chandra/HETG observations of XTE J1710$-$281 were fit using a combination of disk blackbody and blackbody additive components, modified by interstellar absorption. We again obtained best-fit continuum parameters that are within a physically acceptable range: for Obs 1 we obtain a blackbody and disk blackbody temperatures of ${kT}_{bb} = {3.0}^{\ddagger}_{-1.1}$ keV and ${kT}_{dbb} = {1.05} \pm 0.05$ keV, respectively, and normalizations that suggest emitting areas with radii in the 2-5 km range. The continuum in Obs 2, however, requires drastically lower blackbody and disk blackbody temperatures of ${1.7}^{+0.5}_{-0.2}$ keV and ${0.8}^{+0.2}_{-0.2}$ keV, and for the blackbody radius to be 5 times larger than in Obs 1.

The best-fit single absorption zone model parameters for Obs 1 of XTE J1710$-$281 (see Figure \ref{fig:XTE_1zone} and \ref{fig:XTE_1_and_2}) listed in Table \ref{tab:all_fits} are well-constrained. In particular, we observe a redshift in the disk atmosphere of ${v}_{z} = {310} \pm 50$ $\text{km}$ $\text{s}^{-1}$, with $3\sigma$ and $5\sigma$ lower bounds of ${v}_{z} > 220$ and $> 180$ $\text{km}$ $\text{s}^{-1}$, respectively. Due to its location and distance in the thick-disk of the Milky Way (see Section \ref{sec:data}), we expect a near zero \emph{mean} relative radial velocity in this region and therefore make direct comparisons to the local velocity dispersion, instead. Given a thick-disk velocity dispersion of $50$ $\text{km}$ $\text{s}^{-1}$ \citep{Pasetto2012}, this roughly corresponds to a $3\sigma$ difference between the measured redshift and the expected kinematics of the galaxy. 

\begin{figure}
\centering
\subfloat{\includegraphics[width=0.44\textwidth,angle=0]{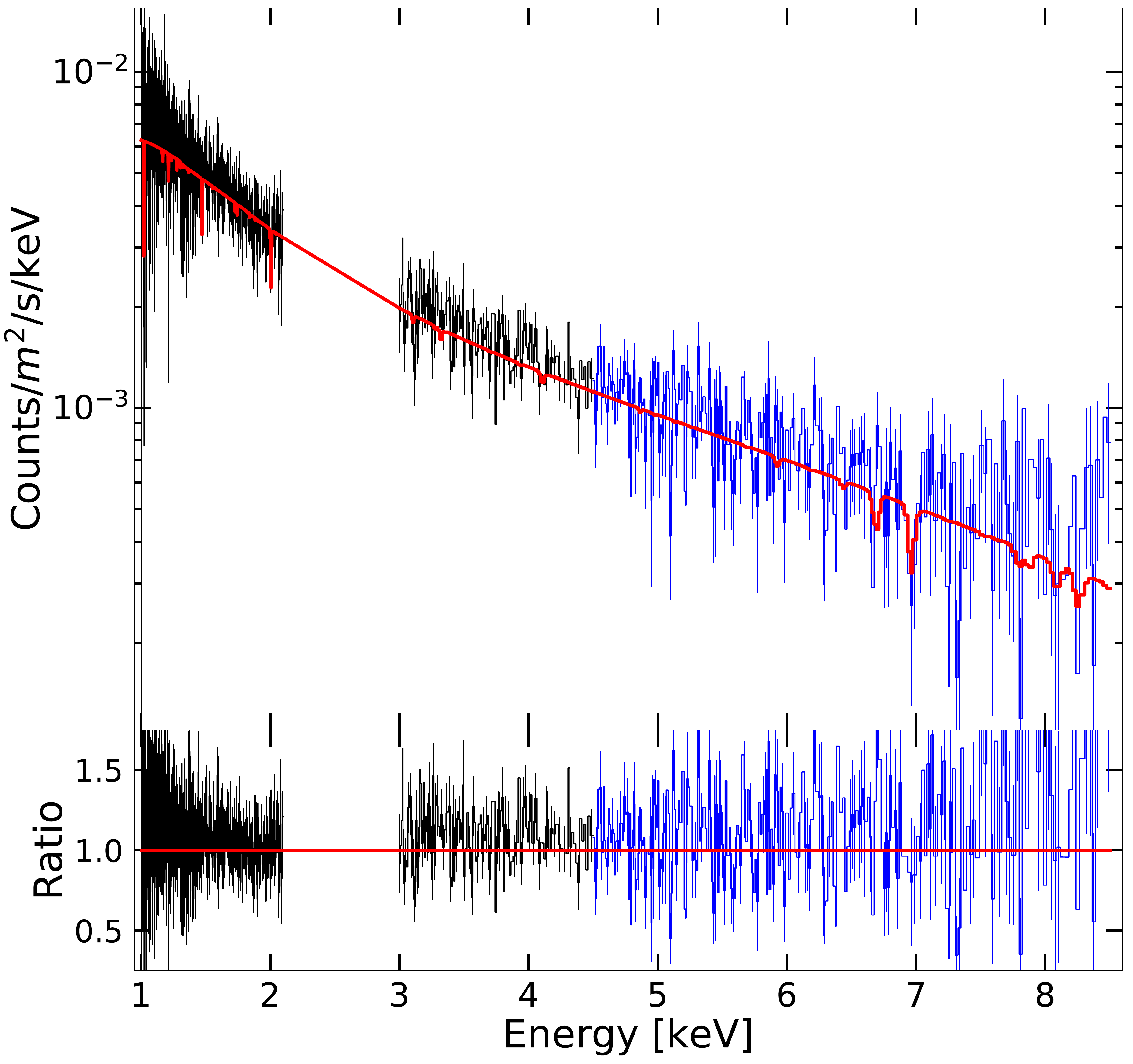}}\\
\subfloat{\includegraphics[width=0.44\textwidth,angle=0]{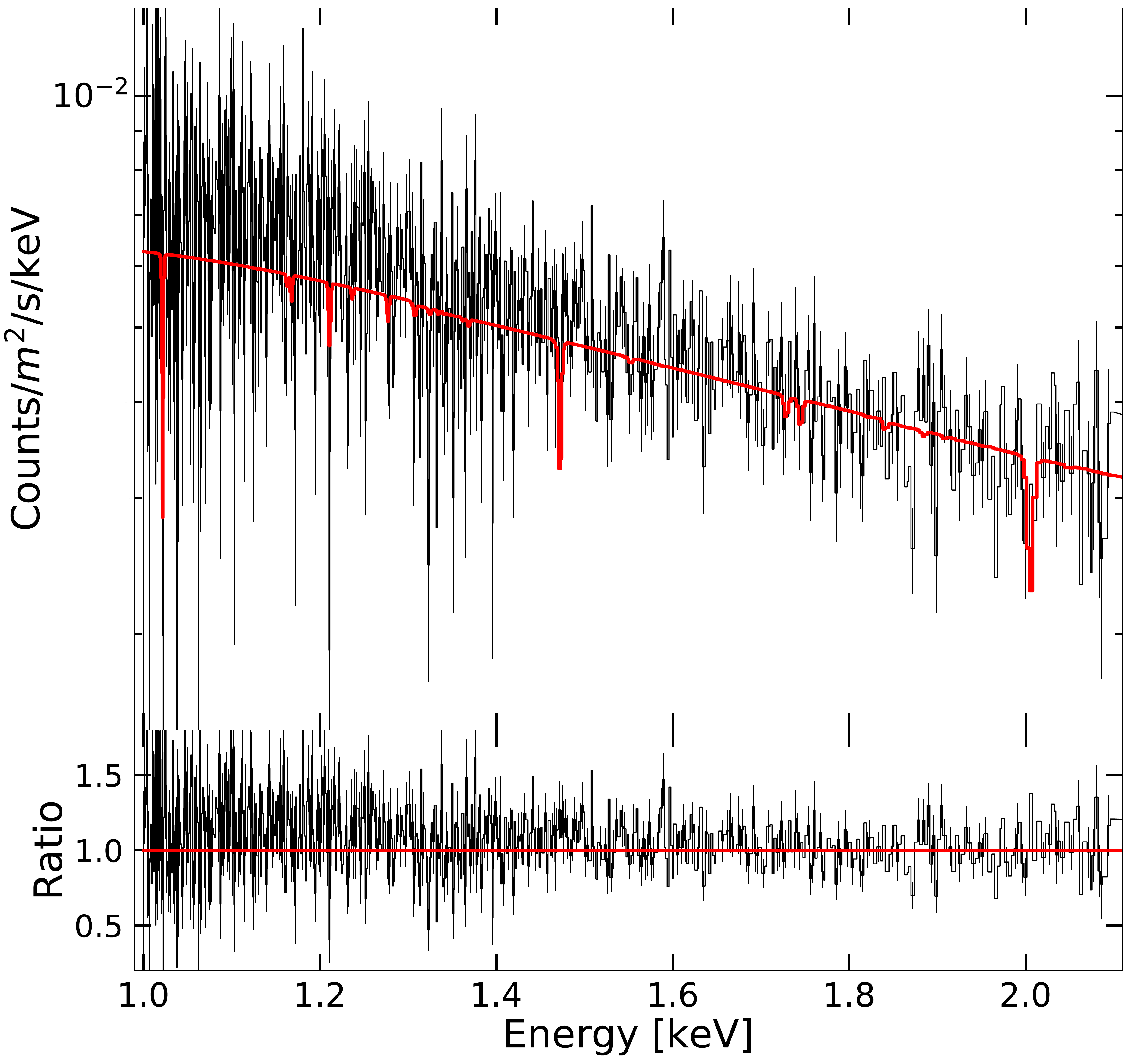}}

  \caption{\footnotesize \emph{Top:} Chandra/HETG spectrum of the alternate observation of XTE J1710$-$281 (observation 2). The MEG portion of the spectrum (1-4.5 keV) is plotted in black, while the HEG portion (4.5-8.5 keV) is plotted in blue. Best-fit model is plotted in red. Fe XXV and XXIV absorption, though tenuous, is more apparent in this observation, however. \emph{Bottom}: The lower energy band of the MEG spectrum. Though the spectrum contains few prominent absorption lines, these were sufficient to constrain the shift for the observed absorption of ${v}_{z} = {-20}^{+120}_{-180}$, distinct from the redshift from observation 1 at the 2$\sigma$ level. The discrepancy suggests that the redshift measured in observation 1 is not due to the radial velocity of the system, consistent with its expected velocity given its location in the galaxy. See text for more detail.}\label{fig:XTE_obs2_low}
\end{figure}

The high significance of the redshift compared to its expected relative radial velocity suggests that the redshift is perhaps not linked to the motion of the system along the line-of-sight; rather it is potentially due gravitational redshift or inflowing gas instead. Unfortunately, we were unable to model the excess absorption during dipping events in order to obtain a direct radial velocity measurement, as was done with 4U 1916$-$053 in \citetalias{Trueba2020}; time-resolved spectra of the few short dips available in the available XTE J1710$-$281 data were too poor in quality for lines to be identified. However, we were able to make direct comparisons to the starkly different absorption found in Obs 2 (Figure \ref{fig:XTE_obs2_low}). Although parameters such as ${N}_{H}$ are very poorly constrained, we do obtain a velocity shift for the absorbing gas of ${v}_{z} = {-20}^{+120}_{-180}$ $\text{km}$ $\text{s}^{-1}$. We find that this velocity shift is distinct to the redshift measured in Obs 1 at the $2\sigma$ level. 

We cannot rule out the possibility that the absorption in Obs 2 could, in principle, be produced by a disk wind with an outflow velocity of $\sim 300$ $\text{km}$ $\text{s}^{-1}$ that is coincidentally cancelled out by the radial velocity of the system - a velocity we found to be in excess of the local mean by a factor of at least 4 times the local velocity dispersion. We argue it is far likelier that this absorption is due to (radially) static gas located at large orbital radii where any gravitational redshift is negligible. The near-zero velocity-shift we measured for Obs 2 therefore reflects the small radial velocity of the system we expect in comparison to the large and well-constrained redshift found in Obs 1, which we argue is likely either gravitational or due to the inflow of gas, though the lack of inverse P-Cygni profiles makes the latter scenario less compelling. 

A gravitational redshift of ${310} \pm 50$ $\text{km}$ $\text{s}^{-1}$ corresponds to a distance from the central neutron star of ${R}_{z} = 970 \pm 160$ $GM/{c}^{2}$ (or, ${1600}^{+400}_{-200}$ $GM/{c}^{2}$ once corrected for the transverse Doppler effect). Using Equation \ref{eq:rup} and assuming a neutron star mass of $1.4{M}_{\odot}$, we derive a radius based on the photoionization using the best-fit parameters in Table \ref{tab:all_fits} of ${R}_{atm} = f\cdot{4.0}^{+7.0}_{-3.0} \times {10}^{5}$ $GM/{c}^{2}$. Assuming the maximal value of $f = 1$ we obtain a radius at which the gravitational redshift is negligible and therefore this quantity has little use as an upper limit, as it does not restrict the location of the absorber to the inner disk. In fact, a filling factor of ($f = 0.01$) is required for these radii to agree at the $1\sigma$ level while preferring a much lower value of $f = 0.002$. Incidentally, the column and ionization values for Obs 2 imply a much smaller upper limit on the radius by comparison, though these are so poorly constrained that they are consistent with radii in the ${10}^{4}$ to ${10}^{5}$ $GM/{c}^{2}$ range.

The small filling factor values, and corresponding high-degrees of clumpiness, derived from our single zone model of Obs 1, are by no means unphysical and have been reproduced in simulations of AGN outflows \citep[see][]{MP2013,Dannen2020}, as well as observed in some LMXB disk winds \citep{Miller2015,Trueba2019}. However, if the observed redshifted absorption in both 4U 1916$-$053 and XTE J1710$-$281 are instances of the same physical phenomenon and are produced by the same physical mechanisms, the apparent large discrepancy in their filling factors requires further examination. Of note, the spectrum in Obs 1 contains many lines in the 1 to 3 keV region that correspond to both low and high ionization absorption, as well as H and He-like Fe absorption in the Fe K band that correspond to high ionization absorption. It is possible that this spectrum contains both high and low ionization absorption, in which case the parameters used to derive the radii and filling factors for Obs 1 may be an average of these absorbers. As with 4U 1916$-$053, we performed a two absorption-zone (2-zone) fit to test this hypothesis.

The best-fit 2-zone model shown in Figure \ref{fig:XTE_1_and_2} and listed in Table \ref{tab:all_fits} results in two \emph{distinct} absorption zones and an f-test derived statistical improvement over the single-zone model above the $5\sigma$ level. The bulk of the absorption is dominated by Zone 1, an apparent disk atmosphere redshifted by ${300} \pm 100$ $\text{km}$ $\text{s}^{-1}$ and with the expected higher ionization and absorbing column of an inner-disk atmosphere, though the latter is poorly constrained. These parameters suggest a radius of ${R}_{atm} = f\cdot{2.5}^{+4.0}_{-2.5\ddagger}$ (${2.5}^{+3.0}_{-2.5\ddagger}$ if uncorrelated) $\times {10}^{4}$ $GM/{c}^{2}$. In this case, the lower bound obtained via standard error propagation is largely uninformative, especially given that the upper error on ${N}_{H}$ is truncated by the fitting range and is of the order of the best-fit value. As a crude estimate of this lower-bound, we calculated ${R}_{atm}$ fixing ${N}_{H}$ at its upper $1\sigma$ value and propagating other errors (${L}_{phot}$ and ${
\xi}$). We find a lower bound of ${R}_{atm} = {1.3}\pm 0.9$ $\times {10}^{4}$ $GM/{c}^{2}$ where a filling factor of $f \sim 0.5$ results in a radius at which the total (gravitational plus transverse) redshift agrees with the observed shift at the $1\sigma$ level, and in much better agreement with our results from 4U 1916$-$053. 

The low ionization and absorbing column for Zone 2, in contrast, suggest absorption at much larger radii. However, this absorption zone appears to be redshifted as well and by a similar amount to Zone 1. The redshift is poorly constrained and significant only slightly above the $1\sigma$ level, suggesting that Zone 2 could still be consistent with absorption in the outer disk. Alternatively, this absorption could represent higher density clumps that are co-spatial with the main absorption from Zone 1, resulting in a two-phase inner disk atmosphere. These clumps could even represent a higher-denser component at lower scale-heights of the disk atmosphere which only narrowly intercepts our line-of-sight, resulting in very low absorbing columns. The latter two scenarios require filling factors ranging from ${10}^{-3}$ down to ${10}^{-4}$ in order to match the observed redshift.

\subsection{AX J1745.6-2901}\label{sec:fAX}

The single Chandra/HETG spectrum of AX J1745.6$-$2901 was observed as a part of a large joint monitoring campaign with XMM-Newton and NuSTAR \citep{Ponti2018a}. Although the Chandra spectrum is omitted, the analysis by \cite{Ponti2018a} clearly established the presence of absorption lines in the Fe K band of the XMM-Newton spectra owing to a recurring, transient, photoionized disk atmosphere which can be observed when the source is a soft state and displays near-constant ionization and absorbing column values. This photoionized absorption disappears during hard spectral states, possibly as a result of (a) over-ionization from hard X-rays, (b) a lower absorbing column correlated to a drop in mass accretion rate during low-luminosity states, or perhaps more likely (c) a combination of both. This is a clear indication that the absorption is local to the system and does not originate in the ISM. Curiously, this state dependence mirrors the well-documented anti-correlation between disk winds and the presence of a jet during hard states \citep{Miller2006,Miller2008,Neilsen2009,King2012,Miller2012,Ponti2012}. This specific Chandra observation has been used in previous analyses \citep[e.g.][]{Mossoux2017,Jin2018,Wang2020,Subroweit2020} which focus on the dust-scattering halo around the source, the environment in the galactic center, and/or Sag A* itself, making this the first analysis of the HETG spectrum of AX J1745.6$-$2901.

\begin{figure}
\centering
\subfloat{\includegraphics[width=0.46\textwidth,angle=0]{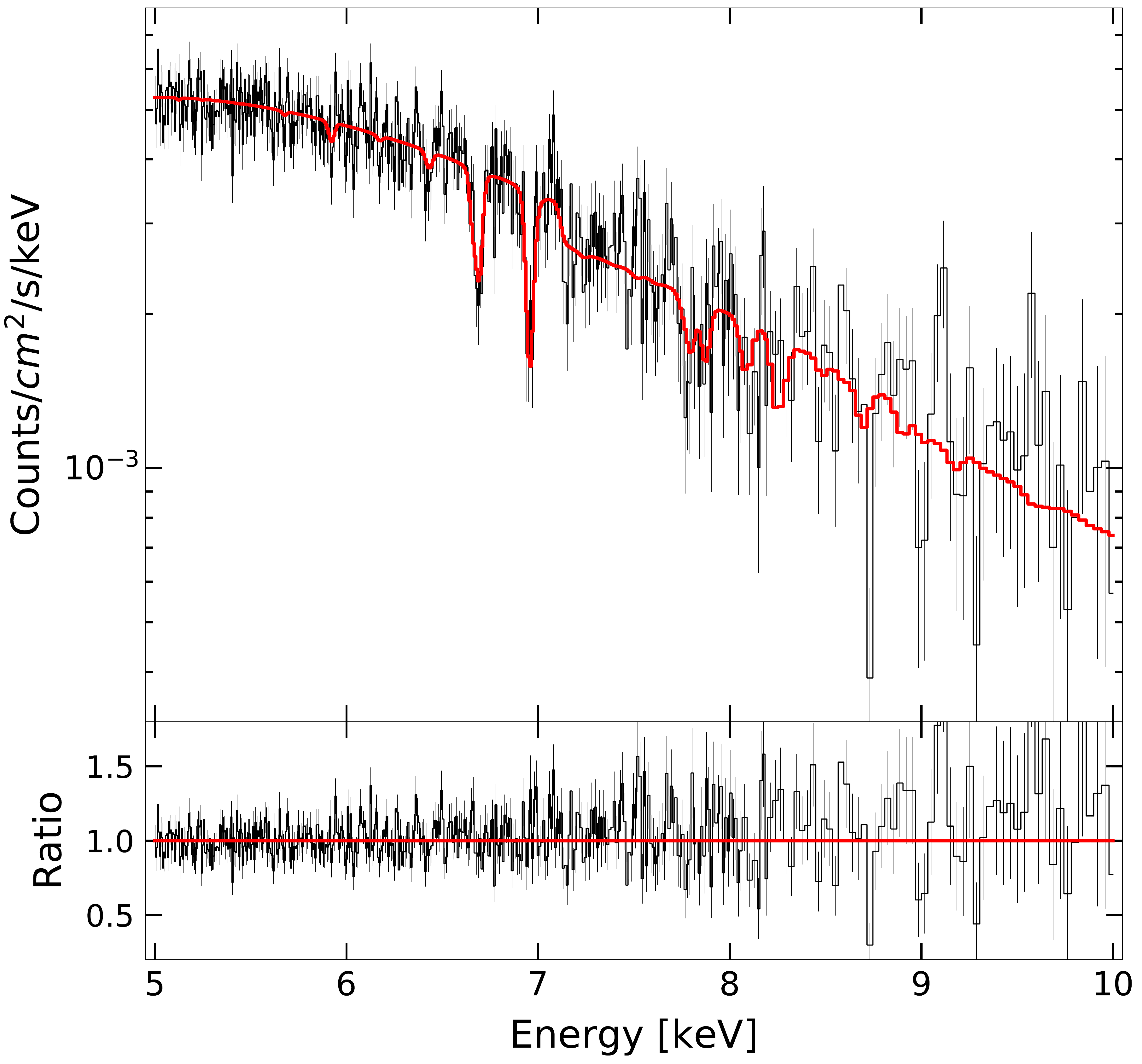}}
  \caption{\footnotesize Chandra/HETG spectrum of AX J1745.6$-$2901, best-fit model plotted in red.}\label{fig:AX_spec}
\end{figure}

AX J1745.6-2901 is located near the galactic center and, unlike 4U 1916$-$053 and XTE J1710$-$281, it is absorbed by a large neutral ISM column (${N}_{H,ISM} \sim 32 \times$ ${10}^{22}$ ${\text{cm}}^{-2}$; \citealt{Ponti2018a}) and therefore a significant amount of the spectrum below 4 keV is lost. We were unable to find statistically significant absorption lines in these regions, although this could be attributed to the low S/N in these portions of the spectra. The spectrum features strong Fe XXV and Fe XXVI absorption lines at $\sim6.7$ and $\sim6.97$ keV, respectively, as well as their corresponding $\beta$ transitions, along with Ni absorption, above 7.5 keV (see Figure \ref{fig:AX_spec}).

In view of the high column density and its effect on the low-energy portion of the spectrum, we restricted our analysis to the 5-10 keV portion of the HEG spectrum. In this range, we are able to describe the continuum using only a blackbody component (${kT}_{bb} = 1.4 \pm 0.1$ and normalization of ${K}_{bb} ={2.5}^{+0.9}_{-0.1} \times {10}^{12}$ ${\text{cm}}^2$), modified by a neutral absorber. 

Our best-fit model shown in Figure \ref{fig:AX_spec} listed in Table \ref{tab:all_fits} largely agree with results from \cite{Ponti2018a} on the XMM-Newton spectrum of AX J1745.6$-$2901 during soft states. The best-fit ${N}_{H}$ value of ${96}_{-56}^{+4\dagger} \times$ ${10}^{22} {\text{cm}}^{-2}$ is poorly constrained and is truncated by the upper limit on its fitting range (Compton-thick regime), yet it is well within $1\sigma$ of those found in the XMM-Newton CCD spectrum ($\sim {40} \pm 30 \times$ ${10}^{22} {\text{cm}}^{-2}$), as is the case with our well-constrained log $\xi$ values compared to most fits reported in \cite{Ponti2018a}. These ionizations and columns suggest a filling factor scaled radius (assuming an NS mass of $1.4{M}_{\odot}$) of ${R}_{atm} = f\cdot{4500}^{+4800}_{-2500\ddagger}$ (${}^{+3500}_{-2500\ddagger}$ if uncorrelated) $GM/{c}^{2}$ (this work) and ${R}_{atm} = f\cdot 1.1 \pm 1.0 \times {10}^{4}$ $GM/{c}^{2}$ (\citealt{Ponti2018a} values), though the latter assumes the same photoionizing luminosity as the Chandra observation and were included simply for comparison. These upper limits place the disk atmosphere within the innermost $\sim {10}^{4}$ $GM/{c}^{2}$ of the disk, consistent with a possible gravitational redshift. The previously mentioned limitations involving this spectrum result in comparatively poor velocity shift constraints. As reported in Table \ref{tab:all_fits}, our best-fit velocity shift of ${v}_{z} = {270} \pm 240$ is significant only slightly above the $1\sigma$ level. In addition, dispersed photons from an extended dust-scattering halo \citep{Jin2018} could also be responsible for the lack of sensitivity and loss of spectral resolution in this spectrum.

We briefly note that we tested an alternative continuum model using a thermal comptonization model (COMT in SPEX) with a plasma temperature fixed above 50 keV in order to determine whether our choice of continuum had an effect on the ionization. This resulted in a worse statistical fit (${\chi}^{2}/\nu = 502/431$), though the best-fit ionization (log $\xi$ $\sim 3.84$) is within errors of the model listed in Table \ref{tab:all_fits}.

\subsection{Physical Implications}

It is notable that, at the time of writing, these redshifted atmospheres are found exclusively in short period systems. Although redshifted absorption has been reported in the Chandra/HETG spectra of some BH LMXBs such as GRS 1915$+$105 \citep{Miller2020} and MAXI J1305$-$704 \citep{Miller2014}, the former likely represents a ``failed wind'' while the latter suffers from instrumental issues (though we note that MAXI J1305$-$704 has a relatively short orbital period of $\sim9.36$ hours, \citealt{Mata2021}).

A possible explanation likely involves a selection effect with the geometry of these systems: though some numerical models predict observable ionized Fe absorption from disk atmospheres originating much closer to the compact object than $1000$ $GM/{c}^{2}$ (see \citealt{Roz2011}), nominally hydrostatic atmospheres may have intrinsically smaller scale-heights \citep[e.g.][in AGN]{Roz2015} compared to the outflowing disk winds observed in other NS and BH LMXBs and, therefore, may require higher viewing angles in order to intercept the observer's line-of-sight. A near edge-on viewing angle may be relatively unobstructed in short-period systems, while absorption in the outer radii of sources with much larger disks (or the disk itself) may obstruct the line-of-sight \citep[see][]{Jimenez2002}. The disk atmospheres in these lower luminosity ($L < {10}^{37}~\text{erg~s}^{-1}$) short-period sources share a similar state dependence to disk winds in more luminous BH LMXBs ($L \sim {10}^{38-39}~\text{erg~s}^{-1}$; see \citealt{Ponti2014,Miller2015,Bianchi2017}), suggesting perhaps a similar physical origin that is sensitive to the mass accretion rate. 

\section{The Central Engine Model}\label{sec:RCE}

\begin{figure*}
\begin{center}
\subfloat{\includegraphics[width=0.5\textwidth,angle=0]{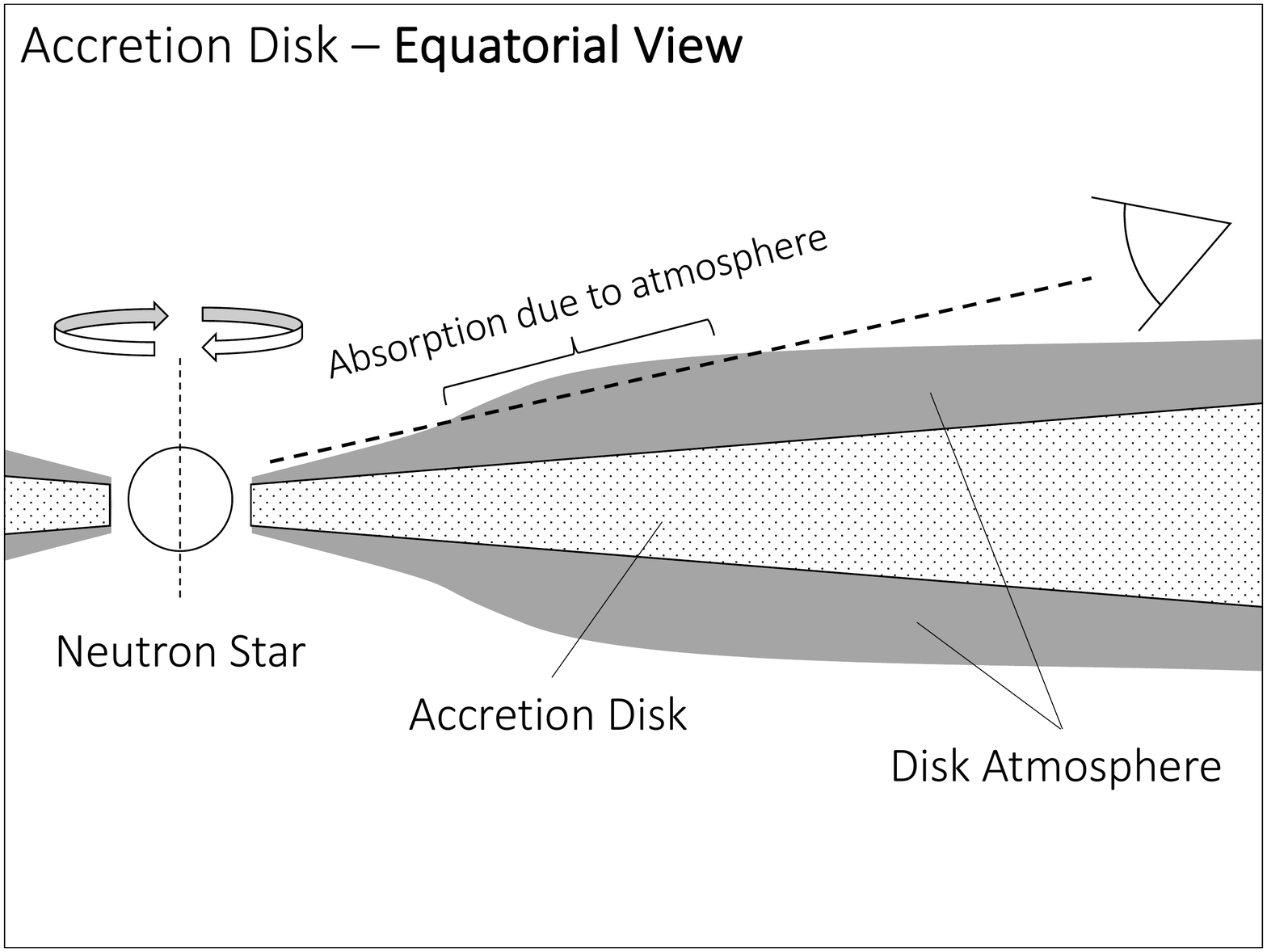}}
\subfloat{\includegraphics[width=0.5\textwidth,angle=0]{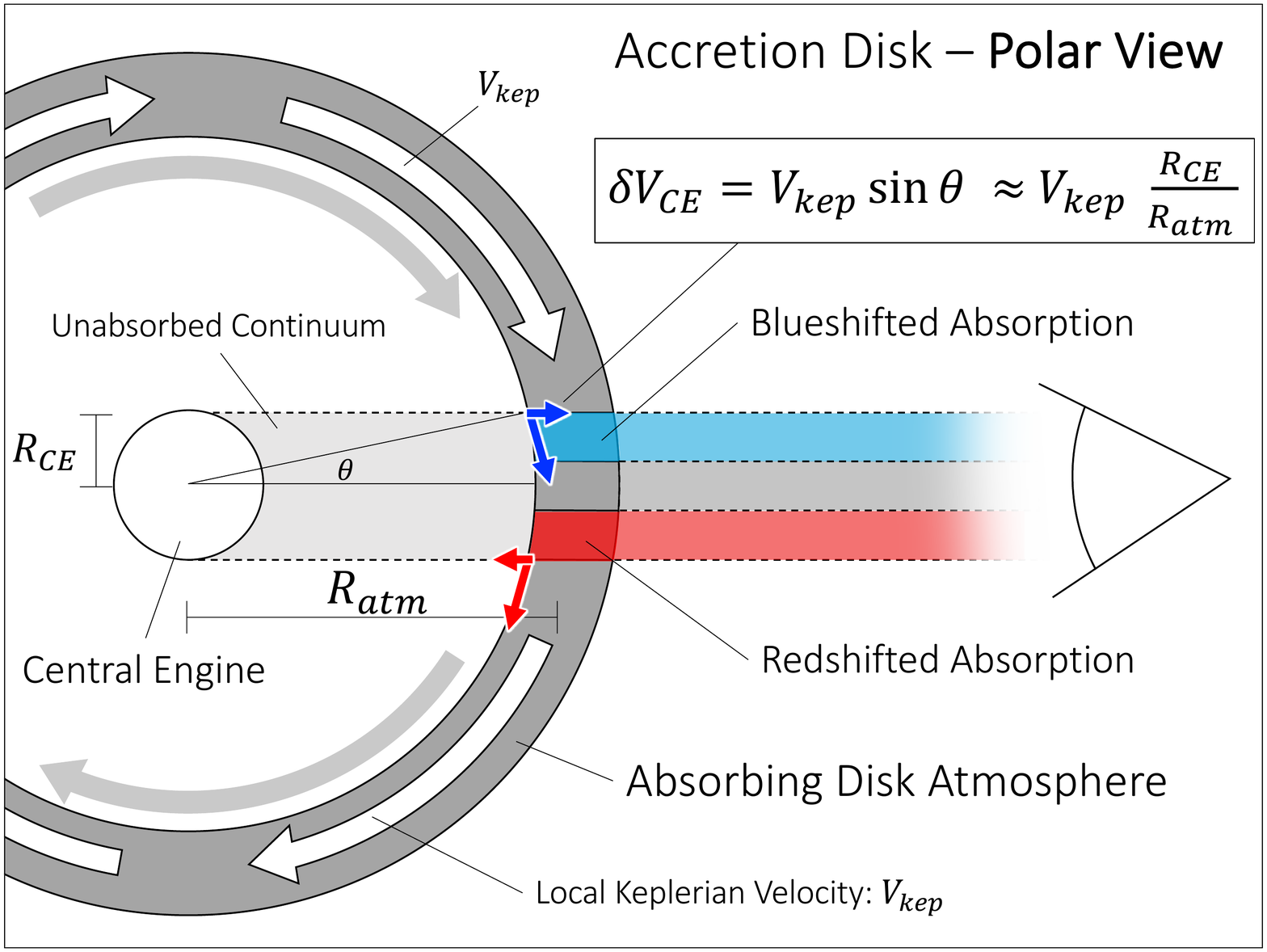}}
\end{center}
\vspace{-0.2in}
\caption{Schematic of the line broadening effects caused by the size of the central engine. The left panel shows a cross$-$section of a plausible, albeit exaggerated geometry for the absorbing disk atmosphere. The right panel shows a polar view of the same disk geometry. The gray ring represents the absorbing portion of the disk atmosphere that is along the line of sight (as seen in the left panel). The circular region labelled as the central engine in right panel encompasses all of the central X-ray emitting components of the system, including the emitting surface of the neutron star, the inner radii of the disk, and/or any corona that may be present and, therefore, ${R}_{CE}$ represents a weighted average (see Section \ref{sec:discussions} for more detail).}\label{fig:RCE}

\end{figure*}

Studies of absorption phenomena, especially in accreting compact objects, typically treat the central emitting region as a point source. The underlying assumption is that that the entire ``face'' of the emitting area is absorbed by gas with near constant properties in the plane orthogonal to the line-of-sight, though still allowing variation along the line-of-sight. Partial covering absorbers in AGN \citep[e.g.][]{Reeves2009,Gallo2015} are a notable exception in which only fraction of the total emitting area is absorbed while rest of the emission passes unabsorbed, though for the most part this effect is noticeable only as it pertains to the shape of the continuum. 

The methods and results from \cite{Calvet1993} on the absorption from outflows in FU Ori are particularly relevant to this work: they developed physically and geometrically motivated wind models which accounted for the different velocity components of the absorbing gas along the line-of-sight to different portions of the emitting area (consisting of the extended inner regions of the disk; many stellar radii in size). They found that disk wind and stellar wind models produced noticeably different absorption line profiles as a product of these geometric effects, and that the disk wind was a better description of the observed line profiles. 

As in FU Ori systems, certain absorption phenomena in X-ray binaries (disk winds and atmospheres, in particular) can produce similar, albeit subtle observational effects due to their geometry. These absorbers originate from the surface of the disk and therefore retain most, if not all, of their Keplerian motion. Naturally, in cases in which the absorber is located at large orbital radii (such as absorption from the outer disk; e.g. X-ray dips) the central engine can be treated as a point source - the Keplerian motion of the absorber is entirely orthogonal to the line-of-sight and therefore produces no observational signature. If this separation between the absorber and central engine is significantly smaller, however, portions of the absorbing gas will have some small component of their Keplerian motion passed along line-of-sight, resulting in some fraction of the emitting area being absorbed by blueshifted gas, while another (equal) fraction by redshifted gas. This effect is illustrated in the schematic shown in Figure \ref{fig:RCE}. The figure shows a disk geometry as seen from above, where the axis of rotation is pointing towards the page. As the separation between the central engine and the rotating absorber decreases, opposite portions of the emitting area will be absorbed by gas rotating towards (blueshifted) and away from the observer (redshifted).

Ultimately, this effect manifests itself as a form velocity broadening on \emph{absorption} lines. Crucially, the degree of this \emph{geometric} line-broadening effect (${\delta V}_{CE}$) depends only on the orbital radius of the absorbing disk-atmosphere (${R}_{abs}$, or ${R}_{atm}$ in the specific case of a disk atmosphere), its Keplerian velocity (${V}_{kep}$), and the size of the central engine (${R}_{CE}$), based on a simple geometric relationship
\begin{equation}
{\delta V}_{CE} = {V}_{kep}~sin \theta.
\end{equation}
The inclination of the source was omitted from this expression as all our sources are observed nearly edge-on, as evidenced by the presence of X-ray dips and eclipses (Figures \ref{fig:lc_1916}-\ref{fig:lc_XTE}). A factor of $sin~i$ should be included for lower inclination sources.

Assuming small angles, $sin~\theta$ becomes ${R}_{CE}/{R}_{atm}$. Re-arranging terms allows us to define the size of the central engine size as

\begin{equation}\label{eq:RCE0}
{R}_{CE} = {R}_{atm} \frac{{\delta V}_{CE}}{{V}_{kep}}.
\end{equation}

This simple expression is the central motivation for this work: if an absorber is located at an orbital distance at which this effect is non-negligible, then the size of the central engine can be constrained simply by measuring the degree of line broadening and constraining the orbital radius and velocity of the absorber. In this case, the central engine may be composed of the emitting regions of the NS, the inner radii of the disk, and/or a corona; ${R}_{CE}$, therefore represents a weighted average (see Section \ref{sec:discussions} for more detail).

An obvious limitation on the sensitivity of this method is the fact that there are likely multiple sources of line-broadening acting simultaneously on the spectrum, the most important of which is turbulent motion in the absorbing gas, itself. Moreover, the possible presence of additional absorbers with different mean  line-of-sight velocities or perhaps even a modest velocity gradient within a ``single'' absorber could affect the sensitivity of this method. As we discuss in Section \ref{sec:rce_model}, \ref{sec:discussions}, and Appendix \ref{sec:EW}, this specific type of geometric line-broadening (${\delta V}_{CE}$) displays very different behavior as compared to turbulent broadening, especially in regards to how it affects line ratios. Although we argue in the following sections that these forms of line broadening have little effect on our sensitivity, we still report our results strictly as upper limits. 

Extraneous sources of line broadening aside, the sensitivity of this method is limited primarily by our ability to constrain (a) the distance and Keplerian velocity of the absorber, and (b) the degree of line-broadening due to this geometric effect. The photoionization parameter formalism can be used to estimate this distance (as discussed in Section \ref{sec:ind}); however, this typically requires an independent constraint on the gas density of the absorber ($r=\sqrt{L/n\xi}$). Deriving a Keplerian velocity from this radius estimate would only compound this source of uncertainty when using equation \ref{eq:RCE0}, and requires a measurement of the mass of the compact object.

Gravitationally redshifted inner-disk atmospheres in ultra-compact and short-period X-ray binaries are ideal laboratories in which to apply this method. First, the location of the absorption is determined \emph{solely} by the magnitude of the measured redshift. At radii larger than $few\times10$ $GM/{c}^{2}$, the gravitational redshift is well described by the approximation $z \simeq 1/R$, where R is given in units of $GM/{c}^{2}$. Having measured a gravitational redshift, the orbital distance of the disk atmosphere is given in units of $GM/{c}^{2}$ by 
\begin{equation}\label{eq:Ratm_uncorr}
{R}_{atm} = \frac{c}{{v}_{z}},
\end{equation}
where the uncertainty in this quantity depends only on the uncertainty of the measured redshift. This expression, however, neglects an additional redshift term corresponding to the transverse Doppler effect (or, TDE) that arises from the orbital motion of the gas, ${z}_{TDE} = 0.5\cdot {v}_{\perp}^{2}/{c}^{2}=0.5\cdot{z}_{grav}$. Corrected for this effect, expression \ref{eq:Ratm_uncorr} becomes 
\begin{equation}\label{eq:Ratm}
{R}_{atm} = 1.5\frac{c}{{v}_{z}}.
\end{equation}
It is important to note that this expression is only valid in the limit that the redshift, $z$, is small; however, this approximation only begins to break down when the absorber is located at $\sim 10$ $GM/{c}^{2}$ (where the deviation is at $\sim 3\%$).

In addition, the corresponding Keplerian velocity can be derived directly from the orbital radius, provided it is given in gravitational units; the Keplerian velocity of an absorbing disk atmosphere can therefore be described \emph{solely} in terms of the measured redshift as
\begin{equation}\label{eq:Vkep}
{V}_{Kep} = \frac{c}{\sqrt{{R}_{atm}}} = \sqrt{c\cdot {v}_{z}/1.5},
\end{equation}
and therefore the uncertainty in this velocity depends only on the uncertainty of the measured redshift. Equation \ref{eq:RCE0} can be re-written in terms of the measured redshift and geometric velocity-broadening, ${\delta V}_{CE}$, as
\begin{equation}\label{eq:RCE}
{R}_{CE} = {\delta V}_{CE} \cdot\frac{{R}_{atm}^{3/2}}{c} = {\delta V}_{CE} \cdot\frac{{1.5}^{3/2}\cdot{c}^{1/2}}{{v}_{z}^{3/2}},
\end{equation}
where ${1.5}^{3/2} \simeq 1.84$.

This expression is powerful in that it reduces our sources of uncertainty to only two parameters that are directly measurable in our spectra, allowing us to constrain the size of the central engine (in units of $GM/{c}^{2}$) without significant model dependencies and uncertainties regarding the mass and luminosity of the source. Re-arranging this expression to 
\begin{equation}\label{eq:sensitivity}
{\delta V}_{CE} = {R}_{CE} \cdot\frac{c}{{R}_{atm}^{3/2}}
\end{equation}
highlights the fact that, for a constant ${R}_{CE}$, the magnitude of this effect decreases dramatically with larger values of ${R}_{atm}$.
The redshifts measured in our sources suggest orbital radii of the order of 1000 $GM/{c}^{2}$; despite their lower quality, spectra of gravitationally redshifted disk atmosphere absorption are likely much more sensitive to this effect. This direct dependence on ${R}_{atm}$ does require for this distance to be fairly constant throughout any exposures being added to produce a spectrum. This is not likely to be problematic in short observations displaying little variation in persistent flux (i.e. excluding dips and bursts), though we advise caution when adding multiple exposures across different epochs. In the case of 4U 1916$-$053, the constancy in ${N}_{He}$, log $\xi$, and (most importantly) ${v}_{z}$ throughout three epochs strongly suggests a narrow range in ${R}_{atm}$ (see \citetalias{Trueba2020}).

\subsection{Model construction}\label{sec:rce_model}

Although $\delta{V}_{CE}$ arises from a simple geometric relationship (see Figure 10), extracting this quantity from the data requires a procedure that can account for other sources of line broadening likely present in the data. The most important of these is turbulent broadening, a mechanism that can noticeably affect the equivalent widths (or, EWs) of strong absorption lines with saturated cores. These two forms of velocity broadening are not equivalent and cannot be used interchangeably in any model: Turbulent broadening is accrued by integrating multiple velocity components radially along the line of sight, whereas geometric broadening - $\delta{V}_{CE}$ - arises from multiple velocity components absorbing separate parts of the emitter.

A complete discussion on the differences between these two forms of line broadening and their effects on EWs can be found in Appendix \ref{sec:EW}. In short, any absorption lines (including saturated lines) that are primarily shaped by geometric broadening retain a relatively constant EW as geometric effects become more important, whereas the EW of saturated lines primarily shaped by turbulent broadening increases (often significantly) as turbulence increases. The ${v}_{turb}$ parameter in PION (or any absorption model) can therefore be extremely sensitive to line ratios between weak and strong (saturated) absorption lines, as the EW of weak lines does not increase with ${v}_{turb}$.
For spectra containing both weak and strong absorption lines (as is the case in all of our spectra),
any linewidth constraints using this parameter as a proxy for $\delta{V}_{CE}$ will mostly reflect the model's ability to achieve the observed line ratios and will significantly underpredict the errors for $\delta{V}_{CE}$ (confirmed via fitting tests in Section \ref{sec:RCE_results}), leading to falsely tight constraints on ${R}_{CE}$. A preferable approach would be one that allowed both forms of line-broadening to be fit simultaneously using two independent parameters, where the errors on $\delta{V}_{CE}$ were decoupled from EWs or line ratios and determined solely by its effect on the shape of the line. 

In this work, we adopted a semi-literal approach to modeling this scenario: We utilized multiple absorption components, with each component absorbing its corresponding portion of the emitting area. All components shared identical gas properties (${N}_{H}$, log $\xi$, ${v}_{turb}$, and ${v}_{z}$), with the exception of an \emph{additional} velocity shift corresponding to this geometric effect (as in Figure \ref{fig:RCE}), unique to each component. This approach allowed us to fit the degree of geometric broadening directly via the $\delta{V}$ parameter while still allowing ${v}_{turb}$ (as well as ${N}_{H}$, log $\xi$, and the mean redshift in the atmosphere) to vary freely.

In the simple case of an absorbed emitter \emph{not} subject to these geometric effects, the velocity-dependent flux, $F(V)$, is given by
\begin{equation}\label{eq:not_rce}
{F}(V) = {F}_{0} {e}^{-\tau(V)},
\end{equation}
where, for simplicity, ${F}_{0}$ represents a flat continuum. The $\tau(V)$ function is a convenient short-hand for $\tau(V) = N\cdot\sigma(V)$, where $\sigma(V)$ is the frequency-dependent (transformed to velocity space) opacity from an arbitrary absorption line, though $\sigma(V)$ can also be used to represent an \emph{entire set} of absorption lines originating from the same absorber. Note that in the following discussion, the same treatment applies regardless of which choice of line-profile (e.g. Lorenz, Gaussian, or Voigt) or absorption model (with multiple lines) is most appropriate for a given scenario; any thermal, turbulent, or natural broadening is accounted for within the $\sigma(V)$ function. 

Including geometric effects in this scenario requires separating the emitter into multiple flux components, ${dF}_{i}(V)$; individual components can be described using Equation \ref{eq:not_rce} with the distinction that each component is subject to an additional velocity shift specific to its location, ${dF}_{i}(V) \sim {dF}_{0} {e}^{-\tau(V+\Delta {V}_{i})}$ (see Appendix \ref{sec:EW} for a more detailed review).  

In an idealized description of this scenario, assuming a circular emitting area, the total absorbed flux, $F(V)$, can be described by
\begin{equation}\label{eq:RCE_real}
{F}(V) = \frac{2}{\pi}\int_{-1}^{1}{F}_{0} {e}^{-\tau(V,x)} \sqrt{1-{x}^{2}}~dx,
\end{equation}
where
\begin{equation}
\tau(V,x) = \tau(V + x\cdot\delta{V}_{CE}).
\end{equation}
The dimensionless parameter, $x$, normalizes the contribution for each absorption component: $x$ ranges from -1 to 1, therefore the velocity divergence for each component appropriately goes from $-\delta{V}_{CE}$ to $+\delta{V}_{CE}$ (see Figure \ref{fig:RCE}), and components which diverge the most in terms of velocity contribute the least in terms of area absorbed. Alternatively, a hypothetical rectangular-shaped central engine would result in a larger contribution from components where the velocity divergence is largest, meaning that that for a fixed velocity width constraint, assuming a rectangular central engine would translate to a tighter constraint on ${R}_{CE}$ compared to a circular geometry. While a rectangular geometry might in some instances be a better physical description of the system (e.g. equatorial bands in the NS surface, or thin concentric emitting rings in a disk), we adopted a circular geometry as it provides more conservative (larger) upper-bounds on ${R}_{CE}$. See Appendix \ref{sec:geometries} for a more complete discussion of different central engine geometries.

A good approximation for integral in Equation \ref{eq:RCE_real} involves the use of multiple absorbers with different velocity components. Unfortunately, photoionized absorption models such as PION require significant computational power; the benefit of adding more absorbers is quickly outweighed by the computational cost. Based on the plausible values of $-\delta{V}_{CE}$ compared to typical line-widths, the resolution of the HETG, as well as fitting experiments, we found that we could adequately describe this effect using only three absorbers (hereafter constituent absorbers): a central absorber with no \emph{additional} velocity shift, and two absorbers each corresponding to the redshifted and blueshifted (relative to the mean redshift) edges of the Keplerian absorber (in essence what is shown in Figure \ref{fig:RCE}). Expression \ref{eq:RCE_real} then becomes
\begin{equation}
{F}(V) = {F}_{0}/3 ({e}^{-\tau(V-\delta{V})}+{e}^{-\tau(V)}+{e}^{-\tau(V+\delta{V})}).
\end{equation}

This approach utilizes equal emitting areas for each absorption velocity component - in this case, dividing a circular area into three equal fractions, as described in Appendix \ref{sec:geometries}.
The fitting parameter $\delta{V}$ can be directly related to the quantity we are interested in, $\delta{V}_{CE}$, via a simple weighted integral which results in
\begin{equation}
\delta{V} \simeq 0.6\cdot\delta{V}_{CE}.
\end{equation}
Notably, assuming a rectangular-shaped central engine would shift this ratio to $0.7\cdot\delta{V}_{CE}$, resulting in ${R}_{CE}$ constraints that would be $\sim15\%$ tighter compared to those reported in Section \ref{sec:RCE_results}.

The model was built in SPEX by adapting our single-zone baseline models from Section \ref{sec:ind}. In XSPEC parlance (for simplicity), a hypothetical baseline model can be written as 
\begin{displaymath} 
\text{powerlaw}(N, \Gamma) \times PION({N}_{H},\xi, {v}_{turb}, {v}_{z}).
\end{displaymath}
We modified this baseline model by splitting the emitting area of each continuum component into three equal components with linked parameters, where the total continuum emission remained unchanged. We then replaced the single PION absorption component with three identical PION components (corresponding to each constituent absorber), with the exception of an additional fitting parameter, $\delta{V}$. The resulting model can now be written as
\begin{eqnarray}
\text{powerlaw}(N/3, \Gamma) \times PION(..., {v}_{z}+\delta{V}) +\nonumber\\
\text{powerlaw}(N/3, \Gamma) \times PION(..., {v}_{z}) +\nonumber\\
\text{powerlaw}(N/3, \Gamma) \times PION(..., {v}_{z}-\delta{V}). \nonumber
\end{eqnarray}

Due to the manner in which model components and parameters are defined in SPEX, the actual construction of the model differed slightly from the description above. In this work, the velocity parameter (${v}_{z}$) was linked between two of the PION components by a factor of $-1$, while the velocity for the remaining PION component was frozen at zero. All of the absorbers were then modified using a redshift model (REDS in SPEX) to account for the mean redshift obtained in Section \ref{sec:ind}. The resulting shifts are, in this velocity shift regime, identical to the XSPEC model description above. Again, the same absorption parameters from the baseline models (${N}_{H}$, log $\xi$, ${v}_{turb}$, and the mean redshift in the atmosphere) where linked between the three constituent absorbers and were allowed to vary as free parameters in the central engine model. 

\begin{table*}[t]
\renewcommand{\arraystretch}{1.3}
\caption{Constraints on the Size of the Central Engine from ${R}_{CE}$-model Fits}
\vspace{-1.0\baselineskip}
\begin{footnotesize}
\begin{center}
\begin{tabular*}{\textwidth}{l   l @{\extracolsep{\fill}}  c c  c  c c c c}
\tableline
\tableline
&& \multicolumn{3}{c}{{Redshift-related quantities}} & & \multicolumn{3}{c}{{Central engine radius constraints}}\\\cline{3-5}\cline{7-9}
Source name & Model & $c\cdot{z}_{atm}$ & $R_{atm}$ & ${V}_{kep}$ & ${\delta V}_{CE}$ & ${R}_{CE}$& \multicolumn{2}{l}{Upper limit on ${R}_{CE}$}\\
&&($\text{km}$ $\text{s}^{-1}$)  & (${GM}/{c}^{2}$) & ($\text{km}$ $\text{s}^{-1}$) & ($\text{km}$ $\text{s}^{-1}$)  & (${GM}/{c}^{2}$) & ($3\sigma$) & ($5\sigma$)\\

\tableline 
XTE J1710$-$281&
1-zone  & 
${290} \pm 40$ &
${1500}^{+300}_{-150}$ &
${7700}^{+500}_{-600}$ &
${180}^{+130}_{-130\ddagger}$ &
${40}^{+20}_{\ddagger}$ &
${<90}$ & ${<130}$\\

&
2-zone  & 
${280} \pm 50$ &
${1600}^{+400}_{-200}$&
${7500}^{+600}_{-700}$&
${50}^{+230}_{\ddagger}$ &
${<60}$ &
${<90}$ & ${<140}$\\

\\ [-3.0ex]
\tableline 

4U 1916$-$053&
1-zone   & 
${210} \pm 60$ &
${2100}^{+900}_{-400}$ &
${6500}^{+900}_{-1000}$ &
${340}^{+150}_{-290}$ &
${110}^{+90}_{-90}$&
${<370}$ & ${<900}$\\

&
2-zone  & 
${490}_{-170}^{+200}$ &
${900}^{+600}_{-200}$&
${9900}^{+1800}_{-1900}$&
${50}^{+420}_{\ddagger}$ &
${<60}$&
${<150}$&
${<520}$\\

\\ [-3.0ex]
\tableline 
\tableline

\end{tabular*}
\vspace*{-1.0\baselineskip}~\\ \end{center} 
\tablecomments{Central engine radius constraints (${R}_{CE}$) obtained from our best-fit ${R}_{CE}$-models; ${R}_{CE}$ was calculated using the best-fit redshift, $c\cdot{z}_{atm} (= {v}_{z})$, and geometric linewidth, ${\delta V}_{CE}$ (for a full list of model parameters, see Table \ref{tab:all_RCE_fits}), per Equation \ref{eq:RCE} (corrected for the transverse Doppler effect). Confidence regions for ${R}_{CE}$ were obtained from $\Delta{\chi}^{2}$-grids in ${z}_{atm}$ vs. $\delta{V}_{CE}$ space by calculating ${R}_{CE}$ values along contour line corresponding to each confidence level (see Figures \ref{fig:cont_XTE} and \ref{fig:cont_1916}).
We also included values for the orbital radius of the disk atmosphere ($R_{atm}$) and Keplerian velocity (${V}_{kep}$) based on the best-fit redshift, per Equations \ref{eq:Ratm} and \ref{eq:Vkep}.} 

\end{footnotesize}
\label{tab:RCE}
\end{table*}
\subsection{Results from the Central Engine Model}\label{sec:RCE_results}

Parameters for the best-fit central engine models (for simplicity, ${R}_{CE}$-model) for XTE J1710$-$281 and 4U 1916$-$053 are listed in Tables \ref{tab:RCE} (those relevant to ${R}_{CE}$) and \ref{tab:all_RCE_fits} (full parameter list). We included two versions of the central engine model for each source based on their 1 and 2 absorption zone baseline models from Section \ref{sec:ind}. 
Compared to their baseline counterparts, we found slight shifts in both continuum and absorption parameters in the ${R}_{CE}$-models (Table \ref{tab:all_RCE_fits}) that are within the 1$\sigma$ errors listed Table \ref{tab:all_fits}, and therefore not significant. The additional degree of freedom, however, did result in larger errors in the mean redshift. The redshifts (and uncertainties) used to calculate the central engine radius (or, ${R}_{CE}$) and reported in Table \ref{tab:RCE} were obtained with the ${R}_{CE}$-model. The ${R}_{CE}$-model results for AX J1745.6$-$2901 were not included in these tables due to their low significance. We briefly comment on the source at the end of this section.

\begin{table}[t]
\renewcommand{\arraystretch}{1.1}
\caption{Absorption Parameters for Best-Fit ${R}_{CE}$-models}
\vspace{-1.0\baselineskip}
\begin{footnotesize}
\begin{center}
\begin{tabular*}{0.47\textwidth}{l @{\extracolsep{\fill}}  c  c  }
\tableline
\tableline
Parameter & Zone 1 & Zone 2  \\
\\ [-3.0ex]
\tableline
\tableline
\multicolumn{3}{c}{4U 1916$-$053}\\
\tableline
\\ [-3.0ex]
\tableline
\tableline
\\ [-3.0ex]
${N}_{He}$ (${10}^{22} {\text{cm}}^{-2}$)  & ${50}_{-36}^{\dagger}$ & $-$  \\
 log ${\xi}$ & ${4.4}_{-0.5}^{+0.1}$ & $-$ \\
${v}_{z}$ ($\text{km}$ $\text{s}^{-1}$) &${210} \pm 60 $ & $-$  \\
${v}_{turb}$ ($\text{km}$ $\text{s}^{-1}$) &${150}_{-30}^{+40}$ & $-$   \\
$\delta{V}_{CE}$ ($\text{km}$ $\text{s}^{-1}$) &${340}_{-290\ddagger}^{+150}$ & $-$ \\
$\chi^{2}/\nu$ & \multicolumn{2}{r}{2153.8/2089 = 1.03}\\

\tableline 
\multicolumn{3}{c}{(2-zone model)}\\
\tableline
${N}_{He}$ (${10}^{22}$ ${\text{cm}}^{-2}$) & ${50}_{-20}^{\ddagger}$& ${5}^{+45\ddagger}_{-3}$  \\
log $\xi$  & ${4.8}_{-0.4}^{+\ddagger}$ & ${3.8}^{+1.0\ddagger}_{-0.3}$\\
${v}_{z}$ ($\text{km}$ $\text{s}^{-1}$) & ${490}_{-170}^{+200}$ & ${0}^{\dagger}$\\
${v}_{turb}$ ($\text{km}$ $\text{s}^{-1}$) & ${100}_{-50\ddagger}^{+120}$ & ${60}^{+80}_{-10\ddagger}$  \\
$\delta{V}_{CE}$ ($\text{km}$ $\text{s}^{-1}$) &${50}_{\ddagger}^{+420}$ \\
$\chi^{2}/\nu$ & \multicolumn{2}{r}{2127.2/2087 = 1.02}\\

\\ [-3.0ex]
\tableline
\tableline
\\ [-3.0ex]
\tableline 
\multicolumn{3}{c}{XTE J1710$-$281}\\
\tableline
\\ [-3.0ex]
\tableline
\tableline
\\ [-3.0ex]
${N}_{H}$ (${10}^{22} {\text{cm}}^{-2}$)  &${8}_{-4}^{+8}$& $-$  \\
log ${\xi}$ & ${3.1} \pm 0.1$ & $-$ \\
${v}_{z}$ ($\text{km}$ $\text{s}^{-1}$) & ${290} \pm 40$ & $-$   \\
${v}_{turb}$ ($\text{km}$ $\text{s}^{-1}$) &${100}_{-30}^{+30}$& $-$   \\
$\delta{V}_{CE}$ ($\text{km}$ $\text{s}^{-1}$) &${180}_{-130\ddagger}^{+130}$& $-$  \\
$\chi^{2}/\nu$ & \multicolumn{2}{r}{2460/2182 = 1.13}\\
\tableline 
\multicolumn{3}{c}{(2-zone model)}\\
\tableline
${N}_{H}$ (${10}^{22} {\text{cm}}^{-2}$)  & ${40}_{-30}^{+60\dagger}$ & ${0.5}^{+0.4}_{-0.3}$ \\
log $\xi$  & ${3.4}_{-0.2}^{+0.3}$ & ${2.6}_{-0.2}^{+0.2}$  \\
${v}_{z}$ ($\text{km}$ $\text{s}^{-1}$) & ${280} \pm 50$  & = Zone 1 \\
${v}_{turb}$ ($\text{km}$ $\text{s}^{-1}$)  & ${50}_{\ddagger}^{+40}$ & ${500}_{-190}^{\ddagger}$  \\
$\delta{V}_{CE}$ ($\text{km}$ $\text{s}^{-1}$) &${50}_{\ddagger}^{+230}$ & = Zone 1\\
$\chi^{2}/\nu$ & \multicolumn{2}{r}{2397/2179 = 1.1}\\

\\ [-3.0ex]
\tableline
\tableline

\end{tabular*}
\vspace*{-1.0\baselineskip}~\\ \end{center} 
\tablecomments{Best-fit parameter values for the ${R}_{CE}$-model that best describes each spectrum. The ${R}_{CE}$-model was applied only on observation 1 of XTE J1710$-$281. All errors are at the 1$\sigma$ level. Errors truncated by the parameter fitting range are marked with a $\ddagger$ symbol, frozen parameters with a $\dagger$ symbol. Parameter values are within errors of the best-fit baseline parameters from Table \ref{tab:all_fits}, with the exception of the additional geometric velocity broadening parameter, $\delta{V}_{CE}$.}
\end{footnotesize}
\label{tab:all_RCE_fits}
\end{table}

In complex models, error estimates based on single-parameter scans or based on propagating errors from connected quantities can under-estimate the true errors. Our solution was simply to compute ${\chi}^{2}$ grids in ${z}_{atm}$ vs $\delta{V}_{CE}$ parameter space (the two parameters required to compute ${R}_{CE}$), and then extracting contour lines at each value of $\Delta{\chi}^{2}$ corresponding to the desired significance level. The largest and smallest values of ${R}_{CE}$ along each contour therefore represent the bounds of our confidence regions, a much better representation of the uncertainty in this constraint. We set a lower limit of 50 \kms on the fitting range of the \dvce parameter for all fits. This \dvce value corresponds to an emitter with a radius of 10 \rce absorbed by a disk atmosphere redshifted located at 1500 \gmcc (or, redshifted by $\sim300$ \kms), per Equations \ref{eq:Ratm}-\ref{eq:sensitivity}. This limit is large compared to most plausible values of neutron star radii though it is appropriate given the sensitivity of the data: the threshold at which the model becomes insensitive to small values of \dvce is well above the 50 \kms limit and, therefore, the smallest possible ${\chi}^{2}$ value should be contained within our ${\chi}^{2}$ grids. Crucially, this means that the positive errors on \rce (calculated via ${\Delta \chi}^{2}$) are unaffected by this limit.

We note that all constraints on \dvce and \rce should likely be treated as upper-limits: The \rce-model treats both \dvce and ${v}_{turb}$ as free parameters and, though we do not expect much degeneracy between these two forms of velocity broadening (see Appendix \ref{sec:EW}), the errors on \dvce may attribute some fraction of the turbulent velocity broadening to the central engine. The positive errors on \dvce provide a robust upper-limit, as they allow \dvce to encompass as much of the velocity broadening as is statistically acceptable, including those not accounted for in the model. Conversely, negative error bars on \dvce may be underestimated if this parameter is responsible for other unmodeled sources of velocity broadening. Although none of our spectra display a preference for larger \dvce values (above the 1$\sigma$ level), we advise caution when interpreting lower bounds on any constraint on \rce.

\begin{table}[t]
\renewcommand{\arraystretch}{1.3}
\caption{${R}_{CE}$-model results for XTE J1710$-$281}
\vspace{-1.0\baselineskip}
\begin{footnotesize}
\begin{center}
\begin{tabular*}{0.47\textwidth}{l  @{\extracolsep{\fill}}  c c c c c}
\tableline 
\tableline

Range & $c\cdot{z}_{atm}$ & ${\delta V}_{CE}$ & ${R}_{CE}$& \multicolumn{2}{c}{Upper limits}\\
 &($\text{km}$ $\text{s}^{-1}$) & ($\text{km}$ $\text{s}^{-1}$)  & (${GM}/{c}^{2}$) & ($3\sigma$) & ($5\sigma$)\\

\tableline 

A&
${300}^{+60}_{-50}$ &
${250}^{+160}_{-200\ddagger}$ &
${50}^{+40}_{\ddagger}$ &
${<150}$ & ${<260}$\\

\textbf{B}&
${290} \pm 40$ &
${180}^{+130}_{-130\ddagger}$ &
${40}^{+20}_{\ddagger}$ &
${<90}$ & ${<130}$\\

C&
${290} \pm {40}$ &
${230}^{+100}_{-120\ddagger}$ &
${50}^{+20}_{-30\ddagger}$ &
${<90}$ & ${<120}$\\

\\ [-3.0ex]
\tableline 
\tableline

\end{tabular*}
\vspace*{-1.0\baselineskip}~\\ \end{center} 
\tablecomments{Constraints from the ${R}_{CE}$-model of XTE J1710$-$281, using three different fitting ranges of the HETG spectrum and binning schemes. Range A is the baseline fitting range, using the MEG from 1.0-2.1 and 3.0-4.5 keV, and the HEG from 4.5-8.5 keV, both optimally binned. Range B is the same as A, except a few HEG bins around the strongest lines in the low energy band were included (also optimally binned). Range C included the entire HEG from 1.0 to 2.1 keV, more coarsely binned to a minimum S/N of 1.}

\end{footnotesize}
\label{tab:RCE_cons2}
\end{table}

The best-fit single absorption zone ${R}_{CE}$-model for XTE J1710$-$281 reported in Table \ref{tab:RCE} was obtained by using the same Chandra/HETG fitting range used in Section \ref{sec:ind} with exception that, instead of entirely omitting the HEG below 4.5 keV, a \emph{few} HEG spectral bins at locations where strong lines known to be present in the MEG spectrum: these are 1.01 to 1.03 keV (Ne X), 1.45 to 1.5 keV (Mg XII) and from 1.98 to 2.02 keV (Si XIV). The entire HEG was \emph{not} included in this lower energy band due to the continuum being relatively noisy in a region that contains multiple weak lines, while binning the data more heavily (compared to optimal binning) partially defeated the purpose of including portions of the higher-resolution HEG arm. However, we did find that the strong Ne X, Mg XII, and Si XIV lines identified in the MEG spectrum to be highly significant in the HEG spectrum, as well. 

Constraints on ${R}_{CE}$ for XTE J1710$-$281 using different fitting ranges are listed in Table \ref{tab:RCE_cons2}. For comparison, we also included a range C fit, using the entire HEG from 1.0-2.1 keV. Here, the HEG is more heavily binned (minimum S/N of 1 using the ``vbin'' command in SPEX) in the lower energy band.

Compared to ${R}_{CE}$-model fits that adhere strictly to the baseline fitting range (range A), we obtained noticeably tighter redshift, $\delta{V}_{CE}$, and, therefore, ${R}_{CE}$ constraints by adding these few HEG bins around strong lines (range B). Again, values of ${N}_{H}$, $\xi$, ${v}_{turb}$, and ${z}_{atm}$ (${v}_{z}$) are well within 1$\sigma$ errors of, \emph{and virtually identical} to, those in the baseline model (see Table \ref{tab:all_RCE_fits}). Including additional (albeit more heavily binned) HEG spectral bins in range C resulted in a slight loss of sensitivity to $\delta{V}_{CE}$, while ultimately having little effect on the final ${R}_{CE}$ constraints. This constancy between B and C however, is also reflected in the parameters which are \emph{not} listed in Table \ref{tab:RCE_cons2}, where we again find that values and errors of ${N}_{H}$, $\xi$, and ${v}_{turb}$ using range C are \emph{virtually identical} to those obtained using both range A and B.

These results are reassuring: all three fits effectively represent the same minimum in parameter space - the improvement between A and B is therefore most likely the result of increasing the sensitivity of our combined data set by including high resolution HEG bins. In the interest of not including a large number of coarsely binned spectral bins, we adopted range B when reporting our results.

The constraints on the central engine for Chandra/HETG spectrum of XTE J1710$-$281 suggest upper bounds on the radius of 60 $GM/{c}^{2}$ and 90 $GM/{c}^{2}$ at the 1 and 3$\sigma$ levels, respectively. These correspond to 1 and 3$\sigma$ limits of 120 and 190 km for a $1.4{M}_{\odot}$ central neutron star. Although we report these values as upper limits due to the possibility of extraneous sources of line broadening that may affect our results, the best-fit ${R}_{CE}$ value itself suggests a point-like central engine. Given the sensitivity of the spectrum, the fit is unable to differentiate between ${R}_{CE}$ values below 40~$GM/{c}^{2}$.

\begin{figure}
\subfloat{\includegraphics[width=0.48\textwidth,angle=0]{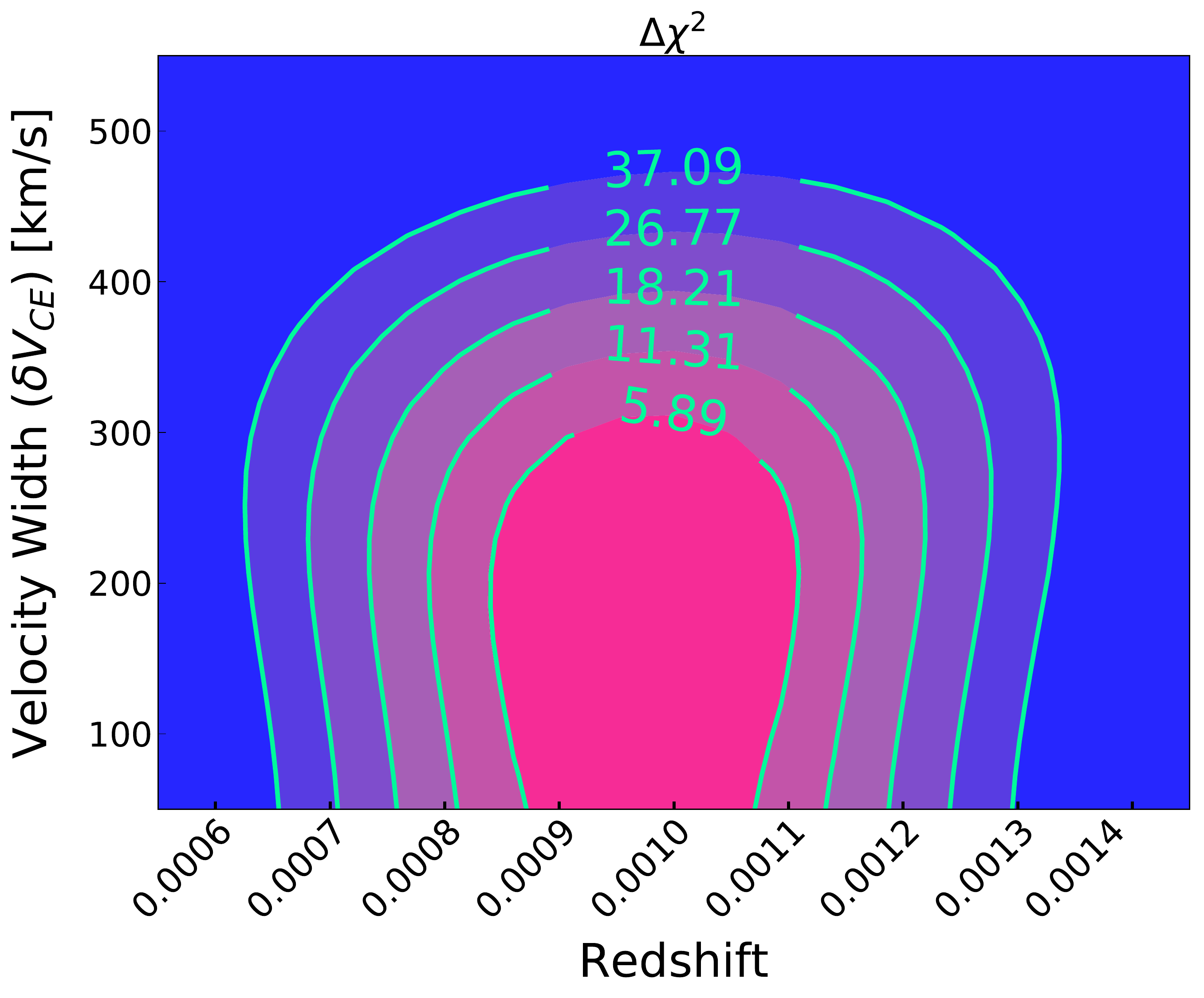}}
\centering
  \caption{\footnotesize Contour plot of $\Delta{\chi}^{2}$ grid for the ${R}_{CE}$-model of XTE J1710$-$281 (single-zone), in redshift vs $\delta{V}_{CE}$ space. Labelled contours indicate the $\Delta{\chi}^{2}$ level corresponding to 1-5$\sigma$ confidence regions. The values and errors for $\delta{V}_{CE}$ and ${z}_{atm}$ listed in Table \ref{tab:RCE} were obtained from this grid. Errors on ${R}_{CE}$ were obtained by calculating ${R}_{CE}$ values along each contour line and extracting the largest and smallest values for each confidence level.}\label{fig:cont_XTE}
\end{figure}

The model's preference for a small central engine ($<40 GM/{c}^{2}$, or $<80$ km for a $1.4{M}_{\odot}$ NS) is unsurprising considering the size scales we are probing, such as a neutron star and/or the inner radii of an accretion disk (especially in weakly magnetized neutron stars that display X-ray bursts). However, it is important to consider whether these results represent meaningful constraints on the size of the central engine, albeit limited by the sensitivity of the data, or whether we are simply probing the underlying uncertainty of other forms of velocity broadening, such as microturbulence.

As a simple test, we used our baseline fits to obtain the uncertainty on the ${v}_{turb}$ parameter to the same 1 to 5$\sigma$ significance levels probed for the ${R}_{CE}$-model, essentially treating the ${v}_{turb}$ parameter as a proxy for the geometric half-width parameter $\delta{V}_{CE}$. We obtain upper bounds on ${v}_{turb}$ of 120, 170, and 260 $\text{km}$ $\text{s}^{-1}$ for 1, 3, and 5$\sigma$ sigma errors, respectively. By comparison, the ${R}_{CE}$-model using the same fitting range (Range A, see Table \ref{tab:RCE_cons2}) suggests upper bounds of 410, 520, and 630 $\text{km}$ $\text{s}^{-1}$ on the velocity half-width. Note that the 5$\sigma$ bounds on ${v}_{turb}$ still fall short of the 1$\sigma$ bounds on $\delta{V}_{CE}$ - this cannot be attributed to increased complexity of the model as only one free parameter was added. It is clear that ${v}_{turb}$ significantly under-predicts the degree of geometric broadening allowed by the spectrum and, as expected, appears to be more sensitive to line-ratios (see Section \ref{sec:rce_model} and Appendix \ref{sec:EW}). Note that even if adding geometric broadening does not result in a statistically improvement in the fit that would suggest a large central engine that is apparent in the data (as appears to be the case in our spectra), these upper-limits are still robust in that they rule out ${R}_{CE}$ values that would \emph{necessarily} produce a statistically observable effect on the spectrum.

\begin{figure*}
\subfloat{\includegraphics[width=0.99\textwidth,angle=0]{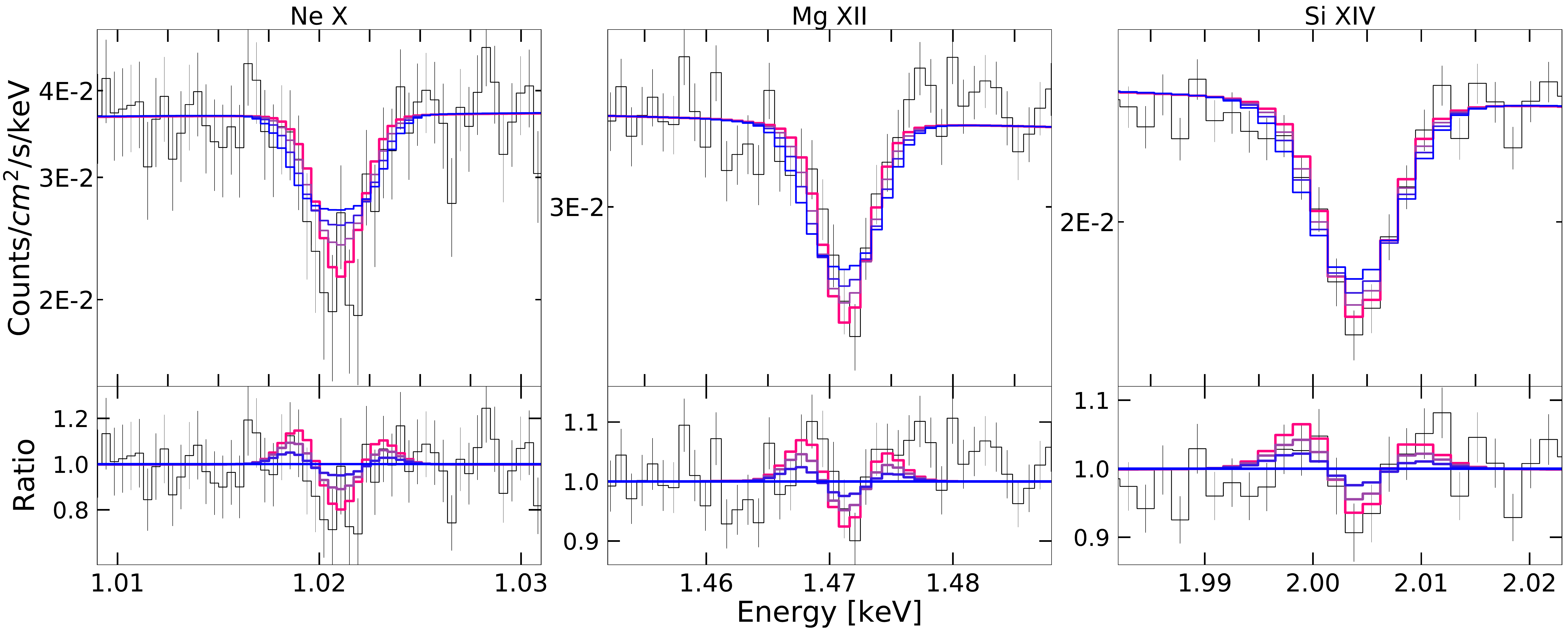}}
\centering
  \caption{\footnotesize ${R}_{CE}$ model for 4U 1916$-$053 for select lines. Best-fit model is plotted in fuchsia, while the positive 5$\sigma$ error on $\delta{V}_{CE}$ plotted is in blue. 1 and 3$\sigma$ errors on $\delta{V}_{CE}$ were plotted in magenta to show intermediate levels of confidence between the best-fit model and those at 5$\sigma$. Bottom panels show residuals to model at the $+5\sigma$ error of $\delta{V}_{CE}$ to illustrate the improvement to the fit as $\delta{V}_{CE}$ approaches its best-fit value.}\label{fig:1916_RCE}
\end{figure*}

\begin{figure}
\subfloat{\includegraphics[width=0.48\textwidth,angle=0]{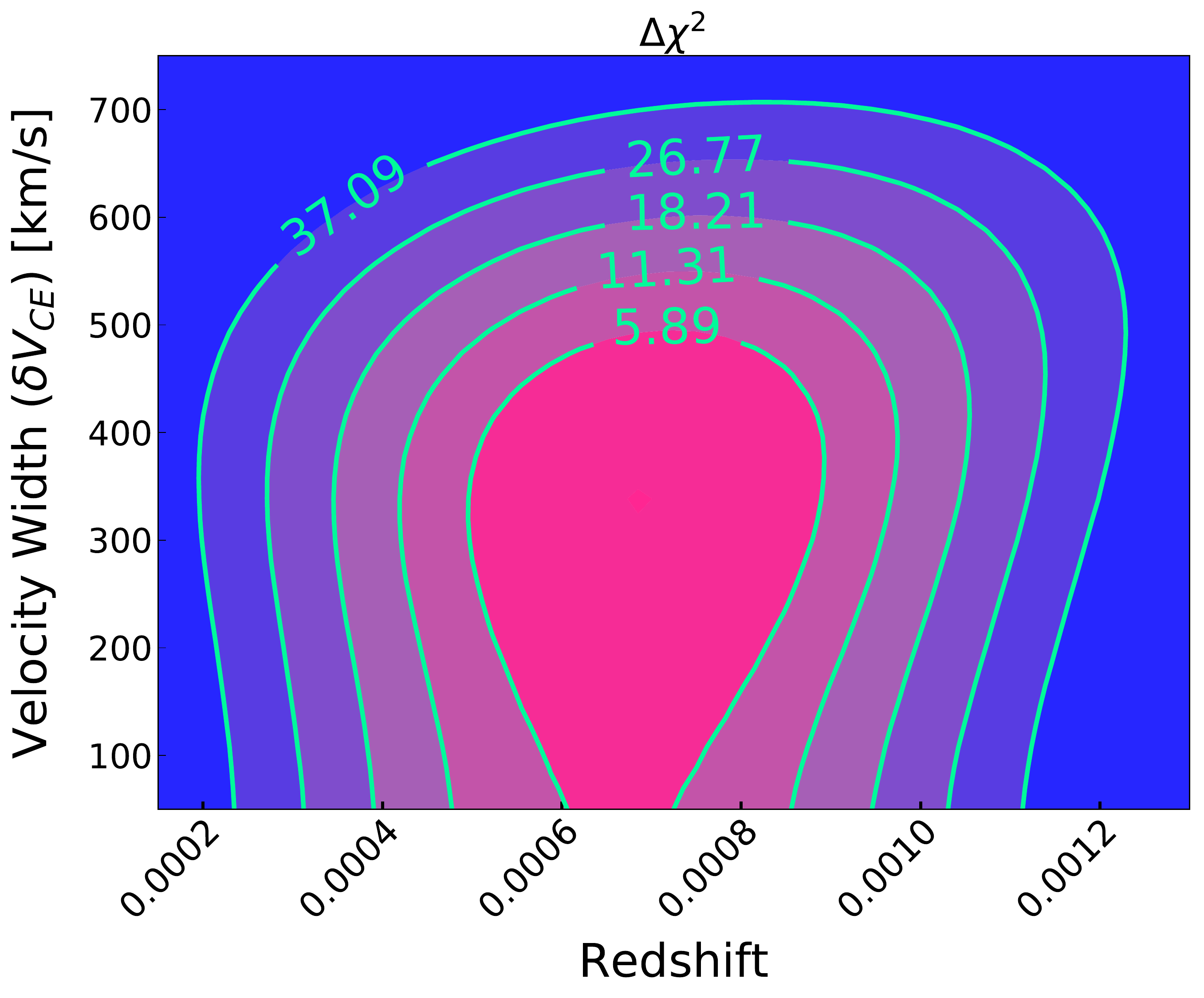}}
\centering
  \caption{\footnotesize Contour plot of $\Delta{\chi}^{2}$ grid for the ${R}_{CE}$ model of 4U 1916$-$053 (single-zone), in redshift vs $\delta{V}_{CE}$ space. Labelled contours indicate the $\Delta{\chi}^{2}$ level corresponding to 1-5$\sigma$ confidence regions. The values and errors for $\delta{V}_{CE}$ and ${z}_{atm}$ listed in Table \ref{tab:RCE} were obtained from this grid. Errors on ${R}_{CE}$ were obtained by calculating ${R}_{CE}$ values along each contour line and extracting the largest and smallest values for each confidence level.}\label{fig:cont_1916}
\end{figure}

Also included in Table \ref{tab:RCE} is the ${R}_{CE}$-model based on the two absorption zone baseline fit of XTE J1710$-$281, as a point of comparison. This model was constructed by assuming both absorption zones are approximately co-spatial and, therefore, each was split into three constituent absorbers (6 in total) in order to model the effect of the central engine for both absorption zones. The mean redshift and velocity width were coupled to the same value for both zones. Naturally, this 2-zone model introduces more model dependencies; given the quality of the spectrum, we advise caution when interpreting these specific results. It is worth noting, however, that the constraints on the central engine are virtually identical between the 1 and 2-zone models. 

The single zone ${R}_{CE}$-model for 4U 1916$-$053, by comparison, yields noticeably worse constraints on ${R}_{CE}$, suggesting upper bounds of 200 $GM/{c}^{2}$ and 370 $GM/{c}^{2}$ for 1 and 3 $\sigma$ errors, respectively. The lower sensitivity of these constraints is primarily driven two factors: First, the lower magnitude of the observed redshift significantly reduces the sensitivity of the method, as given by equations \ref{eq:RCE} and \ref{eq:sensitivity}. Second, the lower quality and fewer absorption lines in this spectrum further reduce our ability to constrain the velocity broadening, resulting in large upper bounds on the $\delta{V}_{CE}$ parameter. Figure \ref{fig:1916_RCE} still shows the model's preference for smaller degrees of geometric velocity broadening.

As with XTE J1710$-$281, we tested the effect of adding portions of the HEG in the 1 to 3 keV region at energies where we identify the strongest lines in the MEG spectrum. We found that, due to the low S/N of the HEG spectrum at these energies, there was little improvement to our constraints by adding these bins. We therefore omit all HEG bins for this source below 6 keV; the fits reported in Table \ref{tab:RCE} for 4U 1916$-$053 were performed using the same baseline fitting range from Section \ref{sec:ind}.

By comparison, we obtained significantly tighter constraints from the two absorption zone ${R}_{CE}$-model for 4U 1916$-$053. Unlike XTE J1710$-$281, this 2-zone fit was performed assuming the outer absorption zone (Zone 2 in Table \ref{tab:all_fits}) originates in the outer disk and has a fixed velocity shift of 0 (see Section \ref{sec:f1916} for details). Velocity widths in this absorber should not be affected by the size of the central engine and, therefore, only the redshifted absorption zone (Zone 1) was split into three constituent absorber. Again, especially in light of the significant improvement in our constraints, we advise caution interpreting this particular result due to the increased model dependency of the 2-zone fit. Moreover, the improvement in our constraints is not the result of improved velocity width constraints: if the large velocity widths observed in our 1-zone fit are indeed the result of some velocity structure, the addition of a second absorption component should afford the model enough flexibility to account for this velocity width. Instead, we see our velocity width constraints \emph{worsen} with the addition of the absorber, due in part to the added degrees of freedom. The improved constraints on ${R}_{CE}$ are the product of the much larger magnitude of the mean redshift of the atmosphere, which increases the sensitivity of the method as described by equation \ref{eq:sensitivity}.

As discussed in Section \ref{sec:f1916}, the larger redshift measured for Zone 1 in the 2-zone fit of 4U 1916$-$053 could indeed be real, in which case the improved constraints on ${R}_{CE}$ would be perfectly valid. However, further Chandra/HETG observations of this source are required to make a more compelling case for this scenario due to the complexity of the model. In this work, we report the results of our 1-zone fit as our primary results for this source.

Finally, we applied the ${R}_{CE}$-model to the low S/N HEG spectrum of AX J1745.6$-$2901. We found that, due to the additional degree of freedom ($\delta{V}_{CE}$) of the more complex ${R}_{CE}$-model, the mean redshift in the disk atmosphere (significant to less than 2$\sigma$ in the baseline fits) yielded a significance below $1\sigma$ when using ${\chi}^{2}$-statistics. This made it impossible to place constraints on ${R}_{CE}$ at the $1\sigma$ level, as ${R}_{atm}$ is unconstrained. Using Cash statistics \citep{Cash1979}, and the $\Delta$C-statistic as a proxy for $\Delta{\chi}^{2}$, we obtained an upper limit of ${R}_{CE} <1300$ $GM/{c}^{2}$ at the $1\sigma$ level. This large upper bound approaches the scale of ${R}_{atm}$ (the distance between the central engine and the disk atmosphere), which is, naturally, the hard upper limit on ${R}_{CE}$. Despite these poor results, we briefly highlight this source as the sensitivity and spectral resolution in the Fe~K band afforded by near-future missions such as XRISM \citep{Tashiro2018} will likely greatly improve constraints on AX J1745.6$-$2901, in particular. 

\section{Discussion}\label{sec:discussions}

We have shown that, at small orbital radii, absorbers originating from the surface of the disk (such as disk atmospheres) are affected by an additional form of line broadening: a geometric effect between the orbital motion of the absorbing gas and the physical size of the central emitting region of the system. We developed a spectral model capable of constraining the magnitude of this velocity broadening effect, taking into account stark differences in behavior compared to most other forms of line broadening (namely turbulent broadening). Applying this model to the Chandra/HETG spectrum of two UCXB NS sources with disk atmosphere absorption, we were able to set upper limits on the size of their central engines. 

An important limitation of this method is the contribution of different emission components of the central engine and their interaction with the Keplerian absorber. Currently, our method outputs a weighted average of the central engine size, where we treat the total emission from different continuum components as a single entity that can be described by a single characteristic radius, ${R}_{CE}$. If, for example, the continuum is composed primarily of a neutron star and the inner radii of the accretion disk, the approximation holds as we expect similar size scales and flux contributions to the continuum from both components. 

This approximation may become problematic if there is a large discrepancy in both the size of the components and the energy bands at which they contribute. For instance, if the continuum flux in the Fe~K band is dominated by emission from a compact neutron star while a larger, radially-extended corona dominates below 3 keV, then the linewidths in these energy bands would be affected differently: the compact neutron star would have little effect on linewidths and would contribute relatively narrow absorption lines across the entire Chandra band; however, in the combined spectrum, its contribution would only be significant in the Fe~K band. Conversely, the much larger corona would display much broader absorption lines, but mostly contribute only in the low energy band. 

The magnitude of this effect is somewhat overstated in this hypothetical example. In the case of XTE J1710$-$281, any discrepancy is likely negligible given that our fits were mostly based on absorption lines located within a narrow band between 1 to 3 keV where dramatic changes in the contribution from different components seem largely implausible. This is likely true even in our fits to 4U 1916$-$053, which were dependent on absorption spanning almost the entire Chandra band: we found that Fe XXV and XXVI absorption lines were largely insensitive to the $\delta{V}_{CE}$ parameter due to the lower resolution of the Chandra/HETG at these energies. Moreover, for these sources in particular, their soft spectra displaying prominent absorption are typically associated with states in which the continuum is dominated by a combination of neutron star and disk blackbody emission. 

These results are the first attempt at constraining the size scales of central engines in both accreting neutron stars and compact objects, in general, using this method. It is therefore important to contextualize these constraints, the scope of the method, and its pitfalls, with those from other established techniques.

To first order, constraints via relativistic reflection lines probe exclusively the size of the ISCO. Although this quantity is particularly useful when constraining BH spin, it can be used in neutron star systems to constrain the innermost radius of the disk (likely larger than the ISCO in NS sources) and therefore set an upper limit on the radius of the neutron star itself. Constraints based on NuSTAR data of three neutron star sources by \cite{Ludlam2019} place upper bounds of 12.0 (GX 3$+$1), 18.6 (4U 1702$-$429), and 40.2 (4U 0614$+$091) $GM/{c}^{2}$ on the inner radius of the disk at a significance of 90\% (assuming near-zero spin). Similar analyses report constraints of 24 $GM/{c}^{2}$ and 15 $GM/{c}^{2}$ for 4U 1636$-$536 \citep{Mondal2021} and Serpens X-1 \citep{Mondal2020}, respectively. Though reflection modeling may allow constraints down to $\sim 10$ $GM/{c}^{2}$ for the most sensitive spectra, our constraints for XTE J1710$-$281, in particular, via linewidths are promising compared to a representative set of results from a well-established method. At the 2$\sigma$ level (95\%), we set an upper-bound of 80 $GM/{c}^{2}$ on the radius of central emitting area of this source. This result does not compare unfavorably considering the limited Chandra/HETG exposures on this class of objects and the comparatively low effective area of CXO, in general; future high-resolution X-ray spectroscopy missions will greatly improve on the sensitivity of this method. Crucially, the ISCO is likely not the benchmark with which to compare ${R}_{CE}$ results, as the peak luminosity contribution of the disk occurs at radii 2-3 times larger \citep{Zhu2012,McClintock2014}. If ${R}_{CE}$ is more representative of the half-light radius (see Appendix \ref{sec:geometries}), then this benchmark should be closer to $20-25$ $GM/{c}^{2}$ for sources with zero spin.
 
Though these initial results are larger compared to \emph{the most} sensitive constraints via modeling the reflection spectrum in neutron stars systems, it is important to note that our constraints comprise a weighted average of the entire central engine, including any X-ray corona that may be present, and not just the scale of the ISCO. Relativistic reflection can, to some extent, probe the geometry of the corona: these models are sensitive to the height of the corona along the rotation axis of a black hole (or, neutron star), while the emissivity slope and cut-off energy of the reflected spectrum often imply compact coronae \citep{Wilkins2012,Miller2015c}. Combined with timing measurements, these techniques can be powerful tools for probing the geometry of the corona \citep{Fabian2015,Fabian2017}. Conversely, relativistic reflection models have also shown some degeneracy with (and even preference for) radially extended coronae \citep{Wilkins2011,Wilkins2015}. While it is difficult to quantify the robustness of constraining the geometry of the corona via the relativistic reflection method alone, these constraints are certainly not as robust as those for the ISCO. By comparison, our linewidth method allows us to probe any component of the central engine indiscriminately, whether the dominant emission is due to the disk, a neutron star, or a corona.

Compared specifically to measurements of black hole spin obtained using either the continuum fitting method or relativistic reflection, our constraints are categorically worse: both methods have been able to constrain the size of the ISCO down to scales smaller than 10 $GM/{c}^{2}$, compared to our best constraints for XTE J1710$-$281 of ${R}_{CE} < 60 GM/{c}^{2}$. An important limitation is that both continuum fitting and relativistic reflection methods depend on careful modelling of the source continuum. In the case of relativistic reflection, difficulty with modelling the hard X-ray tail due to degeneracies with other continuum components can significantly affect the spin constraint. For continuum fitting, some model dependencies exist in terms of the radiative transfer through the disk surface atmosphere (among others), though these have shown to have little effect on spin measurements \citep{Reynolds2014, Reynolds2019}. Degeneracies with other continuum components is typically a non-issue with continuum fitting in the context of spin measurements, as these are limited to soft, geometrically thin-disk states where the inner radius disk extends down to the ISCO and where other continuum components (such as a power-law) make up less that 25$\%$ of the observed luminosity \citep{Steiner2009,Gou2014}, though this method is limited to BH systems. More importantly, continuum fitting requires independent constraints on the distance and inclination of the system, as well as a measurement of the mass of the black hole, limiting the continuum fitting method to only a handful of BH LMXBs \citep{Reynolds2019,Reynolds2020}. Linewidths may eventually contribute meaningful constraints on black hole spin as more sensitive spectra from both current and future high-resolution spectroscopy missions become available, finding a niche in select sources and/or spectral states in which the sensitivity of other methods may be reduced. The method, however, probes the entire central engine and not just the scale of the ISCO, a feature that might instead significantly limit its applicability in this context. 

Both reverberation mapping and microlensing in quasars provide direct measurements of the geometry of the X-ray corona, although these methods can only be applied in AGN. Among the more notable results of quasar microlensing, an analysis of the lensed quasar PG 1115$+$080* by \cite{Morgan2008} derived a half-light radius for the X-ray emitting region of log($R/\text{cm}$) = ${15.6}^{+0.6}_{-0.9}$. In gravitational radii (adopting the same mass estimate of $1.2\times{10}^{9} {M}_{\odot}$) this translates to a 1$\sigma$ upper-bound of $\sim 80$ $GM/{c}^{2}$. These are larger compared to our ${R}_{CE}$ constraints on XTE J1710$-$281, and are more comparable in their sensitivity to our results for 4U 1916$-$053, which set the scale of the central engine at ${R}_{CE} = {110} \pm 90$ $GM/{c}^{2}$. 

Naturally, our constraints using linewidths will improve with further Chandra/HETG exposures of these sources, as well as exposure from future microcalorimeter missions such as XRISM \citep{Tashiro2018} and ATHENA \citep{ATHENA}. Both 4U 1916$-$053 and especially AX J1745.6$-$2901 will benefit from the higher effective area and spectral resolution in the Fe~K band from these missions. At lower energies, both ATHENA and future grating spectrometer mission ARCUS \citep{ARCUS} will provide 2-3 times the resolving power with a significant increase in effective area, allowing for better constraints on sources such as XTE J1710$-$281.

\section{Conclusions}

We have presented a new method for determining the size of the central emitting regions in accreting neutron stars via measurements of widths of absorption lines. We applied this method on three ultra-compact and/or short period LMXBs in which we identified redshifted absorption lines originating in an inner disk atmosphere; both in this work and in \citetalias{Trueba2020} we present evidence that these are gravitational redshifts from atmospheres originating in the inner disk. The main results of this work are as follows:

\begin{enumerate}
\item Spectral fits to the ionized disk atmosphere absorption show highly significant redshifts (above $5\sigma$) in two of three sources: XTE J1710$-$281 ($310 \pm 50$ $\text{km}$ $\text{s}^{-1}$) and 4U 1916$-$053 ($200 \pm 50$ $\text{km}$ $\text{s}^{-1}$). The shifts are largely inconsistent with various estimates of the relative radial velocity of these systems, while the ionization and absorbing columns of the atmosphere suggest the absorption originates in the inner disk, consistent with a gravitational redshift. The redshifts suggest the absorption originates at ${1500}^{+200}_{-200}$ ${R}_{g}$ (XTE J1710$-$281), ${2200}^{+600}_{-300}$ ${R}_{g}$ (4U 1916$-$053), and ${1700}^{+9500}_{-800}$ ${R}_{g}$ (AX J1745.6$-$2901), once corrected for the transverse Doppler effect.
\item By using the gravitational redshift in the disk atmospheres to infer a radius, our constraints for the size of the central emitting regions in these systems were obtained independent of mass and distance to the source. However, the robustness of the methods developed and presented in this work is independent of the validity of the gravitational redshift result. 
\item We developed a method that correctly accounts for the effect the size of the central engine has on linewidths, allowing for constraints on ${R}_{CE}$.
\item We found that the large number of absorption lines in our spectra allowed for better constraints on the magnitude of the observed redshift, as well as allowed us to constrain the degree of velocity broadening due to microturbulence. In turn, the degeneracy between turbulent velocities and the line-widths produced by the size of the central engine was greatly reduced. In contrast, the line-poor spectrum of AX J1745.6$-$2901 resulted in poor constraints on ${R}_{CE}$.
\item We are able to constrain the (weighted averaged) radius of the central engine of XTE J1710$-$281 down to $<60$ ${R}_{g}$ (at the $1\sigma$-level). This initial effort approaches the sensitivity of constraints on the size of the innermost radius of many accreting neutron stars via modeling of relativistic reflection, though our constraints are not limited to setting the scale of the inner radius of the disk and encompass the entire central emitting area. These constraints also compare favorably to constraints via microlensing on black hole systems. In the case of the NS UCXB 4U 1916$-$053, we are able to constrain the central engine down to $<200$ ${R}_{g}$.
\end{enumerate}

The scientific results reported in this work are based on observations made by the Chandra X-ray Observatory and data obtained from the Chandra Data Archive. N. T. acknowledges helpful discussions with Kevin Whitley and Nuria Calvet. R.M.L. acknowledges the support of NASA through Hubble Fellowship Program grant HST-HF2-51440.001. We thank the anonymous referee for comments and suggestions that improved this work. N.T. acknowledges the support of the Horace H. Rackham School of Graduate Studies through the Rackham Predoctoral Fellowship. 

\appendix
\section{Treatment of Continuum Parameters}\label{sec:cont_par}

In Section \ref{sec:ind} we discussed how continuum parameters (such as the temperatures and normalizations of the DBB and BB additive models) must be set as free parameters when fitting the photoionized absorption in order to account for some continuum absorption in the PION model. These continuum parameters, however, were treated as nuisance parameters when calculating confidence regions for the photoionized absorption model parameters: the $\Delta {\chi}^{2}$ value corresponding to a particular significance level (say, 1$\sigma$) was based on the number of free PION parameters (${N}_{H}$, log $\xi$, ${v}_{z}$, and ${v}_{turb}$; therefore $\Delta {\chi}^{2} = 4.72$) and did not include the number free continuum parameters.

The justification for this based on three points: 1) Our continuum model is flexible enough to adapt and compensate for any amount of continuum absorption from the photoionized absorber and produce an equally good continuum fit throughout the parameter space explored in our fits. The change in ${\chi}^{2}$ we observe is, therefore, determined solely by the absorption model's ability to fit spectral bins which contain absorption lines. 2) The continuum absorption is dominated by electron scattering and is mostly gray. Although the shape of the continuum might shift slightly to compensate for some energy-dependent absorption due to ions that have not been completely stripped of electrons, this change is not significant enough to affect the ionization parameter. 3) The change in the underlying continuum is ``physically uninteresting'' and it is not in conflict with any independent constraints of the continuum parameters.

A quantitative test for point 1) is not straightforward: Though SPEX does not offer the option directly to treat the continuum and line absorption components of absorbers separately, it is possible to effectively remove the absorption line component of the absorber by setting the velocity broadening parameter, ${v}_{turb}$ (v in SPEX), to an arbitrarily large number. We found that for a lower ionization (log $\xi = 2.8$) line-rich absorber with a large column (${N}_{H} = {10}^{24}$ ${\text{cm}}^{-2}$), an unphysically large velocity broadening of $v = {10}^{7}$ \kms{} ($\sim 30 \cdot c$) produces an absorber with \emph{precisely} the same degree of continuum absorption without any absorption lines; this includes the larger amount of energy-dependent absorption at these low ionizations. Alternatively, using a value of $v < c$ resulted in some degree of absorption from ultra-broad lines contributing to the apparent continuum absorption. Given that the continuum absorption is entirely unaffected by the degree of velocity broadening, we adopted $v = {10}^{7}$ \kms{} when testing for continuum effects.

Our test model consisted of taking the best-fit 1-zone model of 4U 1916$-$053 (a model that has both a large column and a large uncertainty in the ionization) and freezing the PION parameters to their best-fit values. The resulting spectrum was then absorbed by an \emph{additional} PION layer with the same best-fit ${N}_{He}$ and log $\xi$ values, except the velocity broadening parameter was set to $v = {10}^{7}$ \kms{} in order to convert this layer exclusively into a continuum absorber. Finally, the continuum was fit again to account for the additional continuum absorption. The resulting model should be nearly identical to the best-fit model listed in Table \ref{tab:all_fits}, with the exact same degree of \emph{line} absorption, with the exception that the continuum optical depth is doubled and, therefore, the underlying continuum is brighter to compensate. Indeed, despite this significant increase in continuum absorption, we observed a minimal $\Delta{\chi}^{2}<0.2$ compared to the baseline model. 

This model allowed us vary the degree continuum absorption independently without affecting the original PION layer which produces the absorption lines. By maintaining the line-producing layer frozen at best-fit parameters, our test model was nominally able to track the quality of the continuum fit while keeping the absorption lines constant: if the continuum model is truly flexible enough to account for any degree of continuum absorption, then the change in ${\chi}^{2}$ should be negligible regardless of choice of ${N}_{He}$ and log $\xi$ in the additional continuum absorption layer. Crucially, however, if the shape of the underlying continuum were to change enough to affect the ionization parameter, this should also present itself as a change in ${\chi}^{2}$ in the test model: the PION layer frozen at the best-fit log $\xi$ value would no longer produce the same absorption line spectrum as in the baseline model. As a result, our test model is able to quantitatively test both points 1) and 2).

\begin{figure}

\begin{center}
\subfloat{\includegraphics[width=0.42\textwidth,angle=0]{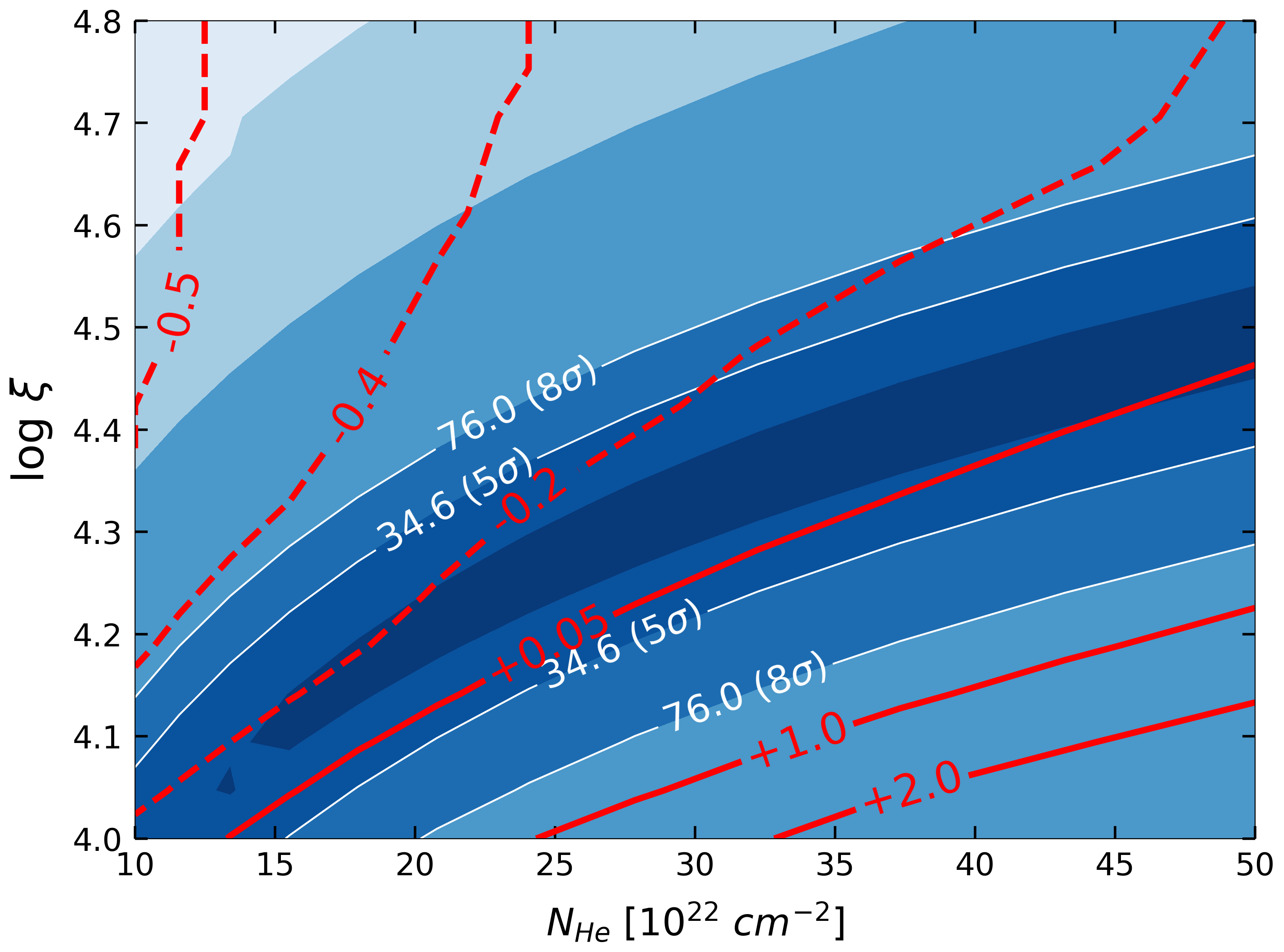}}\\
\subfloat{\includegraphics[width=0.42\textwidth,angle=0]{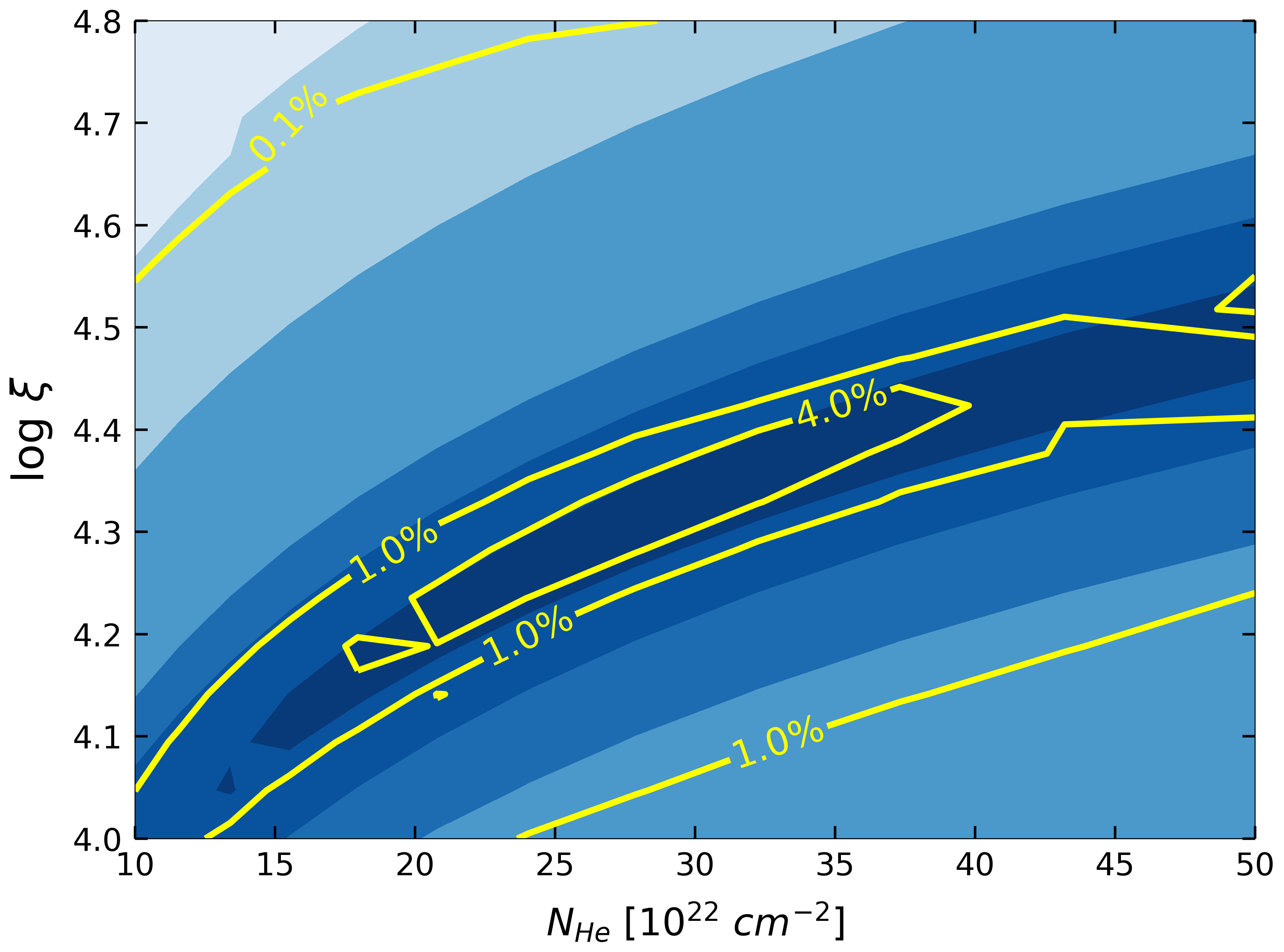}}\\
\subfloat{\includegraphics[width=0.42\textwidth,angle=0]{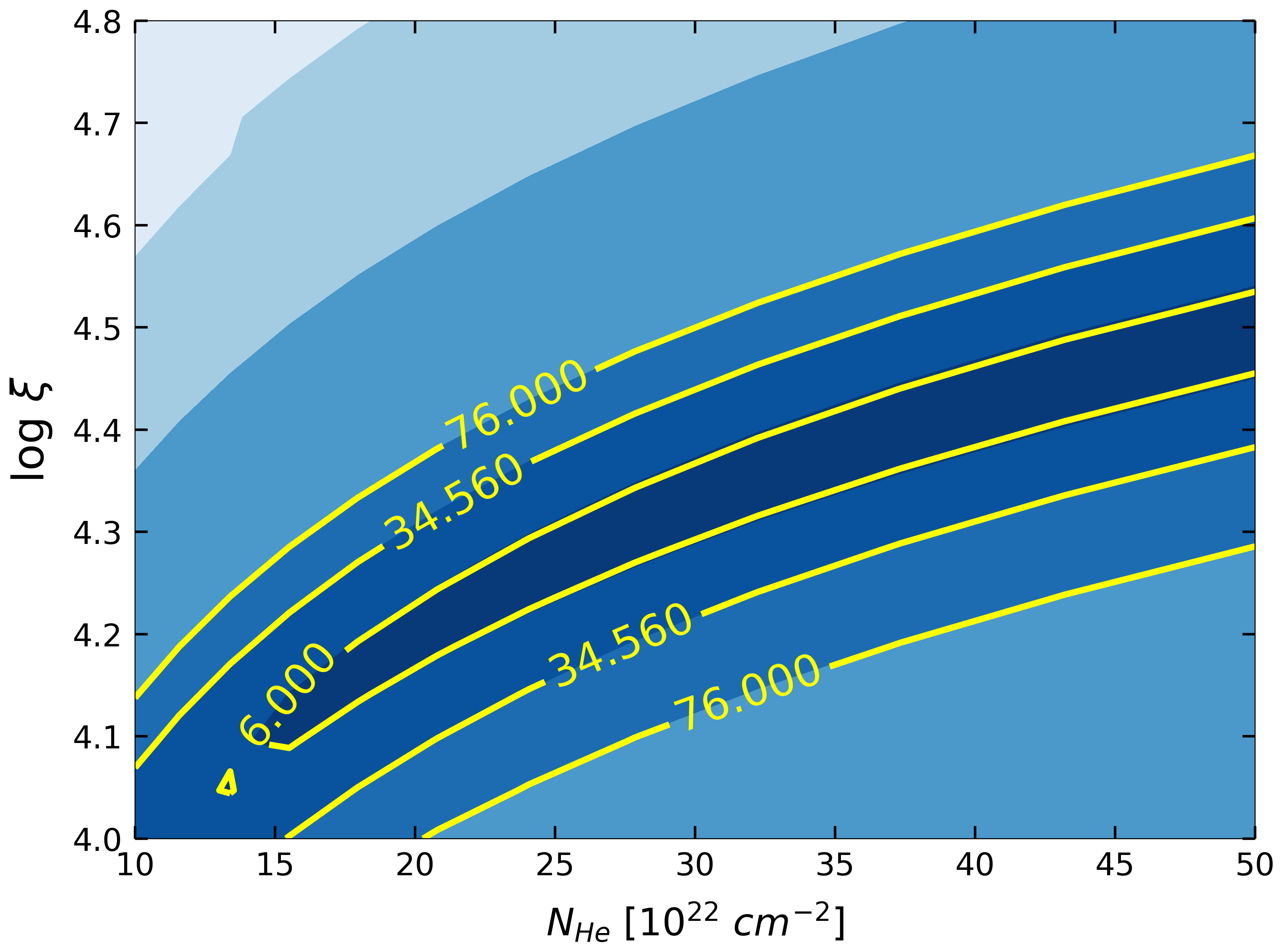}}\\

\end{center}
\vspace{-0.2in}
\caption{The ${N}_{He}$ vs log $\xi$ $\Delta {\chi}^{2}$ surface for the 1-zone baseline fit for 4U 1916$-$053, plotted in blue with white contour lines. \emph{Top:} Red contours plot the $\Delta {\chi}^{2}$ of the test model in response to the changing underlying continuum. \emph{Middle:} Same as top panel, plotted as a percentage of the blue contours. \emph{Bottom:} $\Delta {\chi}^{2}$ contours for baseline fit after subtracting the $\Delta {\chi}^{2}$ due to the changing continuum.}\label{fig:cont_chi}
\end{figure}

The blue contour maps (with white contour labels) in Figure \ref{fig:cont_chi} plot the $\Delta {\chi}^{2}$ grid for the 1-zone baseline model of 4U 1916$-$053 in ${N}_{He}$ vs log $\xi$ space. In the top panel, the red contours show the $\Delta {\chi}^{2}$ grid of the test model over the same parameter space, while in the middle panel we plot the $\Delta {\chi}^{2}$ grid of the test model as a percentage of the $\Delta {\chi}^{2}$ of the baseline model in yellow contours. 

Although the $\Delta {\chi}^{2}$ grid for the test model is evidently not perfectly flat, the change is negligible relative to the baseline model: we only see a change in ${\chi}^{2}$ larger than 1.0 well outside the $8\sigma$ confidence region of the baseline model, while the $\sim1-2\sigma$ innermost contours appear to be affected by less than $\Delta {\chi}^{2} < 0.2$. As a percentage, $\Delta {\chi}^{2}$ from the test model is only as large as $4\%$ in a small region and never exceeds $7\%$, hovering at $1\%$ or lower. Even a $4\%$ change appears to have a little to no effect on the location of contours once we subtract the test model $\Delta {\chi}^{2}$ grid from the baseline grid, as can be seen in the bottom panel of Figure \ref{fig:cont_chi} (yellow contours). 

It is clear from this test that, although the underlying continuum may change to account for different degrees of continuum absorption, this 1) has a negligible effect on the quality of the continuum fit and 2) has little effect on the continuum shape and therefore does not affect the ionization parameter. This, however, does not necessarily mean changes in the underlying continuum can be entirely disregarded. Scanning for errors in the absorbing column of some hypothetical absorber could, for instance, shift the the source luminosity from being well-below Eddington to super-Eddington, shift the NS temperature to unrealistic values, or require a significantly different neutral ISM column compared to the literature. None of these are the case for our three low-luminosity sources: we only see slight modest changes to the continuum parameters (e.g. changes in DBB and BB temperatures under $1\%$) other than normalizations for additive components, and at most a $\sim25\%$ shift in luminosity.

Notably, while this shift in the underling luminosity should increase the uncertainty in the ${L}_{phot}$ values reported in Table \ref{tab:all_fits} for our most heavily absorbed sources (namely, 4U 1916$-$053 and AX J1745.6$-$2901), the quoted errors for quantities such as $r = \frac{L}{{N}_{H}\xi}$ should remain unaffected: The underlying source luminosity can be written in terms of ${L}_{0}$, the luminosity assuming no continuum absorption, as
\begin{equation}
{L}_{u} = {L}_{0} {e}^{{\sigma}{N}_{H}}.    
\end{equation}
For simplicity, we assume gray opacity due only to electron scattering, and therefore ${\sigma} \sim 6.6\cdot {10}^{-25}$ ${\text{cm}}^{2}$. The upper limit on the photoionization radius then becomes $r = \frac{{L}_{0}{e}^{{\sigma}{N}_{H}}}{{N}_{H}\xi}$, which we can then differentiate in terms of ${N}_{H}$ as
\begin{equation}
\frac{dr}{d{N}_{H}} = \frac{{L}_{0}{e}^{{\sigma}{N}_{H}}}{{N}_{H}\xi}(\sigma-\frac{1}{{N}_{H}}),
\end{equation}
simplified as 
\begin{equation}
\frac{dr}{d{N}_{H}} = r(\sigma-\frac{1}{{N}_{H}}).
\end{equation}

For the purpose of this discussion, we ignore linear terms corresponding to the other parameters (namely log $\xi$). The uncertainty on $r$ due solely to ${N}_{H}$ then becomes
\begin{equation}
{\delta}_{r} = \pm|r(\sigma-\frac{1}{{N}_{H}})\cdot{\delta}_{{N}_{H}}|.
\end{equation}
Rewriting ${\delta}_{{N}_{H}}$ as some fraction of ${N}_{H}$, we get 
\begin{equation}\label{eq:delta_r}
{\delta}_{r} = \pm|r(\sigma{N}_{H}-1)\cdot {\delta}_{x}|.
\end{equation}
where ${\delta}_{x} = {\delta}_{{N}_{H}}/{N}_{H}$.

For small values of ${N}_{H}$ ($\sim 17\cdot{10}^{22}$ ${\text{cm}}^{-2}$ or smaller) we find that Equation \ref{eq:delta_r} approaches ${\delta}_{r} \sim \pm|r {\delta}_{x}|$, the error on ${r}$ if we disregard any changes in the underlying continuum ($<10\%$ discrepancy). As expected, changes in the underlying luminosity of sources absorbed by small columns, such as XTE J1710$-$281, are negligible. On the other hand, as ${N}_{H}$ approaches $\sim 150\cdot{10}^{22}$ ${\text{cm}}^{-2}$, then ${\delta}_{r}$ approaches zero. This makes sense: increasing (or decreasing) the value of ${N}_{H}$ in the denominator of $r = \frac{L}{{N}_{H}\xi}$ is counteracted by increasing (or decreasing) the value of $L$, resulting in a constant $r$ in this regime.

Our fits avoid this regime by limiting ${N}_{H}$ to $<100\cdot{10}^{22}$ (or ${N}_{H} < 50\cdot{10}^{22}$, for He) ${\text{cm}}^{-2}$. At this limit, ${\delta}_{r} \sim \pm|r\cdot0.4 {\delta}_{x}|$, meaning that the quoted errors in Section \ref{sec:ind} are actually \emph{overestimated} by not correcting for this change in continuum. However, standard error propagation assumes relatively small Gaussian errors - fits with large ${N}_{H}$ values are accompanied by large minus errors on ${N}_{H}$, meaning that plus errors on ${r}$ are in the regime where ${\delta}_{r} \sim \pm|r {\delta}_{x}|$. 

\subsection{Marginalization}
For clarity, what follows is a formal description of how the fitting and error searching procedure in SPEX effectively results in the marginalization of these nuisance continuum parameters. 

A simplified description of any of our spectral models can be written as 
\begin{equation}\label{eq:F_cont}
{F}_{\nu} = {C}_{\nu}\cdot {e}^{-{\tau}_{\nu}},    
\end{equation}
where ${F}_{\nu}$ is the frequency dependent model flux, ${C}_{\nu}$ is the ``naked'' source continuum, and ${\tau}_{\nu}$ is the frequency dependent optical depth of the photoionized absorber (including both line and continuum absorption). We can separate ${\tau}_{\nu}$ into its absorption line (${\tau}_{\nu,L}$) and continuum (${\tau}_{\nu,C}$) components, and therefore Equation \ref{eq:F_cont} becomes
\begin{equation}\label{eq:F_cont2}
{F}_{\nu} = {C}_{\nu}\cdot {e}^{-{\tau}_{\nu, L}} \cdot {e}^{-{\tau}_{\nu,C}}.    
\end{equation}

The value of both ${\tau}_{\nu, L}$ and ${\tau}_{\nu, C}$ are directly proportional to ${N}_{H}$, while their behavior with respect to frequency depends on the value of log $\xi$ (${v}_{z}$ and ${v}_{turb}$ have a negligible effect on ${\tau}_{\nu, C}$, so we exclude them from this discussion). Equation \ref{eq:F_cont2} can be further simplified by combining the underlying continuum with the continuum absorption into \begin{equation}\label{eq:eta}
{F}_{\nu} = {\eta}_{\nu}\cdot {e}^{-{\tau}_{\nu, L}},
\end{equation}
where ${\eta}_{\nu} = {C}_{\nu} \cdot {e}^{-{\tau}_{\nu,C}}$. This expression is useful as it describes the model flux in terms of the observed continuum, ${\eta}_{\nu}$, and line absorption, ${e}^{-{\tau}_{\nu, L}}$.

After spectral fitting, the best-fit spectral model, ${F}_{\nu}^{*}$, can be written as
\begin{equation}
{F}_{\nu}^{*} = {C}_{\nu}^{*}\cdot {e}^{-{\tau}_{\nu,L}^{*}}\cdot {e}^{-{\tau}_{\nu,C}^{*}},    
\end{equation}
where ${\tau}_{\nu,L}^{*}$ and ${\tau}_{\nu,C}^{*}$ are the optical depths of the absorber at the best-fit ${N}_{H}$ and log $\xi$, and where ${C}_{\nu}^{*}$ is the underlying continuum given the resulting value of ${\tau}_{\nu,C}^{*}$. Simplified, this becomes
\begin{equation}\label{eq:eta_star}
{F}_{\nu}^{*} = {\eta}_{\nu}^{*}\cdot {e}^{-{\tau}_{\nu, L}^{*}},
\end{equation}
where ${\eta}_{\nu}^{*}$ describes the total \emph{observed} continuum (including continuum absorption) for the best-fit model.

Using Equations \ref{eq:eta} and \ref{eq:eta_star} we can describe what happens during a grid search in ${N}_{H}$ vs log $\xi$ space. A set of ${N}_{H,i}$ and log ${\xi}_{j}$ test values are selected and a new continuum absorption, ${e}^{-{\tau}_{\nu,C,i,j}}$, is computed. SPEX then performs a spectral fit of the continuum parameters to produce the new model spectrum
\begin{equation}
{F}_{\nu,i,j} = {\eta}_{\nu,i,j}\cdot {e}^{-{\tau}_{\nu, L,i,j}}
\end{equation}
based on the new continuum absorption. As shown in our test model, this new observed continuum, ${\eta}_{\nu,i,j}$, is nearly identical to the observed continuum in the best-fit model,
\begin{equation}\label{eq:cont_eq}
{\eta}_{\nu,i,j} \simeq {\eta}_{\nu}^{*}.
\end{equation}
Generalizing with Equation \ref{eq:eta}, we can say 
\begin{equation}
{F}_{\nu,i,j} = {\eta}_{\nu}^{*}\cdot {e}^{-{\tau}_{\nu, L,i,j}}.
\end{equation}
The key point here is that, since the continuum fit does not change, the ${\eta}_{\nu}^{*}$ quantity is a constant, and therefore the quality of our fit depends solely on the absorption lines produced by the absorber, ${e}^{-{\tau}_{\nu, L,i,j}}$. During the grid search, the SPEX fitting routines effectively marginalize over these continuum parameters and, therefore, the fit statistic is only tracking changes in the model's ability to describe spectral bins containing absorption lines.

Crucially, we only use the example of a grid search in this discussion as it is a clear and convenient way to demonstrate how the continuum is effectively marginalized in our fits. Statistically, the same principles apply when SPEX is running standard fitting routines and especially error search algorithms. 

\section{The Effects of Velocity Broadening on Equivalent Widths}\label{sec:EW}

\begin{figure*}
\begin{center}
\subfloat{\includegraphics[width=0.99\textwidth,angle=0]{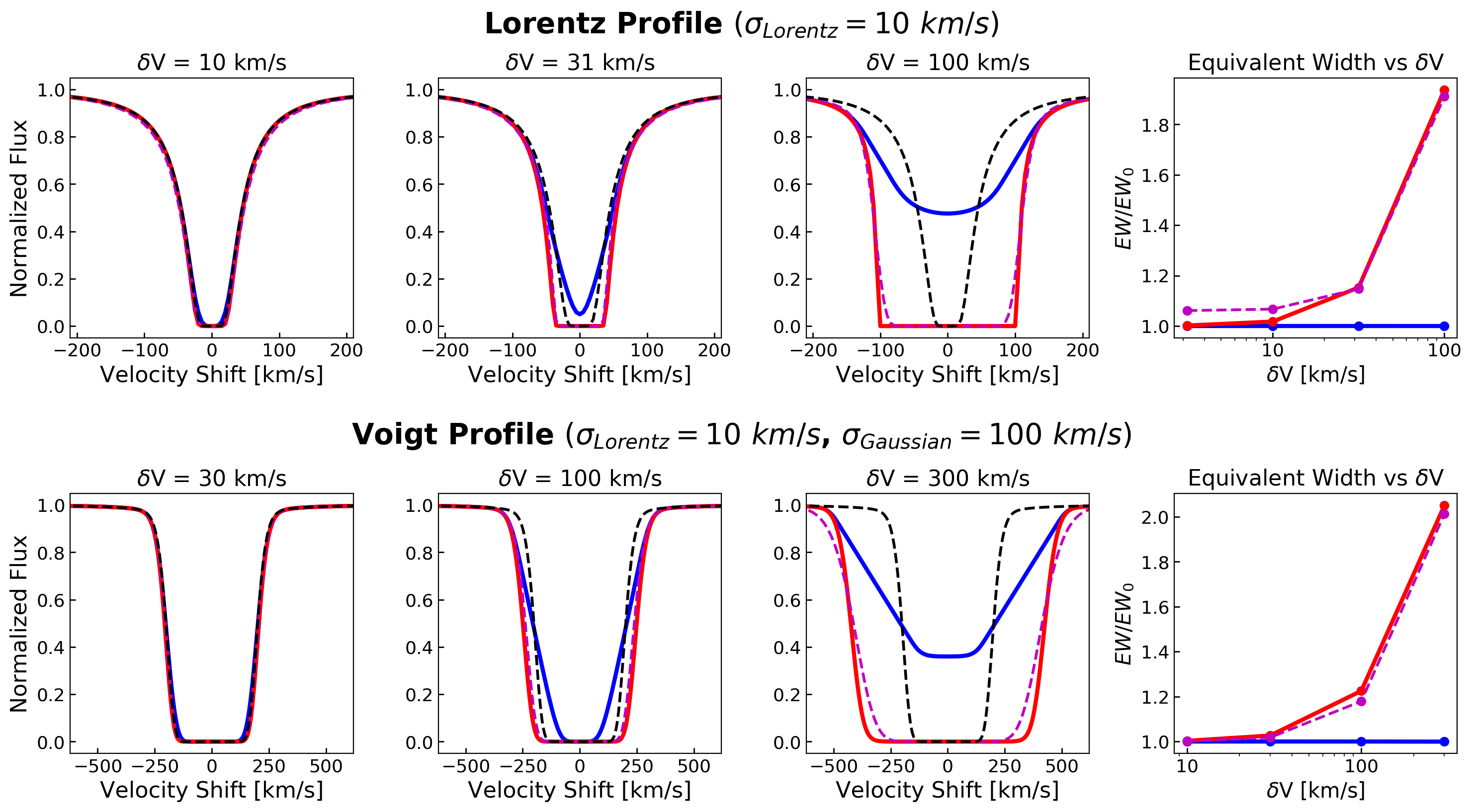}}
\end{center}
\vspace{-0.2in}
\caption{Line profiles produced by applying additional velocity broadening. The base line-profiles (black dashed lines) are subject to both forms of velocity broadening described in the text: multiple velocity components in series (Equation \ref{eq:series}, akin to turbulent broadening; solid red lines), as well as in parallel (Equation \ref{eq:parallel}; solid blue line). In both scenarios, the same uniform velocity distribution was used (see text). 
Panels on the right show the equivalent widths at each velocity broadening value, normalized by the equivalent width of the base profile.}\label{fig:ews}

\end{figure*}

In this section, we describe in detail the differences between the two forms of velocity broadening discussed in this work and their effects on equivalent widths. The purpose of this section is both to clarify why the ${R}_{CE}$-model is designed as described in Section \ref{sec:rce_model}, as well as justify how these two forms of velocity broadening can be decoupled.

The geometric velocity broadening effect described in Section \ref{sec:RCE} arises from the fact that different parts of an emitting area are absorbed by gas with different line-of-sight velocities. This is subtly different to most commonly observed forms of velocity broadening in absorption, such as turbulent broadening, in which different velocity components are located along the line-of-sight yet absorb the entire emitting area equally. To illustrate this, we construct a simple toy model consisting of a uniform emitting area and two absorption scenarios involving an absorber with an arbitrary column $N$. In both scenarios, the properties of the emitting area and absorber are identical with the exception of how the absorber is divided into two velocity components.

In the first scenario, the absorber is divided equally into two velocity components along the line-of-sight. Based on Equation \ref{eq:not_rce}, the resulting spectrum can be described as
\begin{equation}
{F}(\nu) = {F}_{0} \times {e}^{-\tau(\nu)/2} \times {e}^{-\tau(\nu+\Delta\nu)/2},
\end{equation}
which can be simplified as
\begin{equation}\label{eq:series}
{F}(\nu) = {F}_{0} {e}^{-\tau(\nu)/2 - \tau(\nu+\Delta\nu)/2}.
\end{equation}
Here, the frequency dependent optical depth (say, for a line) requires dividing by a factor of 2, since $\tau = N\cdot \sigma(\nu)$ and total column $N$ is divided equally between the two halves of the absorber. 

In the second scenario, the absorber is again divided into two velocity components but now each velocity component only absorbs its corresponding half of the emitting area. Since the absorber is no longer divided along the line-of-sight, however, each half is absorbed by the entire column $N$. The resulting spectrum can be described using
\begin{equation}
{F}(\nu) = \frac{1}{2}{F}_{0}{e}^{-\tau(\nu)}+\frac{1}{2}{F}_{0}{e}^{-\tau(\nu+\Delta\nu)},
\end{equation}
simplified as
\begin{equation}\label{eq:parallel}
{F}(\nu) = \frac{{F}_{0}}{2} ({e}^{-\tau(\nu)}+{e}^{-\tau(\nu+\Delta\nu)}).
\end{equation}

It is clear from Equations \ref{eq:series} and \ref{eq:parallel} that these two scenarios are not equivalent - the most pronounced observable difference is in the behavior of equivalent widths of strong lines with saturated cores. 

The effects of line-of-sight (e.g. turbulent, thermal) velocity broadening on the equivalent widths of strong lines are well-known and are central to discussions on the curve of growth. Using our toy model for a qualitative comparison: if at line center $\tau(\nu)$ is very large and the line core becomes saturated, the entire continuum at $\nu$ would be almost entirely absorbed (Equation \ref{eq:series}). This is the case even when the column is divided by a factor of two. As a result, the EW of lines with saturated cores increases as the degree of velocity broadening increases. Conversely, Equation \ref{eq:parallel} shows how, even if the value of $\tau(\nu)$ approaches infinity, this component will \emph{at most} completely absorb up to half of the continuum. The degree to which the other half of the continuum is absorbed depends solely on the value of $\tau(\nu+\Delta\nu)$. 

Using Equation \ref{eq:parallel} it is straightforward to show how, instead of increasing, the EW of an absorption line remains \emph{exactly} constant regardless of the degree of geometric velocity broadening. Per the definition of the equivalent width, the total flux removed by an absorption lines is equal to the quantity 
\begin{equation}\label{eq:ew}
{F}_{0}\cdot \text{EW} = {F}_{0}\int_{}^{} ~d\nu - \int_{}^{} {F}_{\nu} ~d\nu.
\end{equation}
In the case of an absorption lines NOT subject to geometric broadening, ${F}(\nu) = {F}_{0} {e}^{-\tau(\nu)}$, and therefore the equivalent width is given by 
\begin{equation}\label{eq:ew2}
\text{EW} = \int_{}^{} 1 - {e}^{-\tau(\nu)}~d\nu.
\end{equation}
Adding geometric broadening to this scenario, ${F}(\nu) =\frac{{F}_{0}}{2} ({e}^{-\tau(\nu)}+{e}^{-\tau(\nu+\Delta\nu)})$, and therefore Equation \ref{eq:ew} becomes
\begin{equation}
{F}_{0}\cdot \text{EW} = {F}_{0}\int_{}^{} ~d\nu - \frac{{F}_{0}}{2}\int_{}^{}  ({e}^{-\tau(\nu)}+{e}^{-\tau(\nu+\Delta\nu)}) ~d\nu
\end{equation}
Simplifying this expression, we obtain
\begin{equation}\label{eq:ew3}
\text{EW} = \frac{1}{2}\int_{}^{} 1- {e}^{-\tau(\nu)} ~d\nu + \frac{1}{2}\int_{}^{} 1- {e}^{-\tau(\nu+\Delta\nu)} ~d\nu.
\end{equation}

Comparing Equations \ref{eq:ew2} and \ref{eq:ew3} it is clear that, unless the value of $\int_{}^{} 1- {e}^{-\tau(\nu+\Delta\nu)} ~d\nu$ is noticeably different to $\int_{}^{} 1- {e}^{-\tau(\nu)} ~d\nu$, then equivalent width of the line remains. This makes sense: by adding geometric broadening, we split the emitting area in two halves and simply added a velocity shift to one of them. Since we do not expect the equivalent width of a line to change by simply adding a velocity shift, each half still contributes half of the total equivalent width.

For confirmation, we calculated several line profiles with different degrees of geometric velocity broadening, plotted in Figure \ref{fig:ews}. Unlike our simple 2 component examples, these profiles were calculated using as many as 1000 velocity components, assuming a rectangular velocity profile where $\delta {V}_{i}$ spans linearly from $-\delta {V}$ to $+\delta {V}$ and each velocity component contributes equally. The base profile (black dashed line) remains constant, while the profile subject to geometric broadening is plotted in blue. The same rectangular velocity profile was also applied to the absorber along the line-of-sight (as in Equation \ref{eq:series}, solid red line). This profile is analogous to turbulent broadening, though in the case of turbulent broadening we expect a Gaussian velocity distribution. As a simple point of comparison, we used an ad-hoc scaling actor to convert $\delta {V}$ into a Gaussian velocity width (${\sigma}_{\delta {V}}$) and therefore compute comparable Voigt profiles (dashed magenta line).

The computed equivalent width for each profile is plotted in the rightmost panels of Figure \ref{fig:ews}, normalized by the equivalent width of the base profile. As expected, the equivalent width remains constant in the case of geometric broadening, while it increases when the absorber is slit along the line-of-sight (note that the total column N remains constant for all profiles). 

\section{Different Central Engine Geometries}\label{sec:geometries}

\begin{figure}
\subfloat{\includegraphics[width=0.46\textwidth,angle=0]{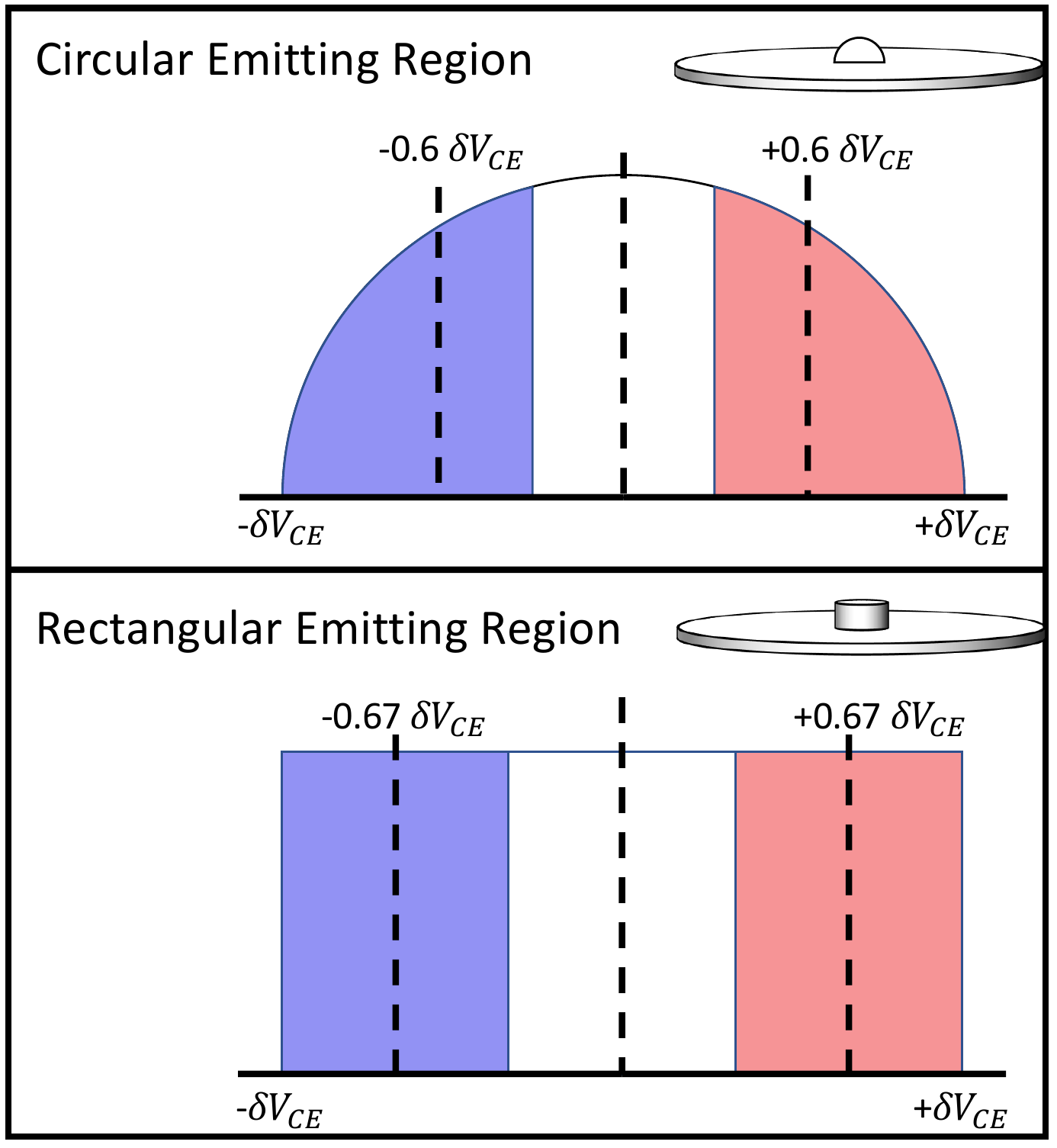}}
\vspace{-0.1in}
\caption{\footnotesize Equatorial view of circular (top) and rectangular (bottom) central engine geometries. The horizontal x-axis corresponds to the projection (along the line-of-sight) of the Keplerian motion of the absorbing gas (see Figure \ref{fig:RCE}). By definition, the extent of each area along the x-axis spans from $-{\delta V}_{CE}$ to $+{\delta V}_{CE}$, and vertical dashed lines plot the location of the (weighted) mean ${\delta V}$ for each area component.}\label{fig:shape}
\end{figure}

As introduced in Section \ref{sec:rce_model}, the assumed ``shape'' of the central engine can affect the interpretation of any ${R}_{CE}$ constraint using the method described in this work. At higher spectral resolutions, different geometries may ultimately affect the shape of the observed line profile; given the sensitivity of our data and how the ${R}_{CE}$-model is designed in this work, this discussion mostly focused on how the fit parameter, ${\delta V}$, can be related to ${\delta V}_{CE}$ quantity (see Figure \ref{fig:RCE}, Section \ref{sec:RCE}). 

The model as described in Section \ref{sec:rce_model} splits the total emitting area into 3 components of equal area: a central region absorbed by gas with no additional velocity shift, and two exterior regions absorbed by gas subject to an additional blueshift ($-{\delta V}$) or redshift ($+{\delta V}$). Figure \ref{fig:shape} shows an equatorial view of two toy model geometries: a circular and rectangular emitting region relative to the axis of rotation. In the case of the rectangular emitter, the relationship between the mean velocity shift for each component is trivial (see Figure \ref{fig:shape}) given that ${\delta V}$ changes linearly along this narrow projection of the orbital velocity (hereafter, the x-axis). Conversely, splitting the circular emitting region equally along the x-axis results in a slender central region that extends to a small fraction of the ${\delta V}_{CE}$ value compared to the exterior regions. Crucially, the mean velocity shift for the exterior regions is calculated via a weighted average, as the portions of the emitter that are subject to the largest velocity shift contribute the least in terms of of emitting area. This can be seen in Figure \ref{fig:shape}, where the mean ${\delta V}$ value (vertical dashed lines) of the exterior regions are smaller compared to those in the rectangular region. 

Using equal-area components in our ${R}_{CE}$-model allows us to account for many central engine geometries in ways that alternative approaches do not. For instance, a best-fit ${\delta V}$ value for a given spectrum could be interpreted as meaning ${\delta V}_{CE} = 1.5 {\delta V}$ when assuming a broadly rectangular central engine geometry, or ${\delta V}_{CE} = 1.7 {\delta V}$ when assuming a circular emitting region. Alternatively, changing the relative contribution of each emitting area would require multiple fits to account for each geometry. 

\subsection{An Optically-Thin Corona}

Thus far this discussion has focused on different central engine geometries for emitting regions that can be treated as ``solid'' surfaces: regions, such as the surface of a neutron star, in which the relative flux contribution of a certain portion of the emitter depends solely on its surface area. If, instead, the central engine in question was composed of optically-thin emitting gas, then the relative flux contribution of different spatial components (in this case, along the x-axis) depends both on the apparent area and the amount of material along the line-of-sight. The purpose of the following discussion is to provide a crude quantitative comparison between an ${R}_{CE}$ value derived from our model and a characteristic size-scale (e.g. the half-light radius) in the case of a simple toy model optically-thin corona. 

\begin{figure}
\subfloat{\includegraphics[width=0.45\textwidth,angle=0]{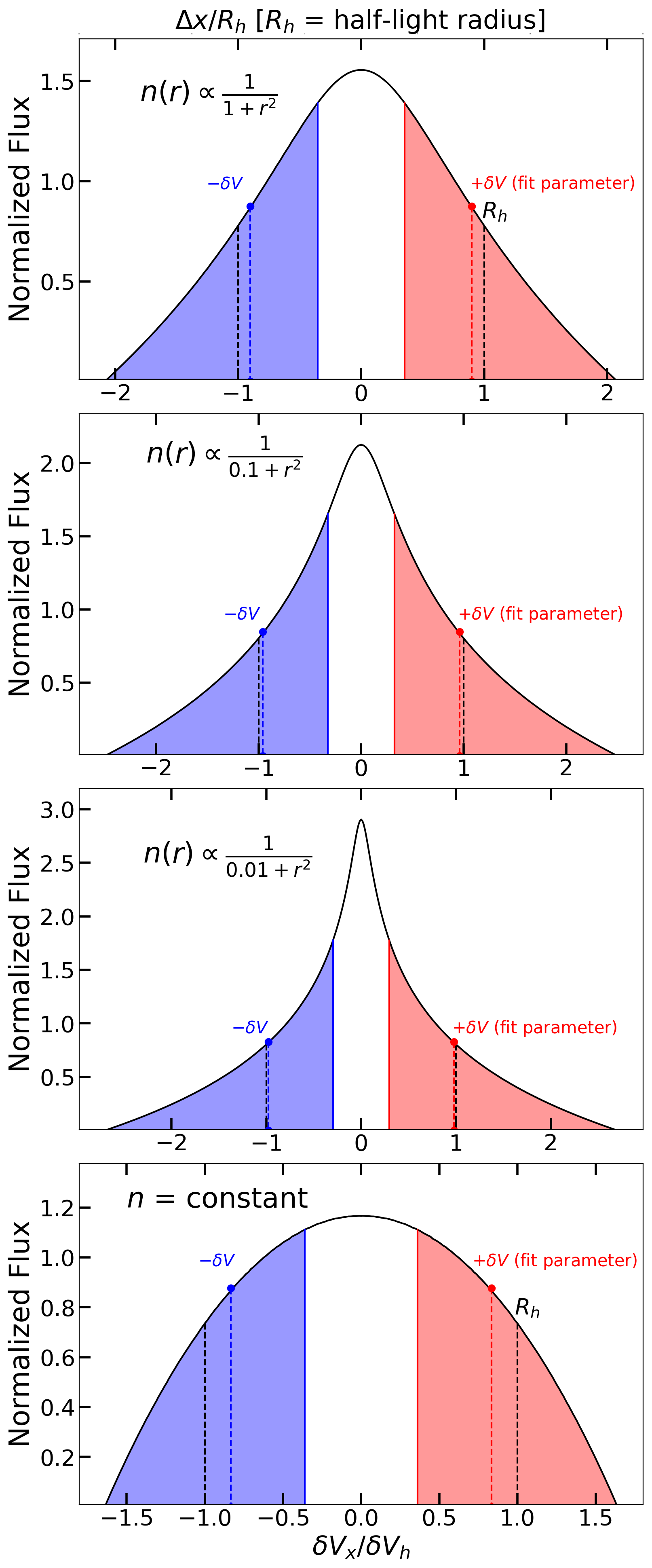}}
\vspace{-0.1in}
\caption{\footnotesize The relative luminosity contribution along the x-axis for a toy model optically-thin corona.}\label{fig:opt_thin}

\end{figure}

Our toy model corona consists of a truncated sphere of gas in which the flux contribution from each area element ($dA$) is proportional only to the total amount of material along the line-of-sight ($dF(x,y) \propto \tau(x,y)$, where the y-axis is the vertical axis in Figures \ref{fig:shape} and \ref{fig:opt_thin}). This quantity is strongly dependent on the density structure of the sphere; however, the absolute value of the gas density is unimportant as we only care about relative flux contribution. This simple toy model also assumes a uniform continuum spectral energy distribution throughout the sphere. It is important to point out that any realistic model of a central corona may diverge significantly from any of these assumptions. The core of the corona, for instance, may be sufficiently dense to become optically-thick. In this case, an analysis of the relative flux contribution of different area elements will most likely yield results that lie somewhere in between the ``solid'' emitting surface and the optically-thin scenarios. Given the scope of this work, we limit this discussion by only exploring these two extreme scenarios. 

We tested two different density profiles for our toy model corona: a constant density sphere and an inverse square profile ($n(r) \propto {r}^{-2}$; often referred to as a singular isothermal sphere profile in hydrostatic systems). In order to avoid the problematic singularity in the latter, we included a softening term to the profile ($n(r) \propto \frac{1}{c+{r}^{2}}$) and tested different degrees of softening relative to the truncation radius imposed by our toy model. 
Figure \ref{fig:opt_thin} shows the relative flux contribution along the x-axis. Although we are no longer dealing solely with area elements, this analysis allows us to treat the relative flux contribution in an analogous manner to that of the solid emitting surfaces (Figure \ref{fig:shape}). As with the rest of this analysis, we divide the emitter into 3 components of equal flux contribution (or, equal normalization during spectral modeling). Instead of the ${\delta V}_{CE}$ parameter, defined as the velocity shift at the edge of the of a well-defined emitting region in Section \ref{sec:RCE} (see Figure \ref{fig:RCE}), we define an analogous quantity for this optically-thin scenario: ${\delta V}_{h}$, or the velocity shift at the half-light radius (or, ${R}_{h}$). Values along the x-axis, as well as the weighted average velocity shift for each component (the fitting parameter $\delta V$), are normalized by the magnitude of ${R}_{h}$ and ${\delta V}_{h}$, respectively, calculated in each scenario. Note that, unlike the rest of this analysis, ${R}_{h}$ is calculated by integrating the flux contribution \emph{radially} (on the image plane), as opposed to along the x-axis. 

The top three panels of Figure \ref{fig:opt_thin} show the relative flux contribution in the case of an inverse square density profile, each with a different degree of softening (spanning three orders of magnitude). The inclusion of the softening factor means this toy model is no longer self-similar relative to the truncation radius. However, the magnitude of the $\delta V$ fitting parameter approaches the velocity shift at the half-light radius, ${R}_{h}$, asymptotically. Even in the case of strong softening (the top panel), $\delta V$ differs from this value by less than $10\%$. Alternatively, the constant density scenario (bottom panel) is completely self-similar. In this case, we find the magnitude of the $\delta V$ parameter is  $0.83\cdot{\delta V}_{h}$.

Taken literally, the results from our various toy models could be used to interpret fits obtained with the ${R}_{CE}$-model. For instance, a best-fit $\delta V$ value of 100 $\text{km}$ $\text{s}^{-1}$ could be interpreted as a $\delta {V}_{CE}$ of either 150 (rectangular) or 170 (circular) $\text{km}$ $\text{s}^{-1}$ for a ``solid'' (optically-thick) emitter, or as a $\delta {V}_{h}$ ranging from 100 (inverse-square density profile) to 120 (constant density) $\text{km}$ $\text{s}^{-1}$ for an optically-thin spherical emitter. It is clear, however, that most of these differences are too subtle relative to the sensitivity of our data and the simplicity of the model. The key takeaway from this discussion is that the $\delta V$ fit-parameter in the ${R}_{CE}$-model (as constructed in Section \ref{sec:rce_model}) under-estimates the $\delta {V}_{CE}$ physical quantity by $\sim 50-70\%$ in the case of solid emitters, while it is representative of the half-light radius in optically-thin emitters.



\end{document}